\documentclass[a4paper]{article}
\usepackage{amssymb}
\usepackage{amsmath}

\input{tcilatex}

\begin{document}

\author{Sawa Manoff \\
\textit{Bulgarian Academy of Sciences,}\\
\textit{Institute for Nuclear Research and Nuclear Energy,}\\
\textit{Department of Theoretical Physics,}\\
\textit{Blvd. Tzarigradsko Chaussee 72,}\\
\textit{1784 Sofia - Bulgaria}}
\date{E-mail address: smanov@inrne.bas.bg}
\title{\textbf{Propagation of signals in spaces with affine connections and
metrics}}
\maketitle

\begin{abstract}
\textit{The propagation of signals in space-time is considered} \textit{on
the basis of the notion of null (isotropic) vector field in spaces with
affine connections and metrics [}$(\overline{L}_{n},g)$\textit{-spaces] as
models of space or space-time. The Doppler effect is generalized for these
types of spaces. The notions of aberration, standard (longitudinal) Doppler
effect, and transversal Doppler effect are introduced. On their grounds, the
Hubble effect appears as Doppler effect with explicit forms of the
centrifugal (centripetal) and Coriolis velocities and accelerations in
spaces with affine connections and metrics. Doppler effect, Hubble effect,
and aberration could be used in mechanics of continuous media and in other
classical field theories in the same way as the standard Doppler effect is
used in classical and relativistic mechanics.}

\end{abstract}

\section{Introduction}

1. Modern problems of relativistic astrophysics as well as of relativistic
physics (dark matter, dark energy, evolution of the universe, measurement of
velocities of moving objects etc.) are related to the propagation of signals
in space or in space-time. The basis of experimental data received as
results of observations of the Doppler effect or of the Hubble effect gives
rise to theoretical considerations about the theoretical status of effects
related to detecting signals from emitters moving relatively to observers
carrying detectors in their laboratories.

2. In classical physics, the Doppler effect is considered only as a
longitudinal effect in a continuous media or in vacuum.\ It is assumed that
the reason for this effect is the relative velocity between emitter and
observer. In acoustics, the signals are considered as propagating in a
continuous media.

3. In relativistic physics, the Doppler effect is considered in special
relativity as (standard) longitudinal and transversal effect caused by the
motion of electromagnetic signals in vacuum with respect to an observer
(detector). In general relativity, the Doppler effect is related to the
propagation of light in astrophysical systems and to the existence of the
red shift relation due to the Hubble effect and Hubble law. The question
arises if the theoretical basis for the use of Doppler effect and Hubble
effect as tools for check-up of theoretical models in astrophysics and
relativistic physics in sophisticated models of space-time such as spaces
with affine connections and metrics [$(\overline{L}_n,g)$-spaces] is
sufficiently work out. It has been recently shown that every classical
(non-quantized) field theory could be considered as a theory of continuous
media \cite{Manoff-1} $\div $ \cite{Manoff-2a}. On this basis, the
propagation of signals in different models of space or of space-time is
worth being investigated. From this point of view, questions arise as how
Doppler effect is related to the Hubble effect from point of view of the
kinematic characteristics of a continuous media and, especially, is there a
relation between the Doppler and Hubble effects and the relative
accelerations between emitters and detectors. In a previous paper \cite%
{Manoff-7} Doppler and Hubble effects are considered on the basis of
dimension preconditions with relations to the relative velocity between an
emitter and an observer. In this paper we will use the properties of a
covariant exponential operator for finding out the change of a null
(isotropic) vector field along the world line of an observer (detector). It
is assumed that the null vector field is related to the propagation of a
signal when at a given time moment in the frame of the observer the emitter
and the observer are at rest. After that moment the observer could detect
the relative motion of the emitter and observe the frequency shifts of its
signals.

4. The notion of null (isotropic) vector field is related to the light
propagation described in relativistic electrodynamics on the basis of
special and general relativity theories \cite{Stephani} $\div $ \cite{Misner}%
. On the other side, the notion of null (isotropic) vector field could be
considered in spaces with (definite) or indefinite metric as a geometric
object (contravariant vector field) with specific properties making it
useful in the description of the propagation of signals in space or in
space-time as well as in geometrical optics based on different mathematical
models. Usually, it is assumed that a signal is propagating with limited
velocity through a continuous media or in vacuum. The velocity of
propagation of signals could be a constant quantity or a non-constant
quantity depending on the properties of the space or the space-time, where
the signals are transmitted and propagated.

5. In the present paper the notion of contravariant null (isotropic) vector
field is introduced and considered in spaces with affine connections and
metrics [$(\overline{L}_n,g)$-spaces]. In Section 2 the properties of null
vector fields are considered on the basis of $(n-1)+1$ representation of
non-null (non-isotropic) vector fields orthogonal to each other. In Section
3 the notions of distance and space velocity are discussed and their
relations to null vector fields are investigated. In Section 4 the kinematic
effects [aberration, longitudinal and transversal Doppler effects, and
Hubble effect] related to the kinematic characteristics of the relative
velocity and the relative acceleration as well as their connections with
null vector fields are considered. In Section 4 the kinematic effects
related to the relative velocity and the relative acceleration are recalled.
In Section 5 the aberration of signals is considered as corollary of the
change of a null vector field along the world line of an observer. In
Section 6 the different types of Doppler effect are introduced and
investigated.. In Section 7 the Hubble effect as Doppler effect with
explicitly given forms of the relative velocities and the relative
accelerations is considered. It is shown that the Hubble effect appears as a
corollary of the standard (longitudinal) and transversal Doppler effects. On
the other side, the Hubble effect is closely related to centrifugal
(centripetal) and Coriolis velocities and accelerations. The results
discussed in the paper could be important from the point of view of the
possible applications of kinematic characteristics in continuous media
mechanics as well as in classical (non-quantum) field theories in spaces
with affine connections and metrics. Section 8 comprises concluding remarks.

The main results in the paper are given in details (even in full details)
for these readers who are not familiar with the considered problems. The
definitions and abbreviations are identical to those used in \cite{Manoff-2}
and \cite{Manoff-2a}. The reader is kindly asked to refer to them for more
details and explanations of the statements and results only cited in this
paper.

\section{Null (isotropic) vector fields. Definition and properties}

\subsection{Definition of a null (isotropic) vector field}

Let us now consider a space with affine connections and metrics [$(\overline{%
L}_n,g)$-space] \cite{Manoff-3}, \cite{Manoff-3a} as a model of a space or
of a space-time. In this space the length $l_v$ of a contravariant vector
field $v\in T(M)$ is defined by the use of the covariant metric tensor field
(covariant metric) $g\in \otimes _{2s}(M)$ as 
\begin{equation}
g(v,v)=\pm l_v^2\,\,\,\,\,\,\text{,\thinspace \thinspace \thinspace
\thinspace \thinspace \thinspace \thinspace \thinspace \thinspace \thinspace
\thinspace \thinspace \thinspace \thinspace \thinspace \thinspace }l_v^2\geq
0\text{ \thinspace .}  \label{1.1}
\end{equation}

\textit{Remark}. The sign before $l_v^2$ depends on the signature $Sgn$ of
the covariant metric $g$. $M$ is differentiable manifold, $dim\,M=n$. A $(%
\overline{L}_n,g)$-space is a differentiable manifold $M$ provided with
contravariant and covariant affine connections (whose components differ not
only by sign) and metrics. $T(M)=\cup _{x\in M}T_x(M)$. $T_x(M)$ is the
tangent space at a point $x\in M$. $\otimes _{2s}(M)$ is the space of
covariant symetric tensors $g$ of second rank with $det\,g\neq 0$ over $M$.

The contravariant vector fields can be divided into two classes with respect
to their lengths:

\begin{itemize}
\item null or isotropic vector fields with length $l_v=0$,

\item non-null or non-isotropic vector fields with length $l_v\neq 0$.
\end{itemize}

In the case of a positive definite covariant metric $g$ ($Sgng=\pm n$, $%
dimM=n$) the null (isotropic) vector field is identically equal to zero,
i.e. if $l_v=0$ then $v=v^i\cdot e_i\equiv \mathbf{0}\in T(M)$,\thinspace
\thinspace $v^i\equiv 0$.

In the case of an indefinite covariant metric $g$ ($Sgng<n$ or $Sgng>-n$, $%
dimM=n$) the null (isotropic) vector field with equal to zero length $l_v=0$
can have different from zero components in an arbitrary given basis, i.e. it
is not identically equal to zero at the points, where it has been defined,
i.e. if $l_v=0$ then $v\neq \mathbf{0}\in T(M)$, $v=v^i\cdot e_i\in T(M)$
and $v^i\neq 0$. In a $(\overline{L}_n,g)$-space the components $g_{ij}$ of
a covariant metric tensor $g$ could be written in a local co-ordinate system
at a given point of the space as $g_{ij}=(\underset{k\,\,\,times}{%
\underbrace{-1,\,-1,\,-1,\,...,}}\,\underset{l\,\,\,\,\,times}{\underbrace{%
+1,\,+1,\,+1,\,...}})$ with $k+l=n$.

The signature $Sgn$ of $g$ is defined as 
\begin{equation}
Sgn\,\,g=-k+l=2\cdot l-n=n-2\cdot k\,\,\,\,\text{, \thinspace \thinspace
\thinspace \thinspace }n,\,k,\,l\in \mathbf{N}\text{,}  \label{1.2}
\end{equation}
\thinspace \thinspace \thinspace where $k=n-l$,\thinspace \thinspace
\thinspace $l=n-k$.

In the relativistic physics for $dim\,M=4$, the number $l$ and $k$ are
chosen as $l=1$, $k=3$ or $l=3$, $k=1$ so that $Sgn\,g=-2\sim $ $%
(-1,\,-1,\,-1,\,+1)$ or $Sgn\,g=+2$ $\sim (+1,\,+1,\,+1,\,-1)$. In general,
a $(\overline{L}_n,g)$-space could be consider as a model of space-time with 
$Sgn\,g<0$ and $(k>l$, $l=1)$ or with $Sgn\,g>0$ and $(l>k$, $k=1)$.

The non-null (non-isotropic) contravariant vector fields are divided into
two classes.

1. For $Sgn\,g<0$

(a) $g(v,v)=+l_v^2>0$ $:=$ time like vector field $v\in T(M)$,

(b) $g(v,v)=-l_v^2<0$ $:=$ space like vector field $v\in T(M)$.

2. For $Sgn\,g>0$

(a) $g(v,v)=-l_v^2<0$ $:=$ time like vector field $v\in T(M)$,

(b) $g(v,v)=+l_v^2>0\,:=$ space like vector field $v\in T(M)$.

Therefore, if we do not fix a priory the signature of the space-time models
we can distinguish a \textit{time like vector field} $u$ with

\begin{center}
\begin{tabular}{ll}
$g(u,u)=+l_u^2\text{ \thinspace \thinspace \thinspace \thinspace \thinspace
for}$ & $Sgn\,g<0$ \\ 
$\,\,\,\,\,\,\,\,\,\,\,\,\,\,\,\,\,\,\,=-l_u^2\,\,\,\,\,\,\,\,$for & $%
Sgn\,g>0$%
\end{tabular}
\end{center}

or $g(u,u)=\pm l_u^2$, and a \textit{space like vector field} $\xi _{\perp }$
with

\begin{center}
\begin{tabular}{ll}
$g(\xi _{\perp },\xi _{\perp })=-l_{\xi _{\perp }}^2\text{ \thinspace
\thinspace \thinspace \thinspace \thinspace for}$ & $Sgn\,g<0$ \\ 
$\,\,\,\,\,\,\,\,\,\,\,\,\,\,\,\,\,\,\,=+l_{\xi _{\perp
}}^2\,\,\,\,\,\,\,\,\,\,\,\,\,\,\,\,\,$for & $Sgn\,g>0$%
\end{tabular}
\end{center}

or $g(\xi _{\perp },\xi _{\perp })=\mp l_{\xi _{\perp }}^2$. This means that
in symbols $\pm l_{\diamond }^2$ or $\mp l_{\diamond }^2$ \thinspace
\thinspace $(\diamond \in T(M))$ the sign above is related to $Sgn\,g<0$ and
the sign below is related to $Sgn\,g>0$.

\textit{Remark}. Since $l_{\diamond }=\pm \sqrt{l_{\diamond }^2}$, the sings
in this case will be denoted as not related to the signature of the metric $%
g $. Usually, it is assumed that $l_{\diamond }=+\sqrt{l_{\diamond }^2}\geq
0 $.

A non-null (non-isotropic) contravariant vector field $v$ could be
represented by its length $l_v$ and its corresponding unit vector $n_v=\frac
v{l_v}$ as $v=\pm l_v\cdot n_v$ in contrast to a null (isotropic) vector
field $\widetilde{k}$ with $l_{\widetilde{k}}=0\,$\thinspace \thinspace (the
sings here are not related to the signature of the metric $g$)$\,$%
\begin{equation*}
v=\pm l_v\cdot n_v\text{ \thinspace \thinspace \thinspace \thinspace
,\thinspace \thinspace \thinspace \thinspace \thinspace \thinspace
\thinspace \thinspace }g(v,v)=l_v^2\cdot g(n_v,n_v)=\pm l_v^2\text{%
\thinspace \thinspace \thinspace \thinspace \thinspace ,\thinspace
\thinspace \thinspace \thinspace \thinspace \thinspace \thinspace \thinspace 
}g(n_v,n_v)=\pm 1\,\,\,\text{,}
\end{equation*}
for time like unit vector field $n_v\,$or 
\begin{equation*}
v=\mp l_v\cdot n_v\text{ \thinspace \thinspace \thinspace \thinspace
,\thinspace \thinspace \thinspace \thinspace \thinspace \thinspace
\thinspace \thinspace }g(v,v)=l_v^2\cdot g(n_v,n_v)=\mp l_v^2\text{%
\thinspace \thinspace \thinspace \thinspace \thinspace ,\thinspace
\thinspace \thinspace \thinspace \thinspace \thinspace \thinspace \thinspace 
}g(n_v,n_v)=\mp 1\,\,\,\text{,}
\end{equation*}
for space like unit vector field $n_v$.

\textit{Remark}. In the experimental physics, the measurements are related
to the lengths and to the directions of a non-null (non-isotropic) vector
field with respect to a frame of reference. Since different types of
co-ordinates could be used in a frame of reference, the components of a
vector field related to these co-ordinates cannot be considered as invariant
characteristics of the vector field and on this grounds the components
cannot be important characteristics for the vector fields.

After these preliminary remarks, we can introduce the notion of a null
(isotropic) vector field

\textit{Definition 1.} A contravariant vector field $\widetilde{k}$ with
length zero is called null (isotropic) vector field, i.e. $\widetilde{k}:=$
null (isotropic) vector field if 
\begin{equation}
g(\widetilde{k},\widetilde{k})=\pm l_{\widetilde{k}}^2=0\,\,\,\,\,\text{%
,\thinspace \thinspace \thinspace \thinspace \thinspace \thinspace
\thinspace \thinspace \thinspace \thinspace \thinspace \thinspace }l_{%
\widetilde{k}}=\mid g(\widetilde{k},\widetilde{k})\mid ^{1/2}=0\,\,\,\,\text{%
.}  \label{1.3}
\end{equation}

\subsection{Properties of a null (isotropic) vector field}

The properties of a null (isotropic) contravariant vector field could be
considered in a $(n-1)+1$ invariant decomposition of a space-time by the use
of two non-isotropic contravariant vector fields $u$ and $\xi _{\perp }$,
orthogonal to each other \cite{Manoff-3}, i.e. $g(u,\xi _{\perp })=0$. The
contravariant vectors $u$ and $\xi _{\perp }$ are essential elements of the
structure of a frame of reference \cite{Manoff-4} in a space-time.

\subsubsection{Invariant representation of a null vector field by the use of
a non-null contravariant vector field}

(a) \textit{Invariant projections of a null vector field along and
orthogonal to a non-null (non-isotropic) contravariant vector field} $u$

Every contravariant vector field $\widetilde{k}\in T(M)$ could be
represented in the form 
\begin{equation}
\widetilde{k}=\frac 1e\cdot g(u,\widetilde{k})\cdot u+\overline{g}[h_u(%
\widetilde{k})]=k_{\parallel }+k_{\perp }\,\,\,\,\text{,}  \label{1.4}
\end{equation}
where 
\begin{eqnarray}
e &=&g(u,u)=\pm l_u^2\text{ \thinspace \thinspace \thinspace \thinspace
,\thinspace \thinspace \thinspace \thinspace \thinspace \thinspace
\thinspace \thinspace \thinspace }  \notag \\
\overline{g} &=&g^{ij}\cdot \partial _i.\partial _j\,\,\,\,\,\text{%
,\thinspace \thinspace \thinspace \thinspace \thinspace \thinspace
\thinspace \thinspace \thinspace \thinspace }\partial _i.\partial _j=\frac
12\cdot (\partial _i\otimes \partial _j+\partial _j\otimes \partial
_i)\,\,\,\,\text{,}  \notag \\
g &=&g_{ij}\cdot dx^i.dx^j\,\,\,\,\text{,\thinspace \thinspace \thinspace }%
dx^i.dx^j=\frac 12\cdot (dx^i\otimes dx^j+dx^j\otimes dx^i)\text{ \thinspace
,}  \notag \\
h_u &=&g-\frac 1e\cdot g(u)\otimes g(u)\,\,\,\,\,\text{,\thinspace
\thinspace \thinspace \thinspace \thinspace }h^u=\overline{g}-\frac 1e\cdot
u\otimes u\,\,\,\,\,\text{,}  \notag \\
g(u) &=&g_{i\overline{j}}\cdot u^j\cdot dx^i\,\,\,\text{,\thinspace
\thinspace \thinspace }  \notag \\
\text{\thinspace \thinspace \thinspace \thinspace }\overline{g}[h_u(%
\widetilde{k})] &=&g^{ij}\cdot h_{\overline{j}\overline{l}}\cdot \widetilde{k%
}\,^l\cdot \partial _i:=k_{\perp }\,\,\,\text{,\thinspace \thinspace
\thinspace \thinspace \thinspace \thinspace }k_{\parallel }:=\frac 1e\cdot
g(u,\widetilde{k})\cdot u\,\,\,\,\text{.}  \label{1.5}
\end{eqnarray}
\begin{equation}
g(k_{\parallel },k_{\perp })=0\,\,\,\,\,\,\text{,\thinspace \thinspace
\thinspace \thinspace \thinspace \thinspace \thinspace \thinspace \thinspace
\thinspace \thinspace \thinspace }g(u,k_{\perp })=0\,\,\,\,\,\,\,\text{.}
\label{1.6}
\end{equation}

Let us now take a closer look at the first term $k_{\parallel }$ of the
representation of $\widetilde{k}$. 
\begin{equation}
k_{\parallel }=\frac 1e\cdot g(u,\widetilde{k})\cdot u=\pm \frac
1{l_u^2}\cdot g(u,\widetilde{k})\cdot u=\pm \frac 1{l_u}\cdot
g(u,k_{\parallel })\cdot \frac u{l_u}\text{ .}  \label{1.7}
\end{equation}

If we introduce the abbreviations 
\begin{equation}
n_{\parallel }=\frac{u}{l_{u}}\text{ \thinspace \thinspace ,\thinspace
\thinspace \thinspace \thinspace \thinspace \thinspace \thinspace \thinspace 
}\omega =g(u,\widetilde{k})=g(u,k_{\parallel })\,\,\,\,\,\,\,\text{,}
\label{1.8}
\end{equation}%
where 
\begin{equation}
g(n_{\parallel },n_{\parallel })=\frac{1}{l_{u}^{2}}\cdot g(u,u)=\frac{1}{%
l_{u}^{2}}\cdot (\pm l_{u}^{2})=\pm 1\text{ \thinspace \thinspace \thinspace
,}  \label{1.9}
\end{equation}%
\begin{eqnarray}
\omega  &=&g(u,\widetilde{k})=g(u,k_{\parallel }+k_{\perp
})=g(u,k_{\parallel })=l_{u}\cdot g(n_{\parallel },k_{\parallel })=
\label{1.10} \\
&=&l_{u}\cdot g(k_{\parallel },n_{\parallel })\,\,\,\text{,}  \notag \\
k_{\parallel } &:&=\,\pm l_{k_{\parallel }}\cdot n_{\parallel }\,\,\,\,\,%
\text{,\thinspace \thinspace \thinspace \thinspace \thinspace \thinspace
\thinspace \thinspace \thinspace }g(k_{\parallel },k_{\parallel
})=l_{k_{\parallel }}^{2}\cdot g(n_{\parallel },n_{\parallel })=
\label{1.10a} \\
&=&l_{k_{\parallel }}^{2}\cdot (\pm 1)=\pm l_{k_{\parallel }}^{2}\,\,\,\text{%
,}  \notag \\
g(k_{\parallel },n_{\parallel }) &=&\pm \,l_{k_{\parallel }}\cdot
g(n_{\parallel },n_{\parallel })=\pm \,l_{k_{\parallel }}\cdot (\pm
1)=l_{k_{\parallel }}=\frac{\omega }{l_{u}}\,\,\,\text{,}  \label{1.10b}
\end{eqnarray}%
then $k_{\parallel }$ could be expressed as (the signs are not related to
the signature of the metric $g$) 
\begin{equation}
k_{\parallel }=\pm \frac{\omega }{l_{u}}\cdot n_{\parallel }\,=\pm
\,l_{k_{\parallel }}\cdot n_{\parallel }\,\,\,\,\text{\thinspace \thinspace .%
}  \label{1.11}
\end{equation}

The vector $n_{\parallel }$ is a unit vector $[g(n_{\parallel },n_{\parallel
})=\pm 1]$ collinear to $u$ and, therefore, tangential to a curve with
parameter $\tau $ if $u=\frac d{d\tau }$.

The scalar invariant $\omega =g(u,\widetilde{k})$ is usually interpreted as
the frequency of the radiation related to the null vector field $\widetilde{k%
}$ and propagating with velocity $u$ with absolute value $l_u$ with respect
to the trajectory $x^i(\tau )$. In general relativity $l_u:=c$ and it is
assumed that the radiation is of electromagnetic nature propagating with the
velocity of light $c$ in vacuum. We will come back to this interpretation in
the next considerations.

The contravariant vector field $k_{\perp }$%
\begin{equation*}
k_{\perp }=\overline{g}[h_u(\widetilde{k})]
\end{equation*}
is orthogonal to $u$ (and $k_{\parallel }$ respectively) part of $\widetilde{%
k}$. Since 
\begin{eqnarray}
g(k_{\parallel },k_{\parallel }) &=&g(\pm \frac \omega {l_u}\cdot
n_{\parallel },\,\pm \frac \omega {l_u}\cdot n_{\parallel })=\frac{\omega ^2%
}{l_u^2}\cdot g(n_{\parallel },n_{\parallel })=\pm \frac{\omega ^2}{l_u^2}%
=\pm l_{k_{\parallel }}^2\,\,\,\,\,\text{,}  \label{1.12} \\
l_{k_{\parallel }} &=&\frac \omega {l_u}\text{ }>0\text{\thinspace
\thinspace \thinspace \thinspace ,\thinspace \thinspace \thinspace
\thinspace \thinspace \thinspace }l_{k_{\parallel }}^2=\frac{\omega ^2}{l_u^2%
}\,\,\,\,\text{,}  \label{1.13}
\end{eqnarray}
and \thinspace 
\begin{equation*}
g(\widetilde{k},\widetilde{k})=0\,\,\,\,\,\,\text{, \thinspace \thinspace
\thinspace \thinspace \thinspace \thinspace \thinspace \thinspace \thinspace 
}g(k_{\parallel },k_{\perp })=0\,\,\,\,\,\,\text{,\thinspace \thinspace }
\end{equation*}
we have for $g(k_{\perp },k_{\perp })$%
\begin{eqnarray}
g(\widetilde{k},\widetilde{k}) &=&0=g(k_{\parallel }+k_{\perp
},\,k_{\parallel }+k_{\perp })=g(k_{\parallel },k_{\parallel })+g(k_{\perp
},k_{\perp })=  \notag \\
&=&\pm \frac{\omega ^2}{l_u^2}+g(k_{\perp },k_{\perp })=\pm \frac{\omega ^2}{%
l_u^2}\mp l_{k_{\perp }}^2=0\,\,\,\,\,\text{,}  \label{1.15}
\end{eqnarray}
and, therefore, 
\begin{equation}
l_{k_{\perp }}^2=\frac{\omega ^2}{l_u^2}\,\,\,\,\,\,\,\,\text{,\thinspace
\thinspace \thinspace \thinspace \thinspace \thinspace \thinspace \thinspace
\thinspace \thinspace \thinspace \thinspace }l_{k_{\perp }}=\frac \omega
{l_u}=l_{k_{\parallel }}\,\,\,\,\text{.}  \label{1.16}
\end{equation}

\textit{Remark}. Since $\omega \geq 0$ and $l_u>0$, and at the same time $%
l_{k_{\perp }}>0$, and $l_{k_{\parallel }}>0$, we have 
\begin{equation*}
l_{k_{\parallel }}=\frac \omega {l_u}=l_{k_{\perp }}\,\,\,\,\text{.}
\end{equation*}

If we introduce the unit contravariant vector $\widetilde{n}_{\perp }$ with $%
g(\widetilde{n}_{\perp },\widetilde{n}_{\perp })=\mp 1$ then the vector $%
k_{\perp }$ could be written as 
\begin{equation}
k_{\perp }:=\mp \,l_{k_{\perp }}\cdot \widetilde{n}_{\perp }\,\,\,\,\,\,%
\text{,}  \label{1.17}
\end{equation}
where 
\begin{equation}
g(k_{\perp },k_{\perp })=l_{k_{\perp }}^2\cdot g(\widetilde{n}_{\perp },%
\widetilde{n})=\mp l_{k_{\perp }}^2=\mp \,\frac{\omega ^2}{l_u^2}\,\,\,\,%
\text{, \thinspace \thinspace \thinspace \thinspace \thinspace \thinspace }%
l_{k_{\perp }}^2=\frac{\omega ^2}{l_u^2}\,\,\,\,\,\text{,\thinspace
\thinspace \thinspace \thinspace \thinspace \thinspace \thinspace }%
l_{k_{\perp }}=\frac \omega {l_u}\,\,\text{.}  \label{1.18}
\end{equation}

Therefore, 
\begin{eqnarray}
k_{\perp } &=&\mp \frac \omega {l_u}\cdot \widetilde{n}_{\perp }\,\,\,\,\,\,%
\text{,\thinspace \thinspace \thinspace \thinspace \thinspace \thinspace
\thinspace \thinspace \thinspace \thinspace \thinspace \thinspace \thinspace
\thinspace }k_{\parallel }=\pm \frac \omega {l_u}\cdot n_{\parallel
}\,\,\,\,\,\,\,\text{,}  \label{1.19} \\
\widetilde{k} &=&k_{\parallel }+k_{\perp }=\pm \frac \omega {l_u}\cdot
(n_{\parallel }-\widetilde{n}_{\perp })\,\,\,\,\,\,\text{,}  \label{1.20}
\end{eqnarray}
where 
\begin{equation}
g(n_{\parallel },\widetilde{n}_{\perp })=0\,\,\,\,\,\text{,\thinspace
\thinspace \thinspace \thinspace \thinspace \thinspace \thinspace \thinspace 
}g(k_{\parallel },\xi _{\perp })=0\,\,\,\,\,\text{,\thinspace \thinspace
\thinspace \thinspace \thinspace \thinspace \thinspace \thinspace \thinspace 
}g(u,k_{\perp })=0\,\,\,\,\,\,\text{,}  \label{1.21}
\end{equation}
\begin{eqnarray}
g(n_{\parallel },k_{\parallel }) &=&\pm l_{k_{\parallel }}\cdot
g(n_{\parallel },n_{\parallel })=l_{k_{\perp }}=\frac \omega
{l_u}\,\,\,\,\,\,\text{,}  \label{1.22} \\
g(\widetilde{n}_{\perp },k_{\perp }) &=&\mp l_{k_{\perp }}\cdot g(n_{\perp
},n_{\perp })=\frac \omega {l_u}\,\,=l_{k_{\parallel }}=l_{k_{\perp
}}\,\,\,\,\,\text{,}  \label{1.23}
\end{eqnarray}
\begin{equation}
g(n_{\parallel },k_{\parallel })=g(\widetilde{n}_{\perp },k_{\perp })=\frac
\omega {l_u}=l_{k_{\parallel }}=l_{k_{\perp }}\,\,\,\,\,\,\text{.}
\label{1.23a}
\end{equation}

\textit{Remark}. The signs, not related to the metric $g$, are chosen so to
be the same with the signs related to the metric $g$.

We have now the relations 
\begin{equation}
\omega =g(u,\widetilde{k})=l_u\cdot g(n_{\parallel },k_{\parallel
})=l_u\cdot g(\widetilde{n}_{\perp },k_{\perp })\,\,\,\,\,\text{.}
\label{1.24}
\end{equation}

If $\widetilde{n}_{\perp }$ is interpreted as the unit vector in the
direction of the propagation of a signal in the subspace orthogonal to the
contravariant vector field $u$ and $l_u$ is interpreted as the absolute
value of the velocity of the radiated signal then $l_u\cdot \widetilde{n}%
_{\perp }$ is the path along $\widetilde{n}_{\perp }$ propagated by the
signal in a unit time interval. Then 
\begin{equation}
\omega =g(u_{\perp },k_{\perp })\text{ \thinspace \thinspace \thinspace
\thinspace ,\thinspace \thinspace \thinspace \thinspace \thinspace
\thinspace \thinspace \thinspace }u_{\perp }:=l_u\cdot \widetilde{n}_{\perp
}\,\,\,\,\,\,\text{,\thinspace \thinspace \thinspace \thinspace \thinspace
\thinspace \thinspace \thinspace \thinspace \thinspace }g(u,u_{\perp
})=0\,\,\,\,\,\text{.}  \label{1.25}
\end{equation}

\subsubsection{Explicit form of the vector field $k_{\perp }$ and its
interpretation}

Let us now consider more closely the explicit form of $k_{\perp }$%
\begin{equation}
k_{\perp }=\mp l_{k_{\perp }}\cdot \widetilde{n}_{\perp }=\mp \frac \omega
{l_u}\cdot \widetilde{n}_{\perp }\text{ \thinspace \thinspace \thinspace .}
\label{1.26}
\end{equation}

(a) In $3$-dimensional Euclidean space (as model of space-time of the
Newtonian mechanics) the wave vector $\overrightarrow{k}$ is defined as 
\begin{equation}
\overrightarrow{k}=\frac{2\pi }\lambda \cdot \overrightarrow{n}\,\,\,\text{,}
\label{1.27}
\end{equation}
where $\overrightarrow{n}$ is the unit $3$-vector in the direction of
propagation of a signal with absolute value of its velocity $l_u=\lambda
\cdot \nu $. If we express $\lambda $ by $\lambda =l_u/\nu $ and put the
equivalent expression in this for $\overrightarrow{k}$ we obtain the
expression 
\begin{equation}
\overrightarrow{k}=\frac{2\pi \cdot \nu }{l_u}\cdot \overrightarrow{n}=\frac
\omega {l_u}\cdot \overrightarrow{n}\,\,\,\,\text{,}  \label{1.28}
\end{equation}
which (up to a sign depending on the signature of the metric $g$) is
identical with the expression for $k_{\perp }$ for $n=3$ if $k_{\perp }=%
\overrightarrow{k}$, $\overrightarrow{n}=\widetilde{n}_{\perp }$,\thinspace
and $\omega =2\cdot \pi \cdot \nu $.

(b) In $4$-dimensional (pseudo) Riemannian space (as a model of space-time
of the Einstein theory of gravitation) $l_u$ is interpreted as the absolute
value of the velocity of light in vacuum (normalized by some authors to $1$%
), i.e. $l_u=c$, $1$. Then 
\begin{equation}
k_{\perp }=\mp \frac \omega c\cdot \widetilde{n}_{\perp }=\mp \frac{2\cdot
\pi \cdot \nu }{\lambda \cdot \nu }\cdot \widetilde{n}_{\perp }=\mp \frac{%
2\cdot \pi }\lambda \cdot \widetilde{n}_{\perp }  \label{1.29}
\end{equation}
and we obtain the expression for the wave vector of light propagation in
general relativity, where $\widetilde{n}_{\perp }$ is the unit vector along
the propagation of light in the corresponding $3$-dimensional subspace of an
observer with world line $x^i(\tau )$ if 
\begin{equation}
u=\frac d{d\tau }=l_u\cdot n_{\parallel }\,\,\,\,\,\,\,\text{,\thinspace
\thinspace \thinspace \thinspace \thinspace \thinspace \thinspace \thinspace
\thinspace \thinspace \thinspace }\,n_{\parallel }=\frac 1{l_u}\cdot \frac
d{d\tau }\text{ \thinspace \thinspace \thinspace .}  \label{1.30}
\end{equation}

$l_u$ is the velocity of light measured by the observer.

(c) In the general case for $k_{\perp }$ as 
\begin{equation}
k_{\perp }=\mp \frac \omega {l_u}\cdot \widetilde{n}_{\perp }\,\,\,\text{%
\thinspace \thinspace \thinspace \thinspace \thinspace ,}  \label{1.31}
\end{equation}
$\omega $ could also be interpreted as the frequency of a signal propagating
with velocity with absolute value $l_u$ in a frame of reference of an
observer with world line $x^i(\tau )$. The unit vector $\widetilde{n}_{\perp
}$ is the unit vector in the direction of the propagation of the signal in
the subspace orthogonal to the vector $u$. The velocity of the observer is
usually defined by the use of the parameter $\tau $ of the world line under
the assumption that $ds=l_u\cdot d\tau $, where $ds$ is the distance of the
propagation of a signal for the proper time interval $d\tau $ of the
observer 
\begin{equation}
u=\frac d{d\tau }=\frac d{\frac 1{l_u}\cdot ds}=l_u\cdot \frac d{ds}\,\,\,%
\text{.}  \label{1.32}
\end{equation}

\textit{Remark}. Usually the velocity of a particle (observer) moving in
space-time is determined by its velocity vector field $u=\frac d{d\tau }$,
where $\tau $ is the proper time of the observer. The parameter $\tau $ is
considered as a parameter of observer's world line $x^i(\tau )$. By the use
of $u$ and its corresponding projection metrics $h_u$ and $h^u$ a
contravariant (non-null, non-isotropic) vector field $\xi $ could be
represented in two parts: one part is collinear to $u$ and the other part is
orthogonal to $u$%
\begin{equation}
\xi =\frac 1e\cdot g(\xi ,u)\cdot u+\overline{g}[h_u(\xi )]=\xi _{\parallel
}+\xi _{\perp }\,\,\,\text{,}  \label{1.33}
\end{equation}
where 
\begin{equation}
\xi _{\parallel }=\frac 1e\cdot g(\xi ,u)\cdot u\,\,\,\,\,\,\,\text{%
,\thinspace \thinspace \thinspace \thinspace \thinspace \thinspace
\thinspace \thinspace \thinspace \thinspace }\xi _{\perp }=\text{\thinspace }%
\overline{g}[h_u(\xi )]\,\,\,\,\,\text{,\thinspace \thinspace \thinspace
\thinspace \thinspace \thinspace \thinspace \thinspace \thinspace \thinspace
\thinspace \thinspace }g(\xi _{\parallel },\,\xi _{\perp })=0\,\,\,\,\text{.}
\label{1.34}
\end{equation}

1. If an observer is moving with velocity $v=\frac d{d\overline{\tau }}$ on
his world line $x^i(\overline{\tau })$ then his velocity, considered with
respect to the observer with velocity $u$ and world line $x^i(\tau )$, will
have two parts $v_{\parallel }$ and $v_{\perp }$, collinear and orthogonal
to $u$ $\,$respectively at the cross point $\tau =\overline{\tau }$ of both
the world lines $x^i(\tau )$ and $x^i(\overline{\tau })$%
\begin{equation}
v=\frac 1e\cdot g(v,u)\cdot u+\overline{g}[h_u(v)]=v_{\parallel }+v_{\perp
}\,\,\,\,\,\,\text{.}  \label{1.35}
\end{equation}

The vector $v_{\parallel }$ describes the motion of the observer with
velocity $v$ along the world line of the first observer with velocity $u$.
The vector $v_{\perp }$ describes the motion of the second observer with
velocity $v$ in direction orthogonal to the world line of the first
observer. The vector $v_{\perp }$ is the velocity of the second observer in
the space of the first observer in contrast to the vector $v_{\parallel }$
describing the change of $v$ in the time of the first observer.

2. If we consider the propagation of a signal, characterized by its null
vector field $\widetilde{k}$, the interpretation of the vector field $u$,
tangential to the world line of an observer, changes. The vector field $%
u=l_u\cdot n_{\parallel }$ is interpreted as the velocity vector field of
the signal, propagating in the space-time and measured by the observer at
its world line $x^i(\tau )$ with proper time $\tau $ as a parameter of this
world line. The absolute value $l_u$ of $u$ is the size of the velocity of
the signal measured along the unit vector field $n_{\parallel }$ collinear
to the tangent vector of the world line of the observer.

3. In Einstein's theory of gravitation (ETG) both interpretations of the
vector field $u$ are put together. On the one side, the vector field $u$ is
interpreted as the velocity of an observer on his world line with parameter $%
\tau $ interpreted as the proper time of the observer. On the other side,
the length $l_u$ of the vector field $u$ is normalized either to $\pm 1$ or
to $\pm c=$ const. The quantity $c$ is interpreted as the light velocity in
vacuum. The basic reason for this normalization is the possibility for
normalization of every non-null (non-isotropic) vector field $u$ in the form 
\begin{equation}
n_u=\frac u{l_u}=n_{\parallel }\text{ \thinspace \thinspace \thinspace
\thinspace \thinspace ,\thinspace \thinspace \thinspace \thinspace
where\thinspace \thinspace \thinspace \thinspace \thinspace \thinspace
\thinspace \thinspace \thinspace \thinspace \thinspace \thinspace \thinspace
\thinspace }l_u=\mid g(u,u)\mid ^{1/2}\,\neq 0\,\,\,\,\,\,\text{,}
\label{1.36}
\end{equation}
by the use of its different from zero length $(l_u\neq 0)$, defined by means
of the covariant metric tensor $g$.

Both the interpretations of the vector field $u$ (as a velocity of an
observer or as velocity of a signal) should be considered separately from
each other for avoiding ambiguities. The identification of the
interpretations should mean that we assume the existence of an observer
moving in space-time with velocity $u$ and emitting (or receiving) signals
with the same velocity. Such assumption does not exist in the Einstein
theory of gravitation. This problem is worth to be investigated and a clear
difference between both interpretations should be found. It is related to
the notions of distance and of velocity in spaces with affine connections
and metrics.

\section{Distance and velocity in a $(\overline{L}_n,g)$-space}

\subsection{Distance in a $(\overline{L}_n,g)$-space and its relations to
the notion of velocity}

\subsubsection{Distance in a $(\overline{L}_n,g)$-space for world and space
lines not depending to each other}

1. The distance in a $(\overline{L}_n,g)$-space between a point $P\in M$
with co-ordinates $x^i$ and a point $\overline{P}\in M$ with co-ordinates $%
\overline{x}^i=x^i+dx^i$ is determined by the length of the ordinary
differential $d$, considered as a contravariant vector field $d=dx^i\cdot
\partial _i$ \cite{Manoff-2}. If we denote the distance as $ds$ between
point $P$ and point $\overline{P}$ then the square $ds^2$ of $ds$ could be
defined as the square of the length of the ordinary differential $d$%
\begin{equation}
ds^2=g(d,d)=\pm l_d^2=g_{\overline{i}\overline{j}}\cdot dx^i\cdot dx^j\text{
\thinspace \thinspace \thinspace \thinspace \thinspace \thinspace
,\thinspace \thinspace \thinspace \thinspace \thinspace \thinspace
\thinspace \thinspace \thinspace \thinspace \thinspace \thinspace }l_d^2\geq
0\,\,\,\,\,\,\,\text{.}  \label{2.1}
\end{equation}

2. Let us now consider a two parametric congruence of curves (a set of not
intersecting curves) in a $(\overline{L}_{n},g)$-space 
\begin{equation}
x^{i}=x^{i}(\tau ,r(\tau ,\lambda ))=x^{i}(\tau ,\lambda )\text{ \thinspace
\thinspace \thinspace ,}  \label{2.2}
\end{equation}%
where the function $r=r(\tau ,\lambda )\in C^{r}(M)$, $r\geq 2$, depends on
the two parameters $\tau $ and $\lambda $, $\tau ,\lambda \in \mathbf{R}$.
Then 
\begin{equation}
dr=\frac{\partial r(\tau ,\lambda )}{\partial \tau }\cdot d\tau +\frac{%
\partial r(\tau ,\lambda )}{\partial \lambda }\cdot d\lambda \,\,\,\,\,
\label{2.2a}
\end{equation}%
and 
\begin{eqnarray}
dx^{i} &=&\frac{\partial x^{i}(\tau ,r(\tau ,\lambda ))}{\partial \tau }%
\cdot d\tau +\frac{\partial x^{i}(\tau ,r(\tau ,\lambda ))}{\partial r}\cdot
(\frac{\partial r(\tau ,\lambda )}{\partial \tau }\cdot d\tau +  \label{2.2b}
\\
&&+\frac{\partial r(\tau ,\lambda )}{\partial \lambda }\cdot d\lambda ) 
\notag \\
&=&[\frac{\partial x^{i}(\tau ,r(\tau ,\lambda ))}{\partial \tau }+\frac{%
\partial x^{i}(\tau ,r(\tau ,\lambda ))}{\partial r}\cdot \frac{\partial
r(\tau ,\lambda )}{\partial \tau }]\cdot d\tau +  \notag \\
&&+\frac{\partial x^{i}(\tau ,r(\tau ,\lambda ))}{\partial r}\cdot \frac{%
\partial r(\tau ,\lambda )}{\partial \lambda }\cdot d\lambda   \notag
\end{eqnarray}%
or 
\begin{equation}
dx^{i}=(u^{i}+\overline{\xi }\,^{i}\cdot l_{v})\cdot d\tau +\overline{\xi }%
\,^{i}\cdot \frac{\partial r}{\partial \lambda }\cdot d\lambda \,\,\,\,\,\,\,%
\text{,}  \label{2.2c}
\end{equation}%
where 
\begin{equation}
u^{i}=\frac{\partial x^{i}(\tau ,r(\tau ,\lambda ))}{\partial \tau }%
\,\,\,\,\,\text{,\thinspace \thinspace \thinspace \thinspace \thinspace
\thinspace \thinspace \thinspace \thinspace \thinspace }l_{v}=\frac{\partial
r(\tau ,\lambda )}{\partial \tau }\,\,\,\,\,\,\,\text{,\thinspace \thinspace
\thinspace \thinspace \thinspace \thinspace \thinspace \thinspace \thinspace 
}\overline{\xi }\,^{i}=\frac{\partial x^{i}(\tau ,r(\tau ,\lambda ))}{%
\partial r}\,\,\,\,\,\,\text{,\thinspace \thinspace }  \label{2.2d}
\end{equation}%
and 
\begin{eqnarray}
d &=&dx^{i}\cdot \partial _{i}=d\tau \cdot (u+l_{v}\cdot \overline{\xi }%
)+(\partial r/\partial \lambda )\cdot d\lambda \cdot \overline{\xi }%
\,\,\,\,\,\,\text{,}  \label{2.2e} \\
\frac{d\lambda }{d\tau } &=&0\,\,\,\,\,\text{,\thinspace \thinspace
\thinspace \thinspace \thinspace \thinspace \thinspace \thinspace }\frac{%
d\tau }{d\lambda }=0\,\,\,\,\,\text{.}  \notag
\end{eqnarray}

\textit{Remark}. Here, the parameters $\tau $ and $\lambda $ are assumed to
be independent to each other functions.

The change of the contravariant vector field $d$ under the change $d\tau $
of the parameter $\tau $ could be expressed in the form 
\begin{equation}
\frac d{d\tau }=\frac{dx^i}{d\tau }\cdot \partial _i=u+l_v\cdot \overline{%
\xi }=\overline{u}\,^i\cdot \partial _i=\overline{u}\,\,\,\,\,\text{%
,\thinspace \thinspace \thinspace \thinspace \thinspace \thinspace
\thinspace \thinspace \thinspace }\overline{u}^i=\frac{dx^i}{d\tau }%
\,\,\,\,\,\text{,}  \label{2.2f}
\end{equation}
where the relations are valid 
\begin{eqnarray}
g(\overline{u},u) &=&g(u,u)+l_v\cdot g(\overline{\xi },u)\,\,\,\,\text{,}
\label{2.2g} \\
g(\overline{u},\overline{\xi }) &=&g(u,\overline{\xi })+l_v\cdot g(\overline{%
\xi },\overline{\xi })\,\,\,\,\,\text{,}  \notag \\
g(\overline{u},\overline{u}) &=&g(u+l_v\cdot \overline{\xi },u+l_v\cdot 
\overline{\xi })=  \notag \\
&=&g(u,u)+2\cdot l_v\cdot g(u,\overline{\xi })+l_v^2\cdot g(\overline{\xi },%
\overline{\xi })\text{ .}  \notag
\end{eqnarray}

The contravariant vector field $\overline{u}=\overline{u}^i\cdot \partial _i$
is usually interpreted as the velocity of an observer moving in a space-time
described by a $(\overline{L}_n,g)$-space as its model. The contravariant
vector $u$ is a tangent vector field to the curve $x^i(\tau ,r(\tau ,\lambda
)=r_0=$ const.$)=x^i(\tau ,\lambda =\lambda _0=$ const.$)$%
\begin{eqnarray}
u &=&u^i\cdot \partial _i=\frac{\partial x^i}{\partial \tau }\cdot \partial
_i\,\,\,\,\text{,}  \label{2.2h} \\
\overline{u} &=&\frac 1{g(u,u)}\cdot g(\overline{u},u)\cdot u+\overline{g}%
[h_u(\overline{u})]\,\,\,\,\text{.}  \notag
\end{eqnarray}

The contravariant vector $\overline{\xi }$ is a collinear vector to the
tangent vector $\xi $ to the curve $x^i(\tau =\tau _0=$ const.$,\,r(\tau
_0,\lambda ))=x^i(\tau =\tau _0=$ const.$,\lambda )$. This is so because of
the relations 
\begin{equation}
\overline{\xi }=\overline{\xi }\,^i\cdot \partial _i=\frac{\partial x^i}{%
\partial r}\cdot \partial _i\,\,\,\,\,\,\,\,\text{,}\,\,  \label{2.2i}
\end{equation}
\begin{equation}
\frac{\partial x^i}{\partial r}=\frac{\partial x^i(\tau ,r(\tau ,\lambda ))}{%
\partial r}=\frac{\partial x^i(\tau ,\lambda (r,\tau ))}{\partial r}=\frac{%
\partial x^i}{\partial \lambda }\cdot \frac{\partial \lambda }{\partial r}%
=\xi ^i\cdot \frac{\partial \lambda }{\partial r}=\overline{\xi }^i\text{
\thinspace ,}  \label{2.2j}
\end{equation}
where 
\begin{eqnarray}
r &=&r(\lambda ,\tau )\,\,\,\,\text{,\thinspace \thinspace \thinspace
\thinspace \thinspace }\lambda =\lambda (\tau ,r)\text{ \thinspace
\thinspace \thinspace \thinspace \thinspace ,}  \notag \\
\overline{\xi } &=&\overline{\xi }\,^i\cdot \partial _i=\frac{\partial
\lambda }{\partial r}\cdot \xi ^i\cdot \partial _i=\frac{\partial \lambda }{%
\partial r}\cdot \xi \,\,\,\,\text{,\thinspace \thinspace \thinspace
\thinspace \thinspace \thinspace \thinspace \thinspace \thinspace }\xi =%
\frac{\partial x^i}{\partial \lambda }\cdot \partial _i\,\,\,\,\text{.}
\label{2.2k}
\end{eqnarray}

3. Further, since we wish to consider the vector field $u$ as the velocity
vector field of an observer moving at the curve $x^i(\tau ,\lambda =\lambda
_0=$ const.$)$, interpreted as his world line, the vector field $\xi $ (and $%
\overline{\xi }$ respectively) could be chosen to lie in the subspace
orthogonal to $u$, i.e. $u$ and $\overline{\xi }$ could obey the condition $%
g(u,\overline{\xi })=0$ and, therefore, $g(u,\xi )=0$, $\xi =\xi _{\perp }=%
\overline{g}[h_u(\xi )]$, and\thinspace \thinspace \thinspace \thinspace $%
\overline{\xi }=\overline{\xi }_{\perp }$.

4. In the next step, we could consider the vector field $\overline{\xi }$ as
a unit vector field in direction of the vector field $\xi $, i.e. 
\begin{eqnarray}
\overline{\xi }_{\perp } &=&n_{\perp }=\frac{\xi _{\perp }}{l_{\xi _{\perp }}%
}\,\,\,\,\,\,\text{,\thinspace \thinspace \thinspace \thinspace \thinspace
\thinspace \thinspace \thinspace \thinspace \thinspace \thinspace \thinspace
\thinspace \thinspace }g(u,n_{\perp })=0\,\,\,\,\,\,\text{,}  \label{2.2l} \\
g(\overline{\xi }_{\perp },\overline{\xi }_{\perp }) &=&g(n_{\perp
},n_{\perp })=\frac{1}{l_{\xi _{\perp }}^{2}}\cdot g(\xi _{\perp },\xi
_{\perp })=\mp \frac{1}{l_{\xi _{\perp }}^{2}}\cdot l_{\xi _{\perp
}}^{2}=\mp 1\text{ \thinspace \thinspace ,}  \label{2.2n} \\
g(\overline{\xi }_{\perp },\overline{\xi }_{\perp }) &=&g(\frac{\partial
\lambda }{\partial r}\cdot \xi _{\perp },\frac{\partial \lambda }{\partial r}%
\cdot \xi _{\perp })=(\frac{\partial \lambda }{\partial r})^{2}\cdot g(\xi
_{\perp },\xi _{\perp })=  \notag \\
&=&\mp (\frac{\partial \lambda }{\partial r})^{2}\cdot l_{\xi _{\perp
}}^{2}=\mp 1\,\,\,\,\,\text{,}  \label{2.2o}
\end{eqnarray}%
\begin{equation}
(\frac{\partial \lambda }{\partial r})^{2}\cdot l_{\xi _{\perp
}}^{2}=1\,\,\,\,\,\,\,\,\text{,\thinspace \thinspace \thinspace \thinspace
\thinspace \thinspace }l_{\xi _{\perp }}^{2}=\text{\thinspace \thinspace
\thinspace }(\frac{\partial \lambda }{\partial r})^{-2}\,\,\,\,\,\,\text{%
,\thinspace \thinspace \thinspace \thinspace \thinspace \thinspace
\thinspace \thinspace \thinspace \thinspace }l_{\xi _{\perp }}=\pm \text{%
\thinspace \thinspace \thinspace \thinspace \thinspace }(\frac{\partial
\lambda }{\partial r})^{-1}\,\,\,\,\,\text{.}  \label{2.2p}
\end{equation}

After all above considerations for $\xi _{\perp }$ and $\overline{\xi }%
_{\perp }$ we obtain the relations 
\begin{eqnarray}
g(\overline{u},u) &=&g(u,u)\,\,\,\,\,\,\text{,}  \label{2.2q} \\
g(\overline{u},\overline{\xi }_{\perp }) &=&l_v\cdot g(\overline{\xi }%
_{\perp },\overline{\xi }_{\perp })=l_v\cdot g(n_{\perp },n_{\perp })=\mp
l_v\,\,\,\,\text{,}  \notag \\
g(\overline{u},\overline{u}) &=&g(u,u)+l_v^2\cdot g(n_{\perp },n_{\perp })= 
\notag \\
&=&\pm l_u^2\mp l_v^2=\frac{ds^2}{d\tau ^2}\,\,\,\,\text{,}  \notag
\end{eqnarray}
\begin{equation}
\frac{ds^2}{d\tau ^2}=g(\frac d{d\tau },\frac d{d\tau })=\pm l_u^2\mp
l_v^2=\pm l_u^2\cdot (1-\frac{l_v^2}{l_u^2})\,\,\,\,\,\text{.}  \label{2.2r}
\end{equation}

Moreover, 
\begin{eqnarray}
dx^i &=&(u^i+l_v\cdot n_{\perp }^i)\cdot d\tau +\frac{\partial r}{\partial
\lambda }\cdot d\lambda \cdot n_{\perp }^i\,\,\,\,\,\text{,}  \label{2.2s} \\
d &=&d\tau \cdot (u+l_v\cdot n_{\perp })+\frac{\partial r}{\partial \lambda }%
\cdot d\lambda \cdot n_{\perp }=  \notag \\
&=&d\tau \cdot u+(d\tau \cdot l_v+\frac{\partial r}{\partial \lambda }\cdot
d\lambda )\cdot n_{\perp }=  \notag \\
&=&d\tau \cdot u+dr\cdot n_{\perp }\,\,\,\,\text{,}  \label{2.2t} \\
dr &=&d\tau \cdot l_v+\frac{\partial r}{\partial \lambda }\cdot d\lambda
\,\,\,\,\,\text{,}  \label{2.2u} \\
\frac{dr(\tau ,\lambda )}{d\tau } &=&l_v\,\,\,\,\,\text{,\thinspace
\thinspace \thinspace \thinspace \thinspace \thinspace \thinspace \thinspace
\thinspace }\frac{d\lambda }{d\tau }=0\,\,\,\,\,\text{,}  \label{2.2v} \\
\overline{u} &=&u+l_v\cdot n_{\perp }\,\,\,\,\,\,\text{,\thinspace
\thinspace \thinspace \thinspace \thinspace \thinspace \thinspace \thinspace
\thinspace }g(\overline{u},u)=g(u,u)\,\,\,\,\text{,}  \label{2.2x} \\
n_{\perp } &=&\overline{g}[h_u(n_{\perp })]\,\,\,\,\,\,\,\,\text{,}
\label{2.2y}
\end{eqnarray}

\textit{Remark}. If $l_v\neq 0$ then the vector $\overline{u}$ would have a
part $l_v\cdot n_{\perp }$ orthogonal to $u$.

\begin{eqnarray*}
ds^2 &=&g(d,d)=d\tau ^2\cdot g(u,u)+(d\tau \cdot l_v+\frac{\partial r}{%
\partial \lambda }\cdot d\lambda )^2\cdot g(n_{\perp },n_{\perp })= \\
&=&\pm d\tau ^2\cdot l_u^2\mp (d\tau \cdot l_v+\frac{\partial r}{\partial
\lambda }\cdot d\lambda )^2=\pm d\tau ^2\cdot l_u^2\mp d\tau ^2\cdot l_v^2=
\\
&=&\pm d\tau ^2\cdot l_u^2\mp dr^2=\pm (l_u^2\cdot d\tau ^2-dr^2)=\pm d\tau
^2\cdot (l_u^2-\frac{dr^2}{d\tau ^2})= \\
&=&\pm d\tau ^2\cdot (l_u^2-l_v^2)\,\,\,\,\,\text{,}
\end{eqnarray*}

Therefore, 
\begin{eqnarray}
ds^2 &=&g(d,d)=\pm d\tau ^2\cdot (l_u^2-l_v^2)\,\,\,\,\,\,\,\text{,}
\label{2.3} \\
\frac{ds^2}{d\tau ^2} &=&g(\frac d{d\tau },\frac d{d\tau })=g(\overline{u},%
\overline{u})=\pm l_u^2\cdot (1-\frac{l_v^2}{l_u^2})\,\,\,\,\,\text{,}
\label{2.4}
\end{eqnarray}
\begin{equation}
ds^2=\pm l_u^2\cdot d\tau ^2\cdot (1-\frac{l_v^2}{l_u^2})\,\,\,\,\,\,\,\,\,%
\text{.}\,\,\,\,  \label{2.4a}
\end{equation}

5. In the non-relativistic field theories the distance between two points $%
P\in M$ and $\overline{P}\in M$ is defined as 
\begin{equation}
ds^2=\mp dr^2\,\,\,\,\,\,  \label{2.5}
\end{equation}
and $l_u=0$. This means that the distance between two neighboring points $P$
and $\overline{P}$ is the \textit{space} distance measured between them in
the rest (proper) reference frame of the observer (with absolute value $l_u$
of his velocity $u$ equal to zero). The time parameter $\tau $ is not
considered as a co-ordinate in space-time, but as a parameter, independent
of the frame of reference of the observer.

6. In the relativistic field theories and especially in the Einstein theory
of gravitation $dr$ is considered as the space distance between two
neighboring points $P$ and $\overline{P}$ and $l_u\cdot d\tau $ is
interpreted as the distance covered by a light signal in a time interval $%
d\tau $, measured by an observer in his proper frame of reference (when the
observer in it is at rest). The quantity $l_u$ is usually interpreted as the
absolute value $c$ of a light signal in vacuum, i.e. $l_u=c$, or $l_u$ is
normalized to $1$, i.e. $l_u=1$, if the proper time interval $d\tau $ is
replaced with the proper distance interval $ds=c\cdot d\tau $, i.e. $%
\overline{u}=\frac d{d\tau }$ is replaced with $\overline{u}^{\prime }=\frac
d{c\cdot d\tau }=\frac d{ds}$, \thinspace \thinspace \thinspace $ds=c\cdot
d\tau $.

Therefore, there is a difference between the interpretation of the absolute
value $l_u$ of the velocity of an observer in classical and relativistic
physics

(a) In classical physics, from the above consideration, it follows that $%
l_u=0$ (the observer is at rest) and $ds=dr$ is the distance as space
distance. The quantity $l_v$ is the absolute value of the velocity between
the observer at rest and a point $\overline{P}$ in his neighborhood.

(b) In relativistic physics $l_u=c$, or $l_u=1$, and $l_u$ is not the
absolute value of the velocity of the observer but the velocity of the light
propagation which the observer could measure in his proper frame of
reference. If we wish to interpret $l_u$ as the absolute value of the
velocity of the observer himself we should assume that $l_u\neq c$ or $1$
(if the observer is not moving with the speed of light).

\begin{itemize}
\item There is the possibility to identify $l_u$ with $l_v$ as the absolute
value of the velocity of the observer at a point $P$ at his world line,
measured with respect to a neighboring point $\overline{P}$ with the same
proper time as the point $P$. Under this assumption, the ordinary
differential becomes a null (isotropic) vector field [$g(d,d)=0$, $l_d=0$, $%
l_u=l_v\neq 0$] in the proper frame of reference of the observer.

\item We could also interpret $l_u$ as the absolute value of the velocity of
the observer with respect to another frame of reference or

\item we could consider $l_u$ as the absolute value of the velocity of a
signal coming to the observer with velocity, different from the velocity of
light. On the basis of the last assumption we can describe the propagation
of signals with propagation velocity different from the velocity of light
(for instance, the propagation of sound signals or (may be) gravitational
signals).
\end{itemize}

If $l_v=0$ then $\overline{u}=u$ and $u$ could be

\begin{itemize}
\item the velocity vector field $u=\frac d{d\tau }$ of an observer ($l_u\neq
0$, $u=l_u\cdot n_{\parallel }$) in his proper frame of reference along his
world line. Since in his proper frame of reference the observer is at rest, $%
u$ could be interpreted as the velocity of a clock measuring the length
(proper time) of the world line by the use of the parameter $\tau $ or

\item the velocity of a signal detected or emitted by the observer.
\end{itemize}

There is another way for considering the ordinary differential as a
contravariant vector field with its relations to the notions of velocity and
of space velocity.

\subsubsection{Distance in a $(\overline{L}_n,g)$-space for world and space
lines depending to each other}

1. Let us consider the ordinary differential $d=dx^i\cdot \partial _i$ as a
contravariant vector field over a two parameter congruence $x^i=x^i(\tau
,\lambda )$. Then 
\begin{equation}
dx^i=\frac{\partial x^i(\tau ,\lambda )}{\partial \tau }\cdot d\tau +\frac{%
\partial x^i(\tau ,\lambda )}{\partial \lambda }\cdot d\lambda =u^i\cdot
d\tau +\xi _{\perp }^i\cdot d\lambda \,\,\,\text{,}  \label{2.6}
\end{equation}
where 
\begin{equation}
u^i=\frac{\partial x^i(\tau ,\lambda )}{\partial \tau }\,\,\,\,\,\,\,\text{%
,\thinspace \thinspace \thinspace \thinspace \thinspace \thinspace
\thinspace \thinspace }\xi _{\perp }^i=\frac{\partial x^i(\tau ,\lambda )}{%
\partial \lambda }\,\,\,\,\,\text{.}  \label{2.7}
\end{equation}

The ordinary differential will have the form 
\begin{equation}
d=dx^i\cdot \partial _i=d\tau \cdot u+d\lambda \cdot \xi _{\perp }\,\,\,\,%
\text{,\thinspace \thinspace \thinspace \thinspace \thinspace \thinspace
\thinspace \thinspace \thinspace \thinspace \thinspace }u=u^i\cdot \partial
_i\,\,\,\,\,\,\text{,\thinspace \thinspace \thinspace \thinspace \thinspace
\thinspace \thinspace }\xi _{\perp }=\xi _{\perp }^i\cdot \partial _i\,\,%
\text{.}  \label{2.8}
\end{equation}

If we impose the additional condition 
\begin{equation}
g(u,\xi _{\perp })=0\,\,\,\,  \label{2.9}
\end{equation}
we will have the relations 
\begin{eqnarray}
g(d,u) &=&d\tau \cdot g(u,u)=e\cdot d\tau =\pm l_u^2\cdot d\tau \,\,\,\,\,%
\text{,}  \label{2.10} \\
g(\xi _{\perp },d) &=&d\lambda \cdot g(\xi _{\perp },\xi _{\perp })=\mp
l_{\xi _{\perp }}^2\cdot d\lambda \,\,\,\text{.}  \label{2.11}
\end{eqnarray}

2. On the other side, if we consider the projections of $d$ collinear and
orthogonal to $u$ we obtain the relations 
\begin{equation}
d=\frac 1e\cdot g(u,d)\cdot u+\overline{g}[h_u(d)]=d_{\parallel }+d_{\perp
}\,\,\,\,\,\text{,}  \label{2.12}
\end{equation}
where 
\begin{equation}
d_{\parallel }=\frac 1e\cdot g(u,d)\cdot u\,\,\,\,\,\,\,\,\text{,\thinspace
\thinspace \thinspace \thinspace \thinspace \thinspace \thinspace \thinspace
\thinspace \thinspace \thinspace \thinspace }d_{\perp }=\overline{g}%
[h_u(d)]\,\,\,\,\,\,\,\text{.}  \label{2.13}
\end{equation}

For the explicit form of $d$ as $d=d\tau \cdot u+d\lambda \cdot \xi _{\perp
} $ we have [under the condition $g(u,\xi _{\perp })=0$]$\,$ 
\begin{eqnarray}
d_{\parallel } &=&\frac 1e\cdot g(u,d)\cdot u\,=\frac 1e\cdot e\cdot d\tau
\cdot u=d\tau \cdot u\,\,\,\,\,\text{,}  \label{2.14} \\
\text{\thinspace }d_{\perp } &=&\overline{g}[h_u(d)]\,=\overline{g}%
[h_u(d\tau \cdot u+d\lambda \cdot \xi _{\perp })]=  \notag \\
&=&\overline{g}[h_u(d\lambda \cdot \xi _{\perp })]=d\lambda \cdot \overline{g%
}[h_u(\xi _{\perp })]=d\lambda \cdot \xi _{\perp }\,\,\,\,\,\,\text{.}
\label{2.15}
\end{eqnarray}

Therefore, the representation of $d$ as $d=d\tau \cdot u+d\lambda \cdot \xi
_{\perp }$ is the representation in its collinear and orthogonal to $u$
parts.

3. Since 
\begin{eqnarray}
g(d,d) &=&ds^2=g(d\tau \cdot u+d\lambda \cdot \xi _{\perp },\,d\tau \cdot
u+d\lambda \cdot \xi _{\perp })=  \notag \\
&=&d\tau ^2\cdot g(u,u)+d\lambda ^2\cdot g(\xi _{\perp },\xi _{\perp })= 
\notag \\
&&\pm l_u^2\cdot d\tau ^2\mp l_{\xi _{\perp }}^2\cdot d\lambda ^2\text{
\thinspace \thinspace \thinspace ,}  \label{2.16}
\end{eqnarray}
it follows for the line element $ds$%
\begin{equation}
ds^2=\pm l_u^2\cdot d\tau ^2\mp l_{\xi _{\perp }}^2\cdot d\lambda ^2=\pm
l_u^2\cdot d\tau ^2\cdot (1-\frac{l_{\xi _{\perp }}^2\cdot d\lambda ^2}{%
l_u^2\cdot d\tau ^2})\,\,\,\,\text{.}  \label{2.17}
\end{equation}

If we chose the contravariant non-isotropic (non-null) vector field $\xi
_{\perp }$ as a unit vector field, i.e. if $l_{\xi _{\perp }}=1$, $\xi
_{\perp }=\widetilde{n}_{\perp }$, $g(\widetilde{n}_{\perp },\widetilde{n}%
_{\perp })=\mp 1=\mp l_{\xi _{\perp }}^2$, then 
\begin{equation}
ds^2=\pm l_u^2\cdot d\tau ^2\cdot (1-\frac 1{l_u^2}\cdot \frac{d\lambda ^2}{%
d\tau ^2})\,\,\,\,\,\,\text{.}  \label{2.18}
\end{equation}

If we, further, interpret $d\lambda $ as a distance along a curve with a
tangent vector $\xi _{\perp }$, orthogonal to $u$, we can define and
interpret the expression 
\begin{equation}
\frac{d\lambda }{d\tau }=l_v\text{ }  \label{2.19}
\end{equation}
as the 3-dimensional space velocity of a material point along the curve $%
x^i(\tau ,\lambda )$ with tangential vector $\xi _{\perp }=\widetilde{n}%
_{\perp }$ along the curve $x^i(\tau _0,\lambda )$. Then 
\begin{equation}
ds^2=\pm l_u^2\cdot d\tau ^2\cdot (1-\frac{l_v^2}{l_u^2})\text{\thinspace
\thinspace \thinspace \thinspace \thinspace .}  \label{2.20}
\end{equation}

\textit{Remark}. Here, it is assumed that the parameter $\lambda $ is
depending on the parameter $\tau $, i.e. $\lambda =\lambda (\tau )$,
\thinspace \thinspace \thinspace $\tau =\tau (\lambda )$, and $d\lambda
/d\tau \neq 0$. In the opposite case, where $\lambda $ and $\tau $ are
parameters independent to each other, $d\lambda /d\tau =0$.

The quantity $l_u$ is interpreted as the absolute value of the velocity of a
signal (in the relativity theory it is interpreted as the absolute value of
the velocity of light in vacuum). The parameter $\tau $ is interpreted as
the proper time of an observer moving on a trajectory $x^i(\tau ,\lambda _0)$
interpreted as his world line $x^i(\tau ,\lambda _0)$. The quantity $l_v$ is
interpreted as the absolute value of the velocity of a material point moving
along a space distance $\lambda $ from the trajectory of the observer $%
x^i(\tau ,\lambda _0)$.

4. If we now turn back to the general case, when $l_{\xi _{\perp }}\neq 1$, $%
\xi _{\perp }\neq \widetilde{n}_{\perp }$, we have the relation 
\begin{equation}
ds^2=\pm l_u^2\cdot d\tau ^2\cdot [1-\frac 1{l_u^2}\cdot (\frac{l_{\xi
_{\perp }}^2\cdot d\lambda ^2}{d\tau ^2})]\,\,\,\,\,\text{.}  \label{2.21}
\end{equation}

Then we can introduce the abbreviation 
\begin{equation}
dl^2:=l_{\xi _{\perp }}^2\cdot d\lambda ^2\,\,\,\,\,\,\text{,}  \label{2.22}
\end{equation}
interpreted as the square of the \textit{space distance} of a point in $n-1$%
-dimensional subspace of $M$ $(dimM=n)$, $(n=4)$ from the trajectory (world
line) of the observer with proper time $\tau $ and tangential vector $u$,
orthogonal to $\xi _{\perp }=l_{\xi _{\perp }}\cdot \widetilde{n}_{\perp }$.
The square $ds^2$ of the distance $ds$ in the $n$-dimensional manifold $M$
will have the form 
\begin{equation}
ds^2=\pm l_u^2\cdot d\tau ^2\cdot [1-\frac 1{l_u^2}\cdot \frac{dl^2}{d\tau ^2%
}]\,\,\,\,\,\text{.}  \label{2.23}
\end{equation}

If we again denote 
\begin{equation}
\frac{dl^2}{d\tau ^2}:=\overline{l}_v^2\,\,\,\,  \label{2.24}
\end{equation}
we obtain 
\begin{equation}
ds^2=\pm l_u^2\cdot d\tau ^2\cdot (1-\frac{\overline{l}_v^2}{l_u^2}%
)\,\,\,\,\,\,\,\,\text{,}  \label{2.25}
\end{equation}
where $\overline{l}_v$ could be interpreted again as the absolute value of
the space velocity of a point moving at a space distance $dl$ from the world
line $x^i(\tau ,\lambda _0)$ of the observer. In this general case, the
parameter $\lambda $ is not interpreted as a space distance. Instead of $%
d\lambda $ the quantity $dl=l_{\xi _{\perp }}\cdot d\lambda $ has this
interpretation.

In general, we do not need to search for interpretation of an expression as
a space velocity in the above considerations if we consider only the
structure of the square $ds^2$ of the space-time distance $ds$%
\begin{equation}
ds^2=\pm (l_u^2\cdot d\tau ^2-l_{\xi _{\perp }}^2\cdot d\lambda ^2)\,=\pm
(l_u^2\cdot d\tau ^2-dl^2)\,\,\text{.}  \label{2.26}
\end{equation}

If a signal with absolute value $l_u$ of its velocity is covering a space
distance $dl$ with $dl^2=l_{\xi _{\perp }}^2\cdot d\lambda ^2$ in the proper
time interval $d\tau $ of the observer then $ds^2=g(d,d)=0$ and the ordinary
differential becomes a null (isotropic) vector field, where 
\begin{eqnarray}
ds^2 &=&0\text{ \thinspace \thinspace \thinspace \thinspace ,\thinspace
\thinspace \thinspace \thinspace \thinspace \thinspace \thinspace \thinspace
\thinspace \thinspace \thinspace }l_u^2\cdot d\tau ^2=l_{\xi _{\perp
}}^2\cdot d\lambda ^2=dl^2\,\,\,\,\,\text{,}  \label{2.28} \\
dl &=&\pm l_{\xi _{\perp }}\cdot d\lambda \text{\thinspace \thinspace }%
\,\,\,\,\,\,\,\,\text{,\thinspace \thinspace \thinspace \thinspace
\thinspace \thinspace \thinspace \thinspace \thinspace \thinspace \thinspace
\thinspace \thinspace \thinspace \thinspace \thinspace }l_u\cdot d\tau =\pm
l_{\xi _{\perp }}\cdot d\lambda =dl\text{\thinspace \thinspace \thinspace
\thinspace \thinspace \thinspace \thinspace \thinspace ,\thinspace
\thinspace \thinspace \thinspace \thinspace \thinspace \thinspace \thinspace
\thinspace \thinspace \thinspace \thinspace \thinspace \thinspace \thinspace
\thinspace \thinspace }  \label{2.29} \\
d\tau &=&\frac{dl}{l_u}=\pm \frac{l_{\xi _{\perp }}}{l_u}\cdot d\lambda
\,\,\,\,\,\text{,\thinspace \thinspace \thinspace \thinspace \thinspace
\thinspace \thinspace \thinspace \thinspace \thinspace \thinspace \thinspace
\thinspace \thinspace \thinspace \thinspace \thinspace \thinspace \thinspace 
}l_{\xi _{\perp }}\cdot d\lambda =\pm dl\,\,\,\,\,\,\,\text{,}  \label{2.30}
\\
l_u &>&0\,\,\,\text{,}\,\,\,\,\,\,l_{\xi _{\perp }}>0\,\,\,\,\,\,\,\,\,\,%
\text{.}  \notag
\end{eqnarray}

\subsection{Measuring a distance in $(\overline{L}_n,g)$-spaces}

A. If the notion of distance is introduced in a space-time, modeled by a $(%
\overline{L}_n,g)$-space, we have to decide \textit{what is the meaning of
the vector field }$u$\textit{\ as tangent vector to a trajectory interpreted
as the world line of an observer}. On the basis of the above considerations,
we have four possible interpretations for the meaning of the vector field $u$
as

1. Velocity vector field of a propagating signal in space-time identified
with the tangent vector field $u$ at the world line of an observer. The
signal is detected or emitted by the observer on his world line and the
absolute value $l_u$ of $u$ is identified with the absolute value of the
velocity of the signal in- or outcoming to the observer.

2. Velocity vector field of an observer moving in space-time. In this case $%
l_u\neq 0$ and the space-time should have a definite metric, i.e. $%
Sgn\,g=\pm n$, $dimM=n$ (for instance, motion of an observer in an Euclidean
space considered as a model of space-time). The observer, moving in
space-time, could consider processes happened in its subspace orthogonal to
his velocity. The observer will move in a flow and consider the
characteristics of the flow from his own frame of reference.

3. Velocity of a clock moving in space-time and determining the proper time
by a periodical process in the frame of reference of an observer. The
velocity $u$ of the periodical process in the clock in space-time is with
fixed absolute value $l_u$, i.e. $l_u=$ const. The time interval $d\tau $
measured by the clock corresponds to the length $ds$ of its world line, i.e. 
$d\tau ^2=\pm \,$const.$\cdot ds^2$. Under the assumption for the constant
velocity of the periodical process in the clock, we consider the periodical
process as indicator for the time interval $d\tau $ in the proper frame of
reference of the clock and of the observer respectively.

4. Velocity $u$ of an $(n-1)$-dimensional subspace moving in time with $%
l_u\neq 0$. If the subspace deforms in some way, the deformations reflect on
the kinematic characteristics of the vector field $u$ and $u$ is used as an
indicator for the changing properties of the subspace, considered as the
space of an observer (laboratory) where a physical system is investigated.
This type of interpretation requires not only the existence of the velocity
vector field $u$ with $l_u\neq 0$ but also the existence of (at least one)
orthogonal to $u$ vector field $\xi _{\perp }$, $g(u,\xi _{\perp })=0$,
lying in the orthogonal to $u$ subspace $T^{\perp u}(M)$.

All indicated interpretations could be used in solving different physical
problems related to motions of physical systems in space-time.

B. After introducing the notion of distance, the question arises \textit{how
a space distance between two points in a space could be measured}. We could
distinguish three types of measurements:

1. Direct measurements by using a measuring device (e.g. a roulette, a
linear (running) meter, yard-measure-stick etc.)

2. Direct measurements by sending signals from a basic point to a fixed
point of space and detecting at the basic point the reflected by the fixed
point signal.

3. Indirect measurement by receiving signals from a fixed point of space
without sending a signal to it.

Let us now consider every type of measurements more closely.

1. \textit{Direct measurements by using a measuring device.} The space
distance between two points $A$ and $B$ in a space could be measured by a
second observer moving from point $A$ (where the first observer is at rest)
to point $B$ in space. At the same time, the second observer moves in time
from point $B$ to point $B^{\prime }$. The space distance measured by the
observer with world line $AA^{\prime }$ could be denoted as $\vartriangle
r=AB$ and the time period passed as $\vartriangle \tau =AA^{\prime }$. This
is a direct measurement of the space distance $AB=\vartriangle r$ from point 
$A$ to point $B$ in the space during the time $AA^{\prime }=\vartriangle
\tau $. It is \textit{assumed} that point $A$ and point $B$ are at rest
during the measurement. Instead of measuring the space distance $AB$ the
observers measure the space distance $A^{\prime }B^{\prime }$ which exists
at the time $\tau +\vartriangle \tau $ if the measurement has began at the
time $\tau $ from the point of the first observer with world line $%
AA^{\prime }$.

2. \textit{Direct measurements by sending signals from a basic point to a
fixed point of space and detecting at the basic point the reflected by the
fixed point signal.} The space distance between two points $A$ and $B$ in a
space could be measured by sending a signal with velocity with absolute
value $l_u\neq 0$. Then $AB$ of the curve $x^i(\tau =\tau _0,r)$ through
point $B$ is the distance $\vartriangle r$ at the time $\tau (A)=\tau _0$
and $\tau (B)=\tau _0$.

$A^{\prime }B^{\prime }$ of the curve $x^i(\tau =\tau _0+\,\vartriangle \tau
,r)$ is the space distance $\vartriangle r^{\prime }$ at the time $\tau
(A^{\prime })=\tau _1$. At this time the signal is received at point $%
B^{\prime }$ which is point $B$ at the time $\tau _1$, i.e. $\tau (B^{\prime
})=\tau _1$. $B^{\prime }A^{\prime }$ is the space distance between $B$ and $%
A$ at the time $\tau (A^{\prime })=\tau _1$, where $\tau (B^{\prime })=\tau
_1$, $\tau (B^{\prime \prime })=\tau _2$. At the time $(\tau _2)$ the point $%
B(\tau _0)$ will be moved in the time to point $B^{\prime \prime }(\tau _2)$%
. The signal will be propagated

(a) for the time interval $AA^{\prime }=\tau _1-\tau _0$ to the point $%
B^{\prime }$ at the time $\tau _1$ at the space distance $\vartriangle
r=l_u\cdot (\tau _1-\tau _0)$, where $l_u$ is the velocity of the signal
measured by the observer with world line $AA^{\prime }$.

(b) for the time interval $A^{\prime }A^{\prime \prime }$ from point $%
B^{\prime }$ at the time $\tau _1$ to the point $A^{\prime \prime }$ at the
time $\tau _2$ at a space distance $l_u\cdot (\tau _2-\tau _1)$. The whole
space distance covered by the signal in the time interval $AA^{\prime
}A^{\prime \prime }=\,\vartriangle \tau =\tau _2-\tau _0$ is $l_u\cdot (\tau
_2-\tau _0)=$ $l_u\cdot (\tau _2-\tau _1)+l_u\cdot (\tau _1-\tau _0)$.

If we now \textit{assume} that point $A$ and point $B$ are at rest to each
other and the space distance between them does not change in the time then 
\begin{equation}
l_u\cdot (\tau _2-\tau _1)=l_u\cdot (\tau _1-\tau _0)  \label{2.31}
\end{equation}
and 
\begin{equation}
l_u\cdot (\tau _2-\tau _0)=2\cdot l_u\cdot (\tau _1-\tau _0)=2\cdot
A^{\prime }B^{\prime }(\tau _1)=2\cdot AB(\tau _0)\,\,\,\text{.}
\label{2.32}
\end{equation}

Therefore, the space distance between point $A$ and point $B$ (at any time,
if both the points are at rest to each other) is 
\begin{equation}
AB=\frac 12\cdot l_u\cdot (\tau _2-\tau _0)\,\,\,\,\text{,}  \label{2.33}
\end{equation}
where $\vartriangle \tau =\tau _2-\tau _0$ is the time interval for the
propagation of a signal from point $A$ to point $B$ and from point $B$ back
to point $A$.

3. \textit{Indirect measurement by receiving signals from a fixed point of
space without sending a signal to it}. If the space distance between point $%
A $ and point $B$ is changing in the time and at point $B$ there is an
emitter then the frequency of the emitter could change in the time related
to the centrifugal (centripetal) or Coriolis' velocities and accelerations
between both the points $A$ and $B$. Therefore, a criteria for no relative
motion between two (space) points (points with one and the same proper time)
could be the lack of change of the frequency of the signals emitted from the
second point $B$ to the basic point $A$. [But there could be motions of an
emitter which could so change its frequency that the changes compensate each
other and the observer at the basic point $A$ could come to the conclusion
that there is no motions between points $A$ and $B$.]

If an emitter at point $B(\tau _0)$ emits a signal with velocity $\overline{u%
}$ and frequency $\overline{\omega }$ then this signal will be received
(detected) at the point $A^{\prime }(\tau _1)$ after a time interval $%
AA^{\prime }=\,\vartriangle \tau =\tau _1-\tau _0$ by an observer (detector)
moving in the time interval $\vartriangle \tau $ from point $A(\tau _0)$ to
point $A^{\prime }(\tau _1)$ on his world line $x^i(\tau )$. If the emitter
is moving relatively to point $A$ with relative velocity $_{rel}v$ and / or
with relative acceleration $_{rel}a$ then the detected at the point $%
A^{\prime }$ frequency $\omega $ will differ from the emitted frequency $%
\overline{\omega }$. If both the points $A$ and $B$ are at rest to each
other then $\overline{\omega }=\omega $.

C. The question arises \textit{how can we find the space distance between
two points }$A$\textit{\ and }$B$\textit{\ lying in such a way in the space
that only signals emitted from the one point (point }$B$\textit{) could be
detected at the basic point (point }$A$\textit{), where an observer detects
the signal from point }$B$\textit{.} First of all, if we knew the
propagation velocity $l_u$ of a signal and the difference $\overline{\omega }%
-\omega $ between the emitted frequency $\overline{\omega }$ and the
detected frequency $\omega $ we can try to find out the relative velocity
(and acceleration) between the emitter (at a point $B$) and the observer (at
a point $A$). For doing that we will need relations between the difference $%
\overline{\omega }-\omega $ and the relative velocity (and acceleration)
between both the points. Such relations could be found on the basis of the
structures of the relative velocity and the relative acceleration and their
decompositions in centrifugal (centripetal) relative velocity and relative
acceleration and Coriolis relative velocity and relative acceleration \cite%
{Manoff-6}.

\section{Kinematic effects related to the relative velocity and to the
relative acceleration}

1. Let us now consider the change of a null vector field $\widetilde{k}$
under the influence of the relative velocity and of the relative
acceleration on the corresponding emitter and its frequency with respect to
an observer detecting the emitted radiation by the emitter.

Let $\overline{k}_{\perp }$ be the orthogonal to $u$ part of the null vector
field $\overline{k}$ corresponding to the null vector field $\widetilde{k}$
after the influence of the relative velocity $_{rel}v$ and / or the relative
acceleration $_{rel}a$%
\begin{eqnarray}
\overline{k} &=&\widetilde{k}+\,_{rel}k\text{ \thinspace ,\thinspace
\thinspace \thinspace \thinspace \thinspace }\overline{k}=\overline{k}%
_{\parallel }+\overline{k}_{\perp }\,\,\,\,\text{,\thinspace \thinspace
\thinspace \thinspace \thinspace }\widetilde{k}=k_{\parallel }+k_{\perp
}\,\,\,\,\text{,\thinspace }  \label{3.1} \\
\text{\thinspace \thinspace \thinspace \thinspace }_{rel}k
&=&\,_{rel}k_{\parallel }+\,_{rel}k_{\perp }\text{ ,}  \label{3.1a} \\
\overline{k}_{\perp } &=&k_{\perp }+\,_{rel}k_{\perp }\text{\thinspace
\thinspace \thinspace ,}  \label{3.2}
\end{eqnarray}%
where $_{rel}k$ could depend on the relative velocity $_{rel}v$ (\cite%
{Manoff-2} - Ch. 10) and on the relative acceleration $_{rel}a$ \thinspace (%
\cite{Manoff-2} - Ch. 11, Ch. 12).

If $\overline{k}=\widetilde{k}+\,_{rel}k$ then 
\begin{eqnarray}
g(\overline{k},\overline{k}) &=&0=g(_{rel}k,\,_{rel}k)+2\cdot g(\widetilde{k}%
,\,_{rel}k)\,\,\,\,\,\text{,}  \notag \\
g(_{rel}k,\,_{rel}k) &=&-2\cdot g(\widetilde{k},\,_{rel}k)\,\,\,\,\,\text{.}
\label{3.2a}
\end{eqnarray}

In the previous sections we have considered the representation of $%
\widetilde{k}$ as $\widetilde{k}=k_{\parallel }+k_{\perp }$, where 
\begin{eqnarray}
k_{\parallel } &=&\pm l_{k_{\parallel }}\cdot n_{\parallel }=\pm \frac
\omega {l_u}\cdot n_{\parallel }\,\,\,\,\text{,\thinspace \thinspace
\thinspace \thinspace \thinspace \thinspace \thinspace \thinspace \thinspace
\thinspace \thinspace \thinspace \thinspace \thinspace \thinspace \thinspace 
}l_{k_{\parallel }}=\frac \omega {l_u}\,\,\,\,\text{,\thinspace \thinspace
\thinspace \thinspace \thinspace \thinspace \thinspace }  \label{3.2b} \\
\text{\thinspace }k_{\perp } &=&\mp l_{k_{\perp }}\cdot \widetilde{n}_{\perp
}=\mp \frac \omega {l_u}\,\cdot \widetilde{n}_{\perp }\,\,\,\text{%
,\thinspace \thinspace \thinspace \thinspace \thinspace }l_{k_{\perp
}}=\frac \omega {l_u}=\text{\thinspace }l_{k_{\parallel }}\,\,\,\,\text{.}
\label{3.2c}
\end{eqnarray}

The unit vector $\widetilde{n}_{\perp }$ is orthogonal to the vector $u$,
i.e. $g(u,\widetilde{n}_{\perp })=0$ because of $g(u,k_{\perp })=\mp
l_{k_{\perp }}\cdot g(u,\widetilde{n}_{\perp })=0$,\thinspace \thinspace
\thinspace $l_{k_{\perp }}\neq 0$, $l_u\neq 0$. Therefore, $g(\widetilde{n}%
_{\perp },\widetilde{n}_{\perp })=\mp 1=\mp l_{\widetilde{n}_{\perp }}^2$, $%
l_{\widetilde{n}_{\perp }}\neq 0$.

We can represent the unit vector $\widetilde{n}_{\perp }$ (orthogonal to $u$%
) in two parts: one part $n_{\perp }$ collinear to the vector field $\xi
_{\perp }$ (orthogonal to $u$) and one part $m_{\perp }$ orthogonal to the
vectors $u$ and $\xi _{\perp }$, i.e. 
\begin{eqnarray}
\widetilde{n}_{\perp } &=&\alpha \cdot n_{\perp }+\beta \cdot m_{\perp
}\,\,\,\,\,\,\,\text{,\thinspace \thinspace \thinspace \thinspace \thinspace
\thinspace \thinspace \thinspace \thinspace \thinspace }  \label{3.13} \\
g(\widetilde{n}_{\perp },\widetilde{n}_{\perp }) &=&\mp 1=\mp l_{\widetilde{n%
}_{\perp }}^2\,\,\,\text{,\thinspace \thinspace \thinspace \thinspace
\thinspace \thinspace }l_{\widetilde{n}_{\perp }}>0\,\,\,\,\,\,\text{%
,\thinspace \thinspace \thinspace \thinspace \thinspace \thinspace
\thinspace \thinspace \thinspace }l_{\widetilde{n}_{\perp }}=1\,\,\,\,\text{,%
}  \label{3.14}
\end{eqnarray}
\begin{eqnarray}
\eta _{\perp } &:&=l_{\xi _{\perp }}\cdot m_{\perp }\,\,\,\,\,\text{%
,\thinspace \thinspace \thinspace \thinspace \thinspace \thinspace
\thinspace \thinspace \thinspace \thinspace \thinspace }m_{\perp }=\frac{%
\eta _{\perp }}{l_{\xi _{\perp }}}\,\,\,\,\,\,\,\text{,\thinspace \thinspace
\thinspace \thinspace \thinspace \thinspace \thinspace \thinspace }%
g(m_{\perp },m_{\perp })=\mp 1\,\,\,\,\text{,}  \label{3.14a} \\
\text{\thinspace }g(\eta _{\perp },\eta _{\perp }) &=&l_{\xi _{\perp
}}^2\cdot g(m_{\perp },m_{\perp })=\mp l_{\xi _{\perp }}^2=g(\xi _{\perp
},\xi _{\perp })\text{ ,}  \label{3.14b} \\
g(\eta _{\perp },\xi _{\perp }) &=&0\,\,\,\text{,\thinspace \thinspace
\thinspace \thinspace \thinspace \thinspace \thinspace \thinspace }%
g(m_{\perp },n_{\perp })=0\,\,\,\,\text{,\thinspace \thinspace \thinspace
\thinspace \thinspace \thinspace \thinspace }  \label{3.14c}
\end{eqnarray}
where 
\begin{eqnarray}
n_{\perp } &=&\frac{\xi _{\perp }}{l_{\xi _{\perp }}}\,\,\,\,\,\text{%
,\thinspace \thinspace \thinspace \thinspace \thinspace \thinspace
\thinspace }g(n_{\perp },u)=0\,\,\,\,\,\text{,\thinspace \thinspace
\thinspace \thinspace \thinspace }g(n_{\perp },n_{\perp })=\mp 1=\mp
l_{n_{\perp }}^2\,\,\text{,}  \label{3.15} \\
l_{n_{\perp }} &>&0\,\,\,\,\text{,\thinspace \thinspace \thinspace
\thinspace \thinspace \thinspace }l_{n_{\perp }}=1  \label{3.16} \\
m_{\perp } &=&\frac{\eta _{\perp }}{l_{\xi _{\perp }}}=\frac{v_c}{l_{v_c}}%
\,\,\,\,\,\text{,\thinspace \thinspace \thinspace \thinspace \thinspace
\thinspace \thinspace }g(m_{\perp },u)=0\,\,\,\,\,\text{,\thinspace
\thinspace \thinspace \thinspace \thinspace \thinspace }g(m_{\perp },\xi
_{\perp })=0\,\,\text{.}  \label{3.17}
\end{eqnarray}

The vector field $v_c$ is the Coriolis velocity vector field \cite{Manoff-6}
orthogonal to $u$ and to the centrifugal (centripetal) velocity $v_z$
collinear to $\xi _{\perp }$. Since 
\begin{equation}
v_c=\mp l_{v_c}\cdot m_{\perp }\text{\thinspace \thinspace \thinspace
\thinspace \thinspace \thinspace \thinspace ,\thinspace \thinspace
\thinspace \thinspace \thinspace \thinspace \thinspace \thinspace \thinspace
\thinspace }g(v_c,v_c)=\mp l_{v_c}^2\text{\ \ \ \ \ ,}  \label{3.18}
\end{equation}
we also have 
\begin{equation}
g(m_{\perp },m_{\perp })=\mp 1=\mp l_{m_{\perp }}^2\,\,\,\,\,\text{%
,\thinspace \thinspace \thinspace \thinspace \thinspace \thinspace
\thinspace \thinspace }l_{m_{\perp }}>0\,\,\,\,\,\,\,\text{,\thinspace
\thinspace \thinspace \thinspace \thinspace \thinspace }l_{m_{\perp }}=1%
\text{ \thinspace \thinspace \thinspace .}  \label{3.19}
\end{equation}

The Coriolis velocity $v_c$ is related to the change of the vector $\xi
_{\perp }$ in direction orthogonal to $u$ and $\xi _{\perp }$.

Since $\widetilde{n}_{\perp }$ is a unit vector as well as the vectors $%
n_{\perp }$ and $m_{\perp }$, and, further, $g(n_{\perp },m_{\perp })=0$, we
obtain 
\begin{eqnarray}
g(\widetilde{n}_{\perp },\widetilde{n}_{\perp }) &=&\mp 1=g(\alpha \cdot
n_{\perp }+\beta \cdot m_{\perp },\alpha \cdot n_{\perp }+\beta \cdot
m_{\perp })=  \label{3.19a} \\
&=&\alpha ^2\cdot g(n_{\perp },n_{\perp })+\beta ^2\cdot g(m_{\perp
},m_{\perp })=  \label{3.19b} \\
&=&\mp \alpha ^2\mp \beta ^2\,\,\,\text{.}  \label{3.19c}
\end{eqnarray}

Therefore, 
\begin{equation}
\alpha ^2+\beta ^2=1\,\,\,\,\text{.}  \label{3.20}
\end{equation}

On the other side, 
\begin{eqnarray}
g(\widetilde{n}_{\perp },n_{\perp }) &=&g(\alpha \cdot n_{\perp }+\beta
\cdot m_{\perp },n_{\perp })=  \notag \\
&=&\alpha \cdot g(n_{\perp },n_{\perp })=\mp \alpha \,\,\,\,\,\text{,}
\label{3.20a} \\
g(\widetilde{n}_{\perp },m_{\perp }) &=&g(\alpha \cdot n_{\perp }+\beta
\cdot m_{\perp },m_{\perp })=  \notag \\
&=&\beta \cdot g(n_{\perp },n_{\perp })=\mp \beta \,\,\,\text{.}
\label{3.20b}
\end{eqnarray}
i.e. 
\begin{eqnarray}
\alpha &=&\mp g(\widetilde{n}_{\perp },n_{\perp })=\mp l_{\widetilde{n}%
_{\perp }}\cdot l_{n_{\perp }}\cdot cos(\widetilde{n}_{\perp },n_{\perp
})=\mp \,cos(\widetilde{n}_{\perp },n_{\perp })\,\,\,\,\text{,}  \label{3.21}
\\
\beta &=&\mp g(\widetilde{n}_{\perp },m_{\perp })=\mp l_{\widetilde{n}%
_{\perp }}\cdot l_{m_{\perp }}\cdot cos(\widetilde{n}_{\perp },m_{\perp
})=\mp cos(\widetilde{n}_{\perp },m_{\perp })\,\,\,\,\text{.}  \label{3.22}
\end{eqnarray}

Therefore, $\alpha $ and $\beta $ appear as direction cosines of $n_{\perp }$
and $m_{\perp }$ with respect to the unit vector $\widetilde{n}_{\perp }$.
Since 
\begin{equation}
cos^2(\widetilde{n}_{\perp },n_{\perp })+cos^2(\widetilde{n}_{\perp
},m_{\perp })=1\,\,\,\,\,\text{,}  \label{3.22a}
\end{equation}
it follows that 
\begin{eqnarray}
cos^2(\widetilde{n}_{\perp },m_{\perp }) &=&1-cos^2(\widetilde{n}_{\perp
},n_{\perp })=1-sin^2(\widetilde{n}_{\perp },m_{\perp })=sin^2\,(\widetilde{n%
}_{\perp },n_{\perp })\,\,\text{,}  \notag \\
sin^2(\widetilde{n}_{\perp },m_{\perp }) &=&cos^2(\widetilde{n}_{\perp
},n_{\perp })\,\,\,\,\,\text{,}  \notag \\
cos(\widetilde{n}_{\perp },m_{\perp }) &=&\pm \,sin\,(\widetilde{n}_{\perp
},n_{\perp })\,\,\,\,\text{,}  \notag \\
\alpha &=&\mp \,cos(\widetilde{n}_{\perp },n_{\perp })\,\,\,\,\text{,}
\label{3.23} \\
\beta &=&\mp sin\,(\widetilde{n}_{\perp },n_{\perp })\,\,\,\,\,\text{,}
\label{3.24}
\end{eqnarray}
\begin{eqnarray}
\widetilde{n}_{\perp } &=&\alpha \cdot n_{\perp }+\beta \cdot m_{\perp }\,= 
\notag \\
&=&\mp [cos(\widetilde{n}_{\perp },n_{\perp })\,\cdot n_{\perp }+sin\,(%
\widetilde{n}_{\perp },n_{\perp })\,\cdot m_{\perp }]\,\,\,\text{.}
\label{3.25}
\end{eqnarray}

If we denote the angle $(\widetilde{n}_{\perp },n_{\perp })$ between the
vectors $\widetilde{n}_{\perp }$ and $n_{\perp }$ as $\theta =(\widetilde{n}%
_{\perp },n_{\perp })$ then the above relations could be written in the form 
\begin{eqnarray}
cos^2(\widetilde{n}_{\perp },m_{\perp }) &=&sin^2\theta \,\,\,\,\text{,} 
\notag \\
sin^2(\widetilde{n}_{\perp },m_{\perp }) &=&cos^2\theta \,\,\,\text{,} 
\notag \\
\alpha &=&\mp \,cos\,\theta \,\,\,\,\text{,}  \label{3.25b} \\
\beta &=&\mp sin\,\theta \,\,\,\text{\thinspace \thinspace ,}  \label{3.25c}
\end{eqnarray}
\begin{equation}
\widetilde{n}_{\perp }=\alpha \cdot n_{\perp }+\beta \cdot m_{\perp }\,=\mp
[cos\,\theta \,\cdot n_{\perp }+sin\,\theta \,\cdot m_{\perp }]\,\,\,\text{.}
\label{3.25a}
\end{equation}

Further, since $k_{\perp }=\mp l_{k_{\perp }}\cdot \widetilde{n}_{\perp }$
then (see above) 
\begin{eqnarray}
g(k_{\perp },k_{\perp }) &=&l_{k_{\perp }}^2\cdot g(\widetilde{n}_{\perp },%
\widetilde{n}_{\perp })=\mp l_{k_{\perp }}^2\,\,\,\,\,\,\text{,\thinspace
\thinspace \thinspace \thinspace \thinspace \thinspace }g(\widetilde{n}%
_{\perp },\widetilde{n}_{\perp })=\mp 1\,\,\,\,\,\text{,}  \label{3.25d} \\
\text{\thinspace }k_{\perp } &=&\mp \frac \omega {l_u}\cdot \widetilde{n}%
_{\perp }\,\,\,\,\,\,\text{,\thinspace \thinspace \thinspace \thinspace
\thinspace \thinspace \thinspace \thinspace \thinspace \thinspace \thinspace
\thinspace \thinspace \thinspace }k_{\parallel }=\pm \frac \omega {l_u}\cdot
n_{\parallel }\,\,\,\,\,\,\,\text{,\thinspace \thinspace \thinspace
\thinspace \thinspace \thinspace \thinspace \thinspace \thinspace \thinspace
\thinspace }l_{k_{\perp }}=\frac \omega {l_u}=l_{k_{\parallel }}\,\,\,\,%
\text{,}  \label{3.25e} \\
g(\widetilde{n}_{\perp },k_{\perp }) &=&\mp l_{k_{\perp }}\cdot g(\widetilde{%
n}_{\perp },\widetilde{n}_{\perp })=\frac \omega {l_u}\,\,=l_{k_{\parallel
}}=l_{k_{\perp }}\,\,\,\,\,\text{.}  \label{3.25f}
\end{eqnarray}

2. For the contravariant null vector field $\overline{k}$ we have analogous
relations as for the contravariant null vector field $\widetilde{k}$ (just
changing $\widetilde{k}$ with $\overline{k}$, $\omega $ with $\overline{%
\omega }$, and $\widetilde{n}_{\perp }$ with $\widetilde{n}_{\perp }^{\prime
}$) 
\begin{equation}
\overline{k}=\overline{k}_{\parallel }+\overline{k}_{\perp }\,\,\,\,\,\,\,%
\text{,\thinspace \thinspace \thinspace \thinspace \thinspace \thinspace
\thinspace \thinspace \thinspace \thinspace \thinspace \thinspace }\overline{%
\omega }=g(u,\overline{k})\,\,\,\,\text{,}  \label{3.26}
\end{equation}
\begin{eqnarray}
\overline{k}_{\parallel } &=&\pm \frac{\overline{\omega }}{l_u}\cdot
n_{\parallel }\,\,\,\,\,\,\,\,\,\,\,\text{,\thinspace \thinspace \thinspace
\thinspace \thinspace \thinspace \thinspace \thinspace \thinspace \thinspace
\thinspace \thinspace }l_{\overline{k}_{\parallel }}=\frac{\overline{\omega }%
}{l_u}\,\,\,\,\text{,}  \label{3.27} \\
\overline{k}_{\perp } &=&\mp \frac{\overline{\omega }}{l_u}\cdot \widetilde{n%
}_{\perp }^{\prime }\,\,\,\,\,\text{,\thinspace \thinspace \thinspace
\thinspace \thinspace \thinspace \thinspace \thinspace \thinspace \thinspace 
}l_{\overline{k}_{\perp }}=\frac{\overline{\omega }}{l_u}=l_{\overline{k}%
_{\parallel }}\,\,\,\,\text{,}  \label{3.28} \\
g(\overline{k}_{\perp },\overline{k}_{\perp }) &=&\frac{\overline{\omega }^2%
}{l_u^2}.g(\widetilde{n}_{\perp }^{\prime },\widetilde{n}_{\perp }^{\prime
})=\mp l_{\overline{k}_{\perp }}^2=\mp \frac{\overline{\omega }^2}{l_u^2}\,\,%
\text{,}  \label{3.28a} \\
g(\widetilde{n}_{\perp }^{\prime },\widetilde{n}_{\perp }^{\prime }) &=&\mp 1
\label{3.28b}
\end{eqnarray}
\begin{equation}
g(\widetilde{n}_{\perp }^{\prime },\overline{k}_{\perp })=\mp l_{\overline{k}%
_{\perp }}\cdot g(\widetilde{n}_{\perp }^{\prime },\widetilde{n}_{\perp
}^{\prime })=\frac{\overline{\omega }}{l_u}\,\,=l_{\overline{k}_{\parallel
}}=l_{\overline{k}_{\perp }}\,\,\,\,\,\,\,\,\text{.}  \label{3.29}
\end{equation}

From $\overline{k}=\widetilde{k}+\,_{rel}k$,\thinspace \thinspace $\overline{%
k}_{\perp }=\widetilde{k}_{\perp }+\,_{rel}k_{\perp }$, and 
\begin{equation}
\mp \frac{\overline{\omega }}{l_u}\cdot \widetilde{n}_{\perp }^{\prime }=\mp
\frac \omega {l_u}\cdot \widetilde{n}_{\perp }+\,_{rel}k_{\perp }\,\,\,\,%
\text{,}  \label{3.30}
\end{equation}
it follows that 
\begin{eqnarray}
g(\widetilde{n}_{\perp },\overline{k}) &=&g(\widetilde{n}_{\perp },k)+g(%
\widetilde{n}_{\perp },\,_{rel}k)\,\,\,\,\,\,\text{,}  \notag \\
g(\widetilde{n}_{\perp },\overline{k}_{\perp }) &=&g(\widetilde{n}_{\perp
},k_{\perp })+g(\widetilde{n}_{\perp },\,_{rel}k_{\perp })\,\,\,\,\text{,}
\label{3.31} \\
\mp \frac{\overline{\omega }}{l_u}\cdot g(\widetilde{n}_{\perp }^{\prime },%
\widetilde{n}_{\perp }) &=&\frac \omega {l_u}+g(_{rel}k_{\perp },\widetilde{n%
}_{\perp })\,\,\,\,\text{.}  \label{3.32}
\end{eqnarray}

The vector $\widetilde{n}_{\perp }^{\prime }$ could be represented by the
use of the vectors $n_{\perp }$ and $m_{\perp }$ in the form 
\begin{equation}
\widetilde{n}_{\perp }^{\prime }=\alpha ^{\prime }\cdot n_{\perp }+\beta
^{\prime }\cdot m_{\perp }\,\,\,\,\,\,\text{,}  \label{3.31a}
\end{equation}
where 
\begin{eqnarray}
g(\widetilde{n}_{\perp }^{\prime },\widetilde{n}_{\perp }^{\prime }) &=&\mp
1=g(\alpha ^{\prime }\cdot n_{\perp }+\beta ^{\prime }\cdot m_{\perp
},\alpha ^{\prime }\cdot n_{\perp }+\beta ^{\prime }\cdot m_{\perp })= 
\notag \\
&=&\alpha ^{\prime \,\,2}\cdot g(n_{\perp },n_{\perp })+\beta ^{\prime
\,\,2}\cdot g(m_{\perp },m_{\perp })=  \notag \\
&=&\mp \alpha ^{\prime \,\,2}\mp \beta ^{\prime \,\,2}\,\,\,\text{.}
\label{3.31b}
\end{eqnarray}

Therefore, 
\begin{equation}
\alpha ^{\prime \,\,2}+\beta ^{\prime \,\,\,2}=1\,\,\,\,\text{.}
\label{3.32a}
\end{equation}

On the other side we have analogous relations as in the case of the vector $%
\widetilde{n}_{\perp }$: 
\begin{eqnarray}
g(\widetilde{n}_{\perp }^{\prime },n_{\perp }) &=&g(\alpha ^{\prime }\cdot
n_{\perp }+\beta ^{\prime }\cdot m_{\perp },n_{\perp })=  \notag \\
&=&\alpha ^{\prime }\cdot g(n_{\perp },n_{\perp })=\mp \alpha ^{\prime
}\,\,\,\,\,\text{,}  \label{3.31c} \\
g(\widetilde{n}_{\perp }^{\prime },m_{\perp }) &=&g(\alpha ^{\prime }\cdot
n_{\perp }+\beta ^{\prime }\cdot m_{\perp },m_{\perp })=  \notag \\
&=&=\beta ^{\prime }\cdot g(n_{\perp },n_{\perp })=\mp \beta ^{\prime }\,\,\,%
\text{.}  \label{3.31d}
\end{eqnarray}%
i.e. 
\begin{eqnarray}
\alpha ^{\prime } &=&\mp g(\widetilde{n}_{\perp }^{\prime },n_{\perp })=\mp
l_{\widetilde{n}_{\perp }^{\prime }}\cdot l_{n_{\perp }}\cdot cos(\widetilde{%
n}_{\perp }^{\prime },n_{\perp })=\mp \,cos(\widetilde{n}_{\perp }^{\prime
},n_{\perp })\,\,\,\,\text{,}  \label{3.32b} \\
\beta ^{\prime } &=&\mp g(\widetilde{n}/_{\perp },m_{\perp })=\mp l_{%
\widetilde{n}_{\perp }^{\prime }}\cdot l_{m_{\perp }}\cdot cos(\widetilde{n}%
_{\perp }^{\prime },m_{\perp })=\mp cos(\widetilde{n}_{\perp }^{\prime
},m_{\perp })\,\,\text{.}  \label{3.32c}
\end{eqnarray}

Therefore, $\alpha ^{\prime }$ and $\beta ^{\prime }$ appear as direction
cosines of $n_{\perp }$ and $m_{\perp }$ with respect to the unit vector $%
\widetilde{n}_{\perp }^{\prime }$. Since 
\begin{equation}
cos^2(\widetilde{n}_{\perp }^{\prime },n_{\perp })+cos^2(\widetilde{n}%
_{\perp }^{\prime },m_{\perp })=1\,\,\,\,\,\text{,}  \label{3.31f}
\end{equation}
it follows that 
\begin{eqnarray}
cos^2(\widetilde{n}_{\perp }^{\prime },m_{\perp }) &=&1-cos^2(\widetilde{n}%
_{\perp }^{\prime },n_{\perp })=1-sin^2(\widetilde{n}_{\perp }^{\prime
},m_{\perp })=sin^2\,(\widetilde{n}_{\perp }^{\prime },n_{\perp })\,\,\text{,%
}  \notag \\
sin^2(\widetilde{n}_{\perp }^{\prime },m_{\perp }) &=&cos^2(\widetilde{n}%
_{\perp }^{\prime },n_{\perp })\,\,\,\,\,\text{,}  \notag \\
cos(\widetilde{n}_{\perp }^{\prime },m_{\perp }) &=&\pm \,sin\,(\widetilde{n}%
_{\perp }^{\prime },n_{\perp })\,\,\,\,\text{,}  \notag \\
\alpha ^{\prime } &=&\mp \,cos(\widetilde{n}_{\perp }^{\prime },n_{\perp
})\,\,\,\,\text{,}  \label{3.32d} \\
\beta ^{\prime } &=&\mp sin\,(\widetilde{n}_{\perp }^{\prime },n_{\perp
})\,\,\,\,\,\text{,}  \label{3.32e}
\end{eqnarray}
\begin{eqnarray}
\widetilde{n}_{\perp }^{\prime } &=&\alpha ^{\prime }\cdot n_{\perp }+\beta
^{\prime }\cdot m_{\perp }\,=  \notag \\
&=&\mp [cos(\widetilde{n}_{\perp }^{\prime },n_{\perp })\,\cdot n_{\perp
}+sin\,(\widetilde{n}_{\perp }^{\prime },n_{\perp })\,\cdot m_{\perp }]\,\,\,%
\text{.}  \label{3.32f}
\end{eqnarray}

If we denote the angle $(\widetilde{n}_{\perp }^{\prime },n_{\perp })$
between the vectors $\widetilde{n}_{\perp }^{\prime }$ and $n_{\perp }$ as $%
\theta ^{\prime }=(\widetilde{n}_{\perp }^{\prime },n_{\perp })$ then the
above relations could be written in the form 
\begin{eqnarray}
cos^2(\widetilde{n}_{\perp }^{\prime },m_{\perp }) &=&sin^2\theta ^{\prime
}\,\,\,\,\text{,}  \notag \\
sin^2(\widetilde{n}_{\perp }^{\prime },m_{\perp }) &=&cos^2\theta ^{\prime
}\,\,\,\text{,}  \notag \\
\alpha ^{\prime } &=&\mp \,cos\,\theta \,^{\prime }\,\,\,\text{,}
\label{3.31g} \\
\beta ^{\prime } &=&\mp sin\,\theta ^{\prime }\,\,\,\text{\thinspace
\thinspace ,}  \label{3.31h}
\end{eqnarray}
\begin{equation}
\widetilde{n}_{\perp }=\alpha \cdot n_{\perp }+\beta \cdot m_{\perp }\,=\mp
[cos\,\theta \,\cdot n_{\perp }+sin\,\theta \,\cdot m_{\perp }]\,\,\,\text{.}
\label{3.32g}
\end{equation}

From the relation 
\begin{equation}
\mp \frac{\overline{\omega }}{l_u}\cdot \widetilde{n}_{\perp }^{\prime }=\mp
\frac \omega {l_u}\cdot \widetilde{n}_{\perp }+\,_{rel}k_{\perp }\,\,\,
\label{3.31i}
\end{equation}
the relations for $_{rel}k_{\perp }$ follow 
\begin{eqnarray}
\mp \frac{\overline{\omega }}{l_u}\cdot g(\widetilde{n}_{\perp }^{\prime
},n_{\perp }) &=&\mp \frac \omega {l_u}\cdot g(\widetilde{n}_{\perp
},n_{\perp })+g(_{rel}k_{\perp },n_{\perp })\,\,\,\,\,\text{,}  \label{3.33}
\\
\mp \alpha ^{\prime }\cdot \frac{\overline{\omega }}{l_u} &=&\mp \alpha
\cdot \frac \omega {l_u}+g(_{rel}k_{\perp },n_{\perp })\,\,\,\text{,}
\label{3.34} \\
\frac{\overline{\omega }}{l_u}\cdot cos\,\theta \,^{\prime } &=&\frac \omega
{l_u}\cdot cos\,\theta +g(_{rel}k_{\perp },n_{\perp })\,\,\,\,\,\text{,}
\label{3.35} \\
\overline{\omega }\cdot cos\,\theta \,^{\prime } &=&\omega \cdot cos\,\theta
+l_u\cdot g(_{rel}k_{\perp },n_{\perp })\,\,\,\,\,\text{,}  \label{3.36}
\end{eqnarray}
\begin{eqnarray}
\mp \frac{\overline{\omega }}{l_u}\cdot g(\widetilde{n}_{\perp }^{\prime
},m_{\perp }) &=&\mp \frac \omega {l_u}\cdot g(\widetilde{n}_{\perp
},m_{\perp })+g(_{rel}k_{\perp },m_{\perp })\,\,\,\,\,\,\text{,}
\label{3.37} \\
\mp \beta ^{\prime }\cdot \frac{\overline{\omega }}{l_u} &=&\mp \beta \cdot
\frac \omega {l_u}+g(_{rel}k_{\perp },m_{\perp })\,\,\,\text{,}  \label{3.38}
\\
\frac{\overline{\omega }}{l_u}\cdot sin\,\theta \,^{\prime } &=&\frac \omega
{l_u}\cdot sin\,\theta +g(_{rel}k_{\perp },m_{\perp })\,\,\,\,\,\,\,\text{,}
\label{3.39} \\
\overline{\omega }\cdot sin\,\theta \,^{\prime } &=&\omega \cdot sin\,\theta
+l_u\cdot g(_{rel}k_{\perp },m_{\perp })\,\,\,\,\,\text{,}  \label{3.40} \\
\omega &=&l_u\cdot g(\widetilde{n}_{\perp },k_{\perp })\,\,\,\,\,\,\,\text{,}
\label{3.41}
\end{eqnarray}
\begin{eqnarray}
\frac{\overline{\omega }}\omega \cdot cos\,\theta \,^{\prime }
&=&cos\,\theta +\frac 1\omega \cdot l_u\cdot g(_{rel}k_{\perp },n_{\perp })=
\notag \\
&=&cos\,\theta +\frac{g(_{rel}k_{\perp },n_{\perp })}{g(\widetilde{n}_{\perp
},k_{\perp })}\,\,\,\,\,\,\,\text{,}  \label{3.42}
\end{eqnarray}
\begin{eqnarray}
\frac{\overline{\omega }}\omega \cdot sin\,\theta \,^{\prime }
&=&sin\,\theta +\frac 1\omega \cdot l_u\cdot g(_{rel}k_{\perp },m_{\perp })=
\notag \\
&=&sin\,\theta +\frac{g(_{rel}k_{\perp },m_{\perp })}{g(\widetilde{n}_{\perp
},k_{\perp })}\,\,\,\,\,\,\text{,}  \label{3.43}
\end{eqnarray}

From the last (above) two relations, it follows for $tg\theta ^{\prime }$%
\begin{equation}
tg\theta ^{\prime }=\frac{sin\,\theta \,^{\prime }}{cos\,\theta \,^{\prime }}%
=\frac{sin\,\theta +\frac{g(_{rel}k_{\perp },m_{\perp })}{g(\widetilde{n}%
_{\perp },k_{\perp })}}{cos\,\theta +\frac{g(_{rel}k_{\perp },n_{\perp })}{g(%
\widetilde{n}_{\perp },k_{\perp })}}\,\,\,\,\,\,\,\text{.}  \label{3.44}
\end{equation}

If we introduce the abbreviations 
\begin{equation}
\overline{S}:=\frac{g(_{rel}k_{\perp },m_{\perp })}{g(\widetilde{n}_{\perp
},k_{\perp })}\,\,\,\,\,\text{,}  \label{3.45}
\end{equation}
\begin{equation}
\overline{C}:=\frac{g(_{rel}k_{\perp },n_{\perp })}{g(\widetilde{n}_{\perp
},k_{\perp })}\,\,\,\,\text{,}  \label{3.46}
\end{equation}

the expression for $tg\theta ^{\prime }$ could be written in the form 
\begin{equation}
tg\theta ^{\prime }=\frac{sin\,\theta \,^{\prime }}{cos\,\theta \,^{\prime }}%
=\frac{sin\,\theta +\overline{S}}{cos\,\theta +\overline{C}}\,\,\,\,\text{.}
\label{3.47}
\end{equation}

The relations for $sin\,\theta ^{\prime }$ and $cos\,\theta ^{\prime }$ will
have the forms respectively: 
\begin{equation}
\frac{\overline{\omega }}\omega \cdot sin\,\theta \,^{\prime }=sin\,\theta +%
\overline{S\,}\,\,\,\,\,\,\,\,\,\,\text{,}  \label{3.48}
\end{equation}
\begin{equation}
\frac{\overline{\omega }}\omega \cdot cos\,\theta \,^{\prime }=cos\,\theta +%
\overline{C}\text{\thinspace \thinspace \thinspace \thinspace \thinspace
\thinspace \thinspace \thinspace \thinspace \thinspace \thinspace \thinspace
.}  \label{3.49}
\end{equation}

The angle $\theta ^{\prime }$ describes the deviation of the direction of
the vector $\overline{k}_{\perp }$ with respect to the vector $k_{\perp }$.
This type of deviation is usually related to the \textit{aberration} of the
wave vector $\widetilde{k}$ during its motion in a time interval. As we will
see below the aberration is depending on the relative velocity and relative
acceleration included implicitly in the terms $\overline{S}$ and $\overline{C%
}$.

From the expressions for $sin\,\theta ^{\prime }$ and $cos\,\theta ^{\prime
} $ the relation between the emitted frequency $\overline{\omega }$ and the
detected frequency $\omega $ follows in the forms 
\begin{equation}
\frac{\overline{\omega }^2}{\omega ^2}\cdot (sin^2\,\theta \,^{\prime
}+cos^2\,\theta \,^{\prime })=\frac{\overline{\omega }^2}{\omega ^2}%
=(sin\,\theta +\overline{S\,})^2+(cos\,\theta +\overline{C})^2\,\,\,\text{,}
\label{3.50}
\end{equation}
\begin{equation}
\overline{\omega }=[(sin\,\theta +\overline{S\,})^2+(cos\,\theta +\overline{C%
})^2]^{1/2}\cdot \omega \,\,\,\,\,\,\,\,\text{.}  \label{3.51}
\end{equation}

The above relation describe the change of the frequency of the emitter
during the motion of the signal from the emitter to the receiver (detector,
observer) and is related to the description of the Doppler effect and the
Hubble effect in spaces with affine connections and metrics considered as
models of space-time. It is assumed that the emitter is at rest at a given
time with respect to the frame of reference of an observer moving in
space-time and receiving signals from an emitter. If at rest with respect to
an observer, the emitter sends a signal with frequency $\overline{\omega }$
detected after a time interval as a frequency $\omega $ measured by the
observer in his proper frame of reference.

The task is now to find out the explicit form for $_{rel}k_{\perp }$. For
this purpose we would consider the change of a vector field $\xi _{\perp }$,
orthogonal to the vector field $u$, i.e. $g(u,\xi _{\perp })=0$, transported
along the vector field $u$.

\subsection{Change of a non-isotropic vector field $\protect\xi _{\perp }$
along a non-isotropic vector field $u$, when $g(u,\protect\xi _{\perp })=0$}

Let us now consider the change of the vector field $\xi _{\perp }$, $g(u,\xi
_{\perp })=0$, along the world line $x^i(\tau ,\lambda _0)$ of an observer
with tangent vector $u$. The vector $\xi _{\perp }$ could be expressed at a
point $A(\tau _0-d\tau ,\,\lambda _0)$ by means of the vector field $\xi
_{\perp }$ at the point $A^{\prime }(\tau _0,\lambda _0)$ by the use of the
covariant exponential operator (\cite{Manoff-2} - pp. 82-85) up to the
second order of $d\tau $%
\begin{eqnarray}
\xi _{\perp (A)} &:&=\xi _{\perp (\tau _0-d\tau ,\,\lambda _0)}:=\overline{%
\xi }_{(\tau _0,\,\lambda _0)}=  \notag \\
&=&\xi _{\perp (\tau _0,\,\lambda _0)}-d\tau \cdot \nabla _u\xi _{\perp \mid
}{}_{(\tau _0,\,\lambda _0)}+\frac 12\cdot d\tau ^2\cdot \nabla _u\nabla
_u\xi _{\perp \mid (\tau _0,\,\lambda _0)}\,\,\,\,\text{.}  \label{3.52}
\end{eqnarray}

The vector $\overline{\xi }_{\perp (\tau _0\,,\,\,\lambda _0)}$ could not
be, in general, collinear to $\xi _{\perp (\tau _0,\,\,\lambda _0)}$ and
orthogonal to $u_{(\tau _0,\,\lambda _0)}$. If we, further, consider the
part of $\overline{\xi }_{(\tau _0,\,\lambda _0)}$, orthogonal to $u_{(\tau
_0,\,\,\lambda _0)}$ at the point $A^{\prime }(\tau _0,\,\lambda _0)$, we
obtain 
\begin{eqnarray}
\overline{\xi }_{\perp (\tau _0,\,\lambda _0)} &:&=\overline{g}[h_u(%
\overline{\xi }_{(\tau _0,\,\lambda _0)})]=\overline{g}[h_u(\xi _{\perp
})]_{(\tau _0,\,\,\lambda _0)}-  \notag \\
&&-d\tau \cdot \overline{g}[h_u(\nabla _u\xi _{\perp })]_{(\tau
_0,\,\,\lambda _0)}+\frac 12\cdot d\tau ^2\cdot \overline{g}[h_u(\nabla
_u\nabla _u\xi _{\perp })]_{(\tau _0,\,\lambda _0)}\,\,\,\,\text{.}
\label{3.53}
\end{eqnarray}

Since, 
\begin{eqnarray}
\overline{g}[h_u(\xi _{\perp })] &=&\xi _{\perp }\,\,\,\text{,}  \label{3.54}
\\
\overline{g}[h_u(\nabla _u\xi _{\perp })] &=&\,_{rel}v\,\,\,\,\text{,}
\label{3.55} \\
\overline{g}[h_u(\nabla _u\nabla _u\xi _{\perp })] &=&\,_{rel}a\text{%
\thinspace \thinspace \thinspace \thinspace ,}  \label{3.56}
\end{eqnarray}
we have the relation 
\begin{equation}
\overline{\xi }_{\perp (\tau _0,\,\lambda _0)}=\xi _{\perp (\tau
_0,\,\lambda _0)}-d\tau \cdot \,_{rel}v_{(\tau _0,\,\lambda _0)}+\frac
12\cdot d\tau ^2\cdot \,_{rel}a_{(\tau _0,\,\lambda _0)}\,\,\,\,\,\text{.}
\label{3.57}
\end{equation}

Therefore, the vector $\xi _{\perp (\tau _0-d\tau ,\,\lambda _0)}$ with $%
g(\xi _{\perp },u)_{(\tau _0-d\tau ,\,\lambda _0)}=0$ could be considered as
the vector $\overline{\xi }_{\perp (\tau _0,\,\lambda _0)}$ with $g(%
\overline{\xi }_{\perp },u)_{(\tau _0,\,\lambda _0)}=0$ if transported from
the point $A(\tau _0-d\tau ,\,\lambda _0)\equiv $ $A_{(\tau _0-d\tau
,\,\lambda _0)}$ to the point $A^{\prime }(\tau _0,\lambda _0)\equiv
A_{(\tau _0,\,\lambda _0)}$ at the world line with parameter $\tau $.

\textit{Remark}. Analogous considerations could be made with the transport
of the vector $\xi _{\perp }$ from the point $A^{\prime \prime }(\tau
_0+d\tau ,\,\lambda _0)$ to the point $A(\tau _0,\,\lambda _0)$: 
\begin{equation}
\xi _{\perp (\tau _0+d\tau ,\,\lambda _0)}:=\overline{\overline{\xi }}%
_{\perp (\tau _0,\,\lambda _0)}=\xi _{\perp (\tau _0,\,\lambda _0)}+d\tau
\cdot \,_{rel}v_{(\tau _0,\,\lambda _0)}+\frac 12\cdot d\tau ^2\cdot
\,_{rel}a_{(\tau _0,\,\lambda _0)}\,\,\,\,\text{.}  \label{3.58}
\end{equation}

If we summarize the expressions for $\overline{\xi }_{\perp (\tau
_0,\,\lambda _0)}$ and $\overline{\overline{\xi }}_{\perp (\tau _0,\,\lambda
_0)}$ we can find a relation between the vectors $\overline{\xi }_{\perp
(\tau _0,\,\lambda _0)}$,\thinspace $\overline{\xi }_{\perp (\tau
_0,\,\lambda _0)}$,\thinspace $\overline{\overline{\xi }}_{\perp (\tau
_0,\,\lambda _0)}$, and the relative acceleration $\,_{rel}a_{(\tau
_0,\,\lambda _0)}$%
\begin{equation}
\overline{\overline{\xi }}_{\perp (\tau _0,\,\lambda _0)}+\overline{\xi }%
_{\perp (\tau _0,\,\lambda _0)}=2\cdot \xi _{\perp (\tau _0,\,\lambda
_0)}+d\tau ^2\cdot \,_{rel}a_{(\tau _0,\,\lambda _0)}\,\,\,\,\,\text{,}
\label{3.59}
\end{equation}
or 
\begin{equation}
d\tau ^2\cdot \,_{rel}a_{(\tau _0,\,\lambda _0)}=\overline{\overline{\xi }}%
_{\perp (\tau _0,\,\lambda _0)}+\overline{\xi }_{\perp (\tau _0,\,\lambda
_0)}-2\cdot \xi _{\perp (\tau _0,\,\lambda _0)}\,\,\,\,\text{.}  \label{3.60}
\end{equation}

If the proper time interval $d\tau $ is expressed by the use of the
relations between the proper time $\tau $ and the space distance $dl$,
covered by a signal propagating with a velocity with absolute value $l_u$
[from point of view of the observer with world line $x^i(\tau ,\,\lambda _0)$%
], as 
\begin{equation}
d\tau =\pm \frac{l_{\xi _{\perp }}}{l_u}\cdot d\lambda =\frac{dl}{l_u}\text{
\thinspace \thinspace \thinspace }  \label{3.61}
\end{equation}
then the relation between $\overline{\xi }_{\perp (\tau _0,\,\lambda _0)}$
and $\xi _{\perp (\tau _0,\,\lambda _0)}$ could be written in the form 
\begin{equation}
\overline{\xi }_{\perp (\tau _0,\,\lambda _0)}=\xi _{\perp (\tau
_0,\,\lambda _0)}-\frac{dl}{l_u}\cdot \,_{rel}v_{(\tau _0,\,\lambda
_0)}+\frac 12\cdot (\frac{dl}{l_u})^2\cdot \,_{rel}a_{(\tau _0,\,\lambda
_0)}\,\,\,\text{.}  \label{3.62}
\end{equation}

For every point $P^{\prime }(\tau ,\,\lambda _0)$ of the world line of the
observer $x^i(\tau ,\lambda _0)$ the above relation between $P^{\prime
}(\tau ,\,\lambda _0)$ and another point $P(\tau -d\tau ,\,\lambda _0)$ is
valid. In the further consideration we will omit the indications for the
corresponding points: 
\begin{equation}
\overline{\xi }_{\perp }=\xi _{\perp }-\frac{dl}{l_u}\cdot \,_{rel}v+\frac
12\cdot (\frac{dl}{l_u})^2\cdot \,_{rel}a\,\,\,\text{.}  \label{3.63}
\end{equation}

The vector $\overline{\xi }_{\perp }$ is, in general, not collinear to the
vector $\xi _{\perp }$. It could be represented by the use of the unit
vectors $n_{\perp }$ and $m_{\perp }$ in the form 
\begin{eqnarray}
\overline{\xi }_{\perp } &=&l_{\overline{\xi }_{\perp }}\cdot \widetilde{n}%
_{\perp }^{\prime }=l_{\overline{\xi }_{\perp }}\cdot (\alpha ^{\prime
}\cdot n_{\perp }+\beta ^{\prime }\cdot m_{\perp })=  \notag \\
&=&\mp l_{\overline{\xi }_{\perp }}\cdot [cos\,\theta ^{\prime }\,\cdot
n_{\perp }+sin\,\theta ^{\prime }\,\cdot m_{\perp }]\,\,\,\,\text{.}
\label{3.64}
\end{eqnarray}

On the other side, the vector $\xi _{\perp }$ has the form $\xi _{\perp
}=l_{\xi _{\perp }}\cdot n_{\perp }$, where $g(\xi _{\perp },m_{\perp
})=l_{\xi _{\perp }}\cdot g(n_{\perp },m_{\perp })=0$. In the same way, we
can express the relative velocity $_{rel}v$ and the relative acceleration $%
_{rel}a$ by means of their structures, related to the vectors $n_{\perp }$
and $m_{\perp }$.

If we project the expression for $\overline{\xi }_{\perp }$ along the unit
vectors $n_{\perp }$ and $m_{\perp }$ correspondingly we obtain 
\begin{eqnarray}
g(\overline{\xi }_{\perp },n_{\perp }) &=&l_{\overline{\xi }_{\perp }}\cdot
cos\,\theta ^{\prime }=  \notag \\
&=&g(\xi _{\perp },n_{\perp })-\frac{dl}{l_u}\cdot g(_{rel}v,n_{\perp
})+\frac 12\cdot (\frac{dl}{l_u})^2\cdot g(_{rel}a,n_{\perp })=  \notag \\
&=&\mp l_{\xi _{\perp }}-\frac{dl}{l_u}\cdot g(_{rel}v,n_{\perp })+\frac
12\cdot (\frac{dl}{l_u})^2\cdot g(_{rel}a,n_{\perp })\,\,\,\text{,}
\label{3.65}
\end{eqnarray}
\begin{eqnarray}
g(\overline{\xi }_{\perp },m_{\perp }) &=&l_{\overline{\xi }_{\perp }}\cdot
sin\,\theta ^{\prime }=  \notag \\
&=&g(\xi _{\perp },m_{\perp })-\frac{dl}{l_u}\cdot g(_{rel}v,m_{\perp
})+\frac 12\cdot (\frac{dl}{l_u})^2\cdot g(_{rel}a,m_{\perp })=  \notag \\
&=&-\frac{dl}{l_u}\cdot g(_{rel}v,m_{\perp })+\frac 12\cdot (\frac{dl}{l_u}%
)^2\cdot g(_{rel}a,m_{\perp })\,\,\,\,\,\text{.}  \label{3.66}
\end{eqnarray}

We can find now the explicit forms of $g(_{rel}v,n_{\perp })$, $%
g(_{rel}v,m_{\perp })$, $g(_{rel}a,n_{\perp })$, and $g(_{rel}a,m_{\perp })$.

\subsubsection{Explicit form of $l_{\overline{\protect\xi }_{\perp }}\cdot
cos\,\protect\theta ^{\prime }$ and $l_{\overline{\protect\xi }_{\perp
}}\cdot sin\,\protect\theta ^{\prime }$}

Let us recall the explicit form of $_{rel}v$ and $_{rel}a$ with respect to
their decompositions in centrifugal (centripetal) and Coriolis velocities
and accelerations respectively \cite{Manoff-6}.

The relative velocity $_{rel}v$ could be represented in the form 
\begin{equation}
_{rel}v=v_z+v_c\,\,\,\,\,\text{,}  \label{3.67}
\end{equation}
where 
\begin{eqnarray}
v_z &=&\mp l_{v_z}\cdot n_{\perp }=H\cdot l_{\xi _{\perp }}\cdot n_{\perp
}=H\cdot \xi _{\perp }\,\,\,\,\,\,\text{,\thinspace \thinspace \thinspace
\thinspace \thinspace \thinspace \thinspace \thinspace \thinspace }n_{\perp
}=\frac{\xi _{\perp }}{l_{\xi _{\perp }}}\,\,\,\,\text{,}  \label{3.68} \\
v_c &=&\mp l_{v_c}\cdot m_{\perp }=H_c\cdot l_{\xi _{\perp }}\cdot m_{\perp
}=H_c\cdot \eta _{\perp }\,\,\,\,\,\text{,\thinspace \thinspace \thinspace
\thinspace \thinspace \thinspace }m_{\perp }=\frac{\eta _{\perp }}{l_{\xi
_{\perp }}}=\frac{v_c}{l_{v_c}}\,\,\,\text{.}  \label{3.69}
\end{eqnarray}

The relative acceleration $_{rel}a$ could be represented in the form 
\begin{equation}
_{rel}a=a_z+a_c\text{ \thinspace \thinspace \thinspace ,}  \label{3.70}
\end{equation}
where 
\begin{eqnarray}
a_z &=&\mp l_{a_z}\cdot n_{\perp }=\overline{q}\cdot l_{\xi _{\perp }}\cdot
n_{\perp }=\overline{q}\cdot \xi _{\perp }\,\,\,\,\,\,\text{,}  \label{3.71}
\\
a_c &=&\mp l_{a_c}\cdot m_{\perp }=\overline{q}_c\cdot l_{\xi _{\perp
}}\cdot m_{\perp }=\overline{q}\,_c\cdot \eta _{\perp }\,\,\,\text{.}
\label{3.72}
\end{eqnarray}

For the expressions $g(_{rel}v,n_{\perp })$ and $g(_{rel}v,m_{\perp })$ we
obtain respectively 
\begin{eqnarray}
g(_{rel}v,n_{\perp }) &=&l_{v_z}=\mp H\cdot l_{\xi _{\perp }}\,\,\,\,\,\text{%
,}  \label{3.73} \\
g(_{rel}v,m_{\perp }) &=&l_{v_c}=\mp H_c\cdot l_{\xi _{\perp }}\,\,\,\text{.}
\label{3.74}
\end{eqnarray}

For the expressions $g(_{rel}a,n_{\perp })$ and $g(_{rel}a,m_{\perp })$ we
obtain respectively 
\begin{eqnarray}
g(_{rel}a,n_{\perp }) &=&l_{a_z}=\mp \overline{q}\cdot l_{\xi _{\perp
}}\,\,\,\,\,\,\,\text{,}  \label{3.75} \\
g(_{rel}a,m_{\perp }) &=&l_{a_c}=\mp \overline{q}_c\cdot l_{\xi _{\perp
}}\,\,\,\,\,\,\text{.}  \label{3.76}
\end{eqnarray}

By means of the above relations, it follows for $l_{\overline{\xi }_{\perp
}}\cdot cos\,\theta ^{\prime }$ and $l_{\overline{\xi }_{\perp }}\cdot
sin\,\theta ^{\prime }$ respectively 
\begin{eqnarray}
l_{\overline{\xi }_{\perp }}\cdot cos\,\theta ^{\prime } &=&\mp l_{\xi
_{\perp }}-\frac{dl}{l_u}\cdot l_{v_z}+\frac 12\cdot (\frac{dl}{l_u})^2\cdot
l_{a_z}=  \notag \\
&=&\mp l_{\xi _{\perp }}\pm \frac{dl}{l_u}\cdot H\cdot l_{\xi _{\perp }}\mp
\frac 12\cdot (\frac{dl}{l_u})^2\cdot \overline{q}\cdot l_{\xi _{\perp }}= 
\notag \\
&=&\mp l_{\xi _{\perp }}\cdot [1-\frac{dl}{l_u}\cdot H+\frac 12\cdot (\frac{%
dl}{l_u})^2\cdot \overline{q}]\,\,\,\text{.}  \label{3.77}
\end{eqnarray}
\begin{eqnarray}
l_{\overline{\xi }_{\perp }}\cdot sin\,\theta ^{\prime } &=&-\frac{dl}{l_u}%
\cdot l_{v_c}+\frac 12\cdot (\frac{dl}{l_u})^2\cdot l_{a_c}=  \notag \\
&=&\pm \frac{dl}{l_u}\cdot H_c\cdot l_{\xi _{\perp }}\mp \frac 12\cdot (%
\frac{dl}{l_u})^2\cdot \overline{q}_c\cdot l_{\xi _{\perp }}=  \notag \\
&=&\pm l_{\xi _{\perp }}\cdot [\frac{dl}{l_u}\cdot H_c-\frac 12\cdot (\frac{%
dl}{l_u})^2\cdot \overline{q}_c]=  \notag \\
&=&\mp l_{\xi _{\perp }}\cdot [-\frac{dl}{l_u}\cdot H_c+\frac 12\cdot (\frac{%
dl}{l_u})^2\cdot \overline{q}_c]  \label{3.78}
\end{eqnarray}

If we introduce the abbreviations 
\begin{eqnarray}
C &=&-\frac{dl}{l_u}\cdot H+\frac 12\cdot (\frac{dl}{l_u})^2\cdot \overline{q%
}\,\,\,\,\,\,\text{,}  \label{3.79} \\
S &=&-\frac{dl}{l_u}\cdot H_c+\frac 12\cdot (\frac{dl}{l_u})^2\cdot 
\overline{q}_c\,\,\,\,\text{,}  \label{3.80}
\end{eqnarray}
the expressions for $l_{\overline{\xi }_{\perp }}\cdot sin\,\theta ^{\prime
} $ and $l_{\overline{\xi }_{\perp }}\cdot cos\,\theta ^{\prime }$ could
also be written in the forms 
\begin{eqnarray}
l_{\overline{\xi }_{\perp }}\cdot sin\,\theta ^{\prime } &=&\mp l_{\xi
_{\perp }}\cdot S\,\,\,\,\text{,}  \label{3.81} \\
l_{\overline{\xi }_{\perp }}\cdot cos\,\theta ^{\prime } &=&\mp l_{\xi
_{\perp }}\cdot (1+C)\,\,\,\text{.}  \label{3.82}
\end{eqnarray}

The change of the direction of the vector $\xi _{\perp }$ in the time
interval $d\tau $ of the proper time $\tau $ of the observer on his world
line can now be represented as 
\begin{equation}
tg\,\theta ^{\prime }=\frac S{1+C}\text{ \thinspace \thinspace .}
\label{3.83}
\end{equation}

The change of the length of the vector $\xi _{\perp }$ in the time interval $%
d\tau $ could be found in the form 
\begin{eqnarray}
l_{\overline{\xi }_{\perp }}^2 &=&l_{\xi _{\perp }}^2\cdot
[(1+C)^2+S^2]\,\,\,\,\text{,}  \label{3.84} \\
l_{\overline{\xi }_{\perp }} &=&[(1+C)^2+S^2]^{1/2}\cdot l_{\xi _{\perp
}}\,\,\,\,\,\text{,}  \label{3.85} \\
l_{\overline{\xi }_{\perp }} &>&0\,\,\,\,\,\,\text{,\thinspace \thinspace
\thinspace \thinspace \thinspace \thinspace \thinspace \thinspace }l_{\xi
_{\perp }}>0\,\,\,\,\text{.}  \label{3.86}
\end{eqnarray}

\textit{Special case:} The vector $\overline{\xi }_{\perp }$ is collinear to
the vector $\xi _{\perp }$. Then $\widetilde{n}_{\perp }^{\prime }=n_{\perp
} $, $sin\,\theta ^{\prime }=0$, \thinspace $cos\,\theta ^{\prime }=\mp 1$, $%
S=0$, and 
\begin{equation}
l_{\overline{\xi }_{\perp }}=(1+C)\cdot l_{\xi _{\perp }}\,\,\,\,\,\,\,\,%
\text{.}  \label{3.87}
\end{equation}

\textit{Special case}: The vector $\overline{\xi }_{\perp }$ is orthogonal
to the vector $\xi _{\perp }$. Then $\widetilde{n}_{\perp }^{\prime
}=m_{\perp }$, $sin\,\theta ^{\prime }=\mp 1$, \thinspace $cos\,\theta
^{\prime }=0$, $C=-1$, and 
\begin{equation}
l_{\overline{\xi }_{\perp }}=S\cdot l_{\xi _{\perp }}\,\,\,\,\,\text{.}
\label{3.88}
\end{equation}

The change of the length of the vector field $\xi _{\perp }$ along the world
line of an observer shows the role of the relative velocity and the relative
acceleration for the deformation of the vector field $\xi _{\perp }$ along
the world line. This deformation is depending on the corresponding Hubble
functions $H$ and $H_c$ and acceleration parameters $\,\overline{q}$ and $%
\overline{q}_c$.

Since the deformation of the contravariant non-isotropic vector field $%
k_{\perp }$ (orthogonal to the vector field $u$) characterizes uniquely the
deformation of the wave vector $\widetilde{k}$ we could consider the change
of $k_{\perp }$ along the world line of the observer in analogous way as it
has been done for the vector field $\xi _{\perp }$.

\subsection{Change of the vector $k_{\perp }$ along the world line of an
observer}

Let us now consider the change of the vector field $k_{\perp }$, $%
g(u,k_{\perp })=0$, along the world line $x^i(\tau ,\lambda _0)$ of an
observer with tangent vector $u$. For this aim we will consider first of all
the wave vector field $\widetilde{k}$ and then $\widetilde{k}$ will be
projected to direction orthogonal to the vector field $u$.

The vector field $\widetilde{k}$ could be expressed at a point $A(\tau
_0-d\tau ,\,\lambda _0)$ by means of the vector field $\widetilde{k}$ at the
point $A^{\prime }(\tau _0,\lambda _0)$ by the use of the covariant
exponential operator (\cite{Manoff-2} - pp. 82-85) up to the second order of 
$d\tau $%
\begin{eqnarray}
\widetilde{k}_{(A)} &:&=\widetilde{k}_{(\tau _0-d\tau ,\,\lambda _0)}:=%
\overline{k}_{(\tau _0,\,\lambda _0)}=  \notag \\
&=&\widetilde{k}_{(\tau _0,\,\lambda _0)}-d\tau \cdot \nabla _{u\,}{}%
\widetilde{k}_{\mid (\tau _0,\,\lambda _0)}+\frac 12\cdot d\tau ^2\cdot
\nabla _u\nabla _u\widetilde{k}_{\mid (\tau _0,\,\lambda _0)}\,\,\,\,\text{.}
\label{3.89}
\end{eqnarray}

The orthogonal to $u$ parts of $\widetilde{k}$ and $\overline{k}$ at the
point $A^{\prime }(\tau _0,\lambda _0)$ could be found after projection of
both the vectors by means of the projection metric $h_u$%
\begin{eqnarray}
\overline{g}[h_u(\widetilde{k})]_{(\tau _0-d\tau ,\,\lambda _0)} &:&=%
\overline{g}[h_u(\overline{k})]_{(\tau _0,\,\lambda _0)}=  \label{3.91} \\
&=&\overline{g}[h_u(\widetilde{k})]_{(\tau _0,\,\lambda _0)}-d\tau \cdot 
\overline{g}[h_u(\nabla _{u\,}{}\widetilde{k})]_{\mid (\tau _0,\,\lambda
_0)}+  \notag \\
&&+\frac 12\cdot d\tau ^2\cdot \overline{g}[h_u(\nabla _u\nabla _u\widetilde{%
k})]_{\mid (\tau _0,\,\lambda _0)}\,\,\,\,\text{.}  \label{3.92}
\end{eqnarray}

Since, 
\begin{eqnarray}
\overline{g}[h_u(\overline{k})]_{(\tau _0,\,\lambda _0)} &=&\overline{k}%
_{\perp (\tau _0,\,\lambda _0)}\,\,\,\,\,\,\text{,}  \label{3.93} \\
\overline{g}[h_u(\widetilde{k})]_{(\tau _0,\,\lambda _0)} &=&k_{\perp (\tau
_0,\,\lambda _0)}\,\,\,\,\,\,\,\text{,}  \label{3.94} \\
\overline{g}[h_u(\nabla _{u\,}{}\widetilde{k})]_{\mid (\tau _0,\,\lambda
_0)} &=&(\nabla _{u\,}{}\widetilde{k})_{\perp \mid (\tau _0,\,\lambda
_0)}\,\,\,\,\,\,\text{,}  \label{3.95} \\
\overline{g}[h_u(\nabla _u\nabla _u\widetilde{k})]_{\mid (\tau _0,\,\lambda
_0)} &=&(\nabla _u\nabla _u\widetilde{k})_{\perp \mid (\tau _0,\,\lambda
_0)}\,\,\,\,\text{,}  \label{3.96}
\end{eqnarray}
\begin{equation}
\overline{\omega }_{(\tau _0,\,\lambda _0)}=g(\overline{k},u)_{(\tau
_0,\,\lambda _0)}\,\,\,\,\,\text{,\thinspace \thinspace \thinspace
\thinspace \thinspace \thinspace \thinspace \thinspace \thinspace \thinspace
\thinspace \thinspace }\omega _{(\tau _0,\,\lambda _0)}=g(\widetilde{k}%
,u)_{(\tau _0,\,\lambda _0)}\,\,\,\,\text{,}  \label{3.97}
\end{equation}
we have the relation 
\begin{eqnarray}
\overline{k}_{\perp (\tau _0,\,\lambda _0)} &=&k_{\perp (\tau _0,\,\lambda
_0)}-d\tau \cdot \overline{g}[h_u(\nabla _{u\,}{}\widetilde{k})]_{\mid (\tau
_0,\,\lambda _0)}+  \notag \\
&&+\frac 12\cdot d\tau ^2\cdot \overline{g}[h_u(\nabla _u\nabla _u\widetilde{%
k})]_{\mid (\tau _0,\,\lambda _0)}\,\,\,\,\text{.}  \label{3.98}
\end{eqnarray}

The vectors $\overline{g}[h_{u}(\nabla _{u}\widetilde{k})]_{(\tau
_{0},\,\,\lambda _{0})}$ and $\overline{g}[h_{u}(\nabla _{u}\nabla _{u}%
\widetilde{k})]_{(\tau _{0},\,\lambda _{0})}$ could be represented by the
use of the kinematic characteristics of the relative velocity and relative
acceleration \cite{Manoff-2} in the forms 
\begin{eqnarray}
\overline{g}[h_{u}(\nabla _{u\,}{}\widetilde{k})] &=&\overline{g}[h_{u}(%
\frac{g(u,\widetilde{k})}{e}\cdot a+\pounds _{u}\widetilde{k})+d(\widetilde{k%
})]=  \notag \\
&=&\overline{g}[h_{u}(\pm \frac{\omega }{l_{u}^{2}}\cdot a+\pounds _{u}%
\widetilde{k})+d(\widetilde{k})]=  \notag \\
&=&\overline{g}[\pm \frac{\omega }{l_{u}^{2}}\cdot h_{u}(a)+h_{u}(\pounds %
_{u}\widetilde{k})+d(\widetilde{k})]\,\,\,\,\,\,\text{,}  \label{3.99}
\end{eqnarray}%
\begin{eqnarray}
\overline{g}[h_{u}(\nabla _{u}\nabla _{u}\widetilde{k})] &=&\overline{g}%
\{h_{u}[\frac{g(u,\widetilde{k})}{e}\cdot \nabla _{u}a+k(g)(\pounds _{u}%
\widetilde{k})+\nabla _{u}(\pounds _{u}\widetilde{k})]+A(\widetilde{k})\}= 
\notag \\
&=&\overline{g}\{\pm \frac{\omega }{l_{u}^{2}}\cdot h_{u}(\nabla
_{u}a)+h_{u}[k(g)(\pounds _{u}\widetilde{k})]+h_{u}[\nabla _{u}(\pounds _{u}%
\widetilde{k})]+  \notag \\
&&+A(\widetilde{k})\}\,\,\text{.}  \label{3.100}
\end{eqnarray}

If we assume that $\pounds _u\widetilde{k}=0$ we obtain the relations 
\begin{eqnarray}
\overline{g}[h_u(\nabla _{u\,}{}\widetilde{k})] &=&\overline{g}[\pm \frac
\omega {l_u^2}\cdot h_u(a)+d(\widetilde{k})]=  \notag \\
&=&\pm \frac \omega {l_u^2}\cdot \overline{g}[h_u(a)]+\overline{g}%
[d(k_{\perp })]=  \notag \\
&=&\pm \frac \omega {l_u^2}\cdot a_{\perp }+\overline{g}[d(k_{\perp
})]\,\,\,\,\,\,\,\text{,}  \label{3.101}
\end{eqnarray}
\begin{eqnarray}
\overline{g}[h_u(\nabla _u\nabla _u\widetilde{k})] &=&\pm \frac \omega
{l_u^2}\cdot \overline{g}[h_u(\nabla _ua)]+\overline{g}[A(k_{\perp })]= 
\notag \\
&=&\pm \frac \omega {l_u^2}\cdot (\nabla _ua)_{\perp }+\overline{g}%
[A(k_{\perp })]\,\,\,\text{,}  \label{3.102}
\end{eqnarray}
where 
\begin{eqnarray}
\frac{g(u,\widetilde{k})}e &=&\frac \omega {\pm l_u^2}=\pm \frac \omega
{l_u^2}\text{ \thinspace \thinspace \thinspace ,}  \label{3.103} \\
a_{\perp } &=&\overline{g}[h_u(a)]\,\,\,\,\,\,\,\text{,}  \label{3.104} \\
(\nabla _ua)_{\perp } &=&\overline{g}[h_u(\nabla _ua)]\,\,\,\,\,\text{,}
\label{3.105} \\
\overline{g}[d(\widetilde{k})] &=&\overline{g}[d(k_{\perp
})]\,\,\,\,\,\,\,\,\,\text{,}  \label{3.106} \\
\overline{g}[A(\widetilde{k})] &=&\overline{g}[A(k_{\perp })]\,\,\,\,\,\,\,\,%
\text{.}  \label{3.107}
\end{eqnarray}

\textit{Remark}. The condition $\pounds _u\widetilde{k}=0$ assures the
possibility for introducing co-ordinates with tangent vectors $u$ and $%
\widetilde{k}$ respectively.

On the other side, the vector $k_{\perp }$ could be represented in its
projections along the vectors $n_{\perp }$ and $m_{\perp }$ in the form 
\begin{eqnarray}
k_{\perp } &=&\mp l_{k_{\perp }}\cdot \widetilde{n}_{\perp }=\mp \frac
\omega {l_u}\,\cdot \widetilde{n}_{\perp }\,\,\,\,\,\,\text{,}  \label{3.108}
\\
\widetilde{n}_{\perp } &=&\alpha \cdot n_{\perp }+\beta \cdot m_{\perp
}\,=\mp [cos\,\theta \,\cdot n_{\perp }+sin\,\theta \,\cdot m_{\perp }]\,\,\,%
\text{.\thinspace \thinspace }  \label{3.109}
\end{eqnarray}

Then 
\begin{equation}
k_{\perp }=\frac \omega {l_u}\cdot [cos\,\theta \,\cdot n_{\perp
}+sin\,\theta \,\cdot m_{\perp }]\,\,\,\text{.}  \label{3.110}
\end{equation}

In analogous way, the vector $\overline{k}_{\perp }$ could be represented in
the form 
\begin{eqnarray}
\overline{k}_{\perp } &=&\mp l_{\overline{k}_{\perp }}\cdot \widetilde{n}%
_{\perp }^{\prime }=\mp \frac{\overline{\omega }}{l_u}\,\cdot \widetilde{n}%
_{\perp }^{\prime }\,\,\,\,\,\,\text{,}  \label{3.111} \\
\widetilde{n}_{\perp }^{\prime } &=&\alpha ^{\prime }\cdot n_{\perp }+\beta
^{\prime }\cdot m_{\perp }\,=\mp [cos\,\theta ^{\prime }\,\cdot n_{\perp
}+sin\,\theta \,^{\prime }\cdot m_{\perp }]\,\,\,\text{,}  \label{3.112} \\
\overline{k}_{\perp } &=&\frac{\overline{\omega }}{l_u}\cdot [cos\,\theta
^{\prime }\,\cdot n_{\perp }+sin\,\theta ^{\prime }\,\cdot m_{\perp }]\,\,\,%
\text{.}  \label{3.113}
\end{eqnarray}

The representation of $\overline{k}_{\perp }$ in the form $\overline{k}%
_{\perp }=k_{\perp }+\,_{rel}k_{\perp }$ will be given now in the form 
\begin{eqnarray}
\overline{k}_{\perp } &=&k_{\perp }+\,_{rel}k_{\perp }=  \label{3.114} \\
&=&k_{\perp }-d\tau \cdot \overline{g}[h_u(\nabla _{u\,}{}\widetilde{k})]+ 
\notag \\
&&+\frac 12\cdot d\tau ^2\cdot \overline{g}[h_u(\nabla _u\nabla _u\widetilde{%
k})]\,\,\,\,\,\text{,}\,  \label{3.115}
\end{eqnarray}
where 
\begin{eqnarray}
_{rel}k_{\perp } &=&-d\tau \cdot (\nabla _{u\,}{}\widetilde{k})_{\perp
}+\frac 12\cdot d\tau ^2\cdot (\nabla _u\nabla _u\widetilde{k})_{\perp }= 
\notag \\
&=&-d\tau \cdot \{\pm \frac \omega {l_u^2}\cdot a_{\perp }+\overline{g}%
[d(k_{\perp })]\}+  \notag \\
&&+\frac 12\cdot d\tau ^2\cdot \{\pm \frac \omega {l_u^2}\cdot (\nabla
_ua)_{\perp }+\overline{g}[A(k_{\perp })]\}\,\,\,\text{,}  \label{3.116}
\end{eqnarray}
\begin{eqnarray}
(\nabla _{u\,}{}\widetilde{k})_{\perp } &=&\overline{g}[h_u(\nabla _{u\,}{}%
\widetilde{k})]\,\,\,\,\,\,\,\text{,}  \label{3.117} \\
(\nabla _u\nabla _u\widetilde{k})_{\perp } &=&\overline{g}[h_u(\nabla
_u\nabla _u\widetilde{k})]\,\,\,\,\text{.}  \label{3.118}
\end{eqnarray}

The terms in $_{rel}k_{\perp }$ could be further represented by means of the
structures of the relative velocity and the relative acceleration
corresponding to the centrifugal (centripetal) and Coriolis velocities and
accelerations.

\subsubsection{Representation of $_{rel}k_{\perp }$ by means of the
centrifugal (centripetal) and Coriolis velocities and accelerations}

1. The orthogonal to $u$ acceleration $a_{\perp }$ could be found after
projection by the use of the projective metrics 
\begin{eqnarray}
h_{\xi _{\perp }} &=&g-\frac 1{g(\xi _{\perp },\xi _{\perp })}\cdot g(\xi
_{\perp })\otimes g(\xi _{\perp })\text{ \thinspace \thinspace \thinspace
\thinspace \thinspace \thinspace ,}  \label{3.119} \\
h^{\xi _{\perp }} &=&\overline{g}-\frac 1{g(\xi _{\perp },\xi _{\perp
})}\cdot \xi _{\perp }\otimes \xi _{\perp }\,\,\,\,\,\,\text{,}
\label{3.120}
\end{eqnarray}
in parts collinear to the vector field $\xi _{\perp }$ and orthogonal to $%
\xi _{\perp }$. At the same time both the parts are orthogonal to the vector
field $u$. 
\begin{eqnarray}
a_{\perp } &=&\frac{g(a_{\perp },\xi _{\perp })}{g(\xi _{\perp },\xi _{\perp
})}\cdot \xi _{\perp }+\overline{g}[h_{\xi _{\perp }}(a_{\perp })]=
\label{3.121} \\
&=&\frac{l_{\xi _{\perp }}^2}{\mp l_{\xi _{\perp }}^2}\cdot g(a_{\perp
},n_{\perp })\cdot n_{\perp }+\overline{g}[h_{\xi _{\perp }}(a_{\perp })]= 
\notag \\
&=&\mp g(a_{\perp },n_{\perp })\cdot n_{\perp }+\overline{g}[h_{\xi _{\perp
}}(a_{\perp })]=  \notag \\
&=&(a_{\perp })_z+(a_{\perp })_c\,\,\,\,\,\text{,}  \label{3.123}
\end{eqnarray}
where 
\begin{eqnarray}
(a_{\perp })_z &=&\mp g(a_{\perp },n_{\perp })\cdot n_{\perp }\,\,\,\,\,\,%
\text{,}  \label{3.122} \\
(a_{\perp })_c\, &=&\overline{g}[h_{\xi _{\perp }}(a_{\perp })]\,\,\,\,\,%
\text{,}  \label{3.124} \\
g(\xi _{\perp },(a_{\perp })_c) &=&0\text{ \thinspace \thinspace ,\thinspace
\thinspace \thinspace \thinspace }  \label{3.125} \\
(a_{\perp })_z &=&\mp \text{\thinspace }l_{(a_{\perp })_z}\cdot n_{\perp }%
\text{ \thinspace \thinspace \thinspace ,}  \label{3.126} \\
(a_{\perp })_c &=&\mp \text{\thinspace }l_{(a_{\perp })_c}\cdot m_{\perp
}\,\,\,\,\,\,\,\,\text{.}  \label{3.127}
\end{eqnarray}

2. The orthogonal to $u$ change $(\nabla _ua)_{\perp }$ of the acceleration $%
a$ along $u$ could be found after projection by the use of the projective
metrics $h_{\xi _{\perp }}$ and $h^{\xi _{\perp }}\,$in parts collinear to
the vector field $\xi _{\perp }$ and orthogonal to $\xi _{\perp }$. At the
same time both the parts are orthogonal to the vector field $u$. 
\begin{eqnarray}
(\nabla _ua)_{\perp } &=&\frac{g((\nabla _ua)_{\perp },\xi _{\perp })}{g(\xi
_{\perp },\xi _{\perp })}\cdot \xi _{\perp }+\overline{g}[h_{\xi _{\perp
}}(\nabla _ua)_{\perp }]=  \notag \\
&=&\mp g((\nabla _ua)_{\perp },n_{\perp })\cdot n_{\perp }+\overline{g}%
[h_{\xi _{\perp }}(\nabla _ua)_{\perp }]=  \notag \\
&=&(\nabla _ua)_{\perp z}+(\nabla _ua)_{\perp c}\,\,\,\,\,\,\,\text{,}\,
\label{3.128}
\end{eqnarray}
where 
\begin{eqnarray}
(\nabla _ua)_{\perp z} &=&\mp g((\nabla _ua)_{\perp },n_{\perp })\cdot
n_{\perp }\,\,\,\,\,\text{,}  \label{3.129} \\
(\nabla _ua)_{\perp c} &=&\overline{g}[h_{\xi _{\perp }}(\nabla _ua)_{\perp
}]\,\,\,\,\,\,\,\text{,}  \label{3.130} \\
g(\xi _{\perp },(\nabla _ua)_{\perp c}) &=&0\,\,\,\,\,\text{,}  \label{3.131}
\\
(\nabla _ua)_{\perp z} &=&\mp \text{\thinspace }l_{(\nabla _ua)_{\perp
z}}\cdot n_{\perp }\,\,\,\,\,\text{,}  \label{3.131a} \\
(\nabla _ua)_{\perp c} &=&\mp \text{\thinspace }l_{(\nabla _ua)_{\perp
c}}\cdot m_{\perp }\,\,\,\,\,\,\,\,\text{.}\,  \label{3.132}
\end{eqnarray}

3. The orthogonal to $u$ deformation velocity vector $\overline{g}%
[d(k_{\perp })]$ could be found after projection by the use of the
projective metrics $h_{\xi _{\perp }}$ and $h^{\xi _{\perp }}$ in parts
collinear to the vector field $\xi _{\perp }$ and orthogonal to $\xi _{\perp
}$. At the same time both the parts are orthogonal to the vector field $u$.
Since $k_{\perp }=\mp l_{k_{\perp }}\cdot \widetilde{n}_{\perp }$ we have
the relations 
\begin{eqnarray}
d(k_{\perp }) &=&d(\mp l_{k_{\perp }}\cdot \widetilde{n}_{\perp })=\mp
l_{k_{\perp }}\cdot d(\widetilde{n}_{\perp })=  \notag \\
&=&\mp l_{k_{\perp }}\cdot d(\alpha \cdot n_{\perp }+\beta \cdot m_{\perp })=
\notag \\
&=&\mp l_{k_{\perp }}\cdot [\alpha \cdot d(n_{\perp })+\beta \cdot
d(m_{\perp })\,]\,\,\,\text{,}  \label{3.133}
\end{eqnarray}
\begin{equation}
\overline{g}[d(n_{\perp })]=\frac{g(\overline{g}[d(n_{\perp })],\xi _{\perp
})}{g(\xi _{\perp },\xi _{\perp })}\cdot \xi _{\perp }+\overline{g}[h_{\xi
_{\perp }}(\overline{g}[d(n_{\perp })])]\,\,\,\,\,\,\text{,}  \label{3.134}
\end{equation}
\begin{eqnarray}
\overline{g}[d(n_{\perp })] &=&\frac{g(\overline{g}[d(n_{\perp })],n_{\perp
})}{\mp l_{\xi _{\perp }}^2}\cdot l_{\xi _{\perp }}^2\cdot n_{\perp }+%
\overline{g}[h_{\xi _{\perp }}(\overline{g}[d(n_{\perp })])]\,\,=  \notag \\
&=&\mp g(\overline{g}[d(n_{\perp })],n_{\perp })\cdot n_{\perp }+\overline{g}%
[h_{\xi _{\perp }}(\overline{g}[d(n_{\perp })])]\,\,\,\,\text{,}
\label{3.135}
\end{eqnarray}
\begin{equation}
g(\overline{g}[d(n_{\perp })],n_{\perp })=g_{\overline{i}\overline{j}}\cdot
g^{ik}\cdot d_{\overline{k}\overline{l}}\cdot n_{\perp }^l\cdot n_{\perp
}^j=d_{\overline{j}\overline{l}}\cdot n_{\perp }^j\cdot n_{\perp
}^l=d(n_{\perp },n_{\perp })\,\,\,\text{,}  \label{3.136}
\end{equation}
\begin{equation}
\mp g(\overline{g}[d(n_{\perp })],n_{\perp })\cdot n_{\perp }=\mp d(n_{\perp
},n_{\perp })\cdot n_{\perp }\,\,\,\,\text{.}  \label{3.137}
\end{equation}

On the other side, the following relations can be proved: 
\begin{eqnarray}
h^{\xi _{\perp }} &=&\overline{g}-\frac 1{g(\xi _{\perp },\xi _{\perp
})}\cdot \xi _{\perp }\otimes \xi _{\perp }=  \notag \\
&=&\overline{g}-\frac 1{\mp l_{\xi _{\perp }}^2}\cdot l_{\xi _{\perp }}\cdot
l_{\xi _{\perp }}\cdot n_{\perp }\otimes n_{\perp }=  \notag \\
&=&\overline{g}\pm n_{\perp }\otimes n_{\perp }=\overline{g}-\frac
1{g(n_{\perp },n_{\perp })}\cdot n_{\perp }\otimes n_{\perp }=  \notag \\
&=&h^{n_{\perp }}\,\,\,\,\,\,\text{,}  \label{3.138}
\end{eqnarray}
\begin{eqnarray}
h^u &=&\overline{g}-\frac 1{g(u,u)}\cdot u\otimes u=  \notag \\
&=&\overline{g}-\frac 1{\pm l_u^2}\cdot l_u\cdot l_u\cdot n_{\parallel
}\otimes n_{\parallel }=  \notag \\
&=&\overline{g}\mp n_{\parallel }\otimes n_{\parallel }=\overline{g}-\frac
1{g(n_{\parallel },n_{\parallel })}\cdot n_{\parallel }\otimes n_{\parallel
}=  \notag \\
&=&h^{n_{\parallel }}\,\,\,\,\,\,\text{,}  \label{3.139}
\end{eqnarray}
\begin{eqnarray}
h_{\xi _{\perp }} &=&g-\frac 1{g(\xi _{\perp },\xi _{\perp })}\cdot g(\xi
_{\perp })\otimes g(\xi _{\perp })=  \notag \\
&=&g-\frac 1{\mp l_{\xi _{\perp }}^2}\cdot l_{\xi _{\perp }}\cdot l_{\xi
_{\perp }}\cdot g(n_{\perp })\otimes (n_{\perp })=  \notag \\
&=&g-\frac 1{g(n_{\perp },n_{\perp })}\cdot g(n_{\perp })\otimes (n_{\perp
})=  \notag \\
&=&h_{n_{\perp }}\,\,\,\,\,\,\text{,}  \label{3.140}
\end{eqnarray}
\begin{eqnarray}
h_u &=&g-\frac 1{g(u,u)}\cdot g(u)\otimes g(u)=  \notag \\
&=&g-\frac 1{\pm l_u^2}\cdot l_u\cdot l_u\cdot g(n_{\parallel })\otimes
g(n_{\parallel })=  \notag \\
&=&g-\frac 1{g(n_{\parallel },n_{\parallel })}\cdot g(n_{\parallel })\otimes
g(n_{\parallel })=  \notag \\
&=&h_{n_{\parallel }}\,\,\,\,\,\,\text{.}  \label{3.141}
\end{eqnarray}

By the use of the above expressions we can find the explicit form of the
term $\overline{g}[h_{\xi _{\perp }}(\overline{g}[d(n_{\perp })])]$: 
\begin{eqnarray}
\overline{g}[h_{\xi _{\perp }}(\overline{g}[d(n_{\perp })])] &=&g^{ij}\cdot
(h_{\xi _{\perp }})_{\overline{j}\overline{k}}\cdot g^{kl}\cdot d_{\overline{%
l}\overline{m}}\cdot n_{\perp }^m\cdot \partial _i=  \notag \\
&=&g^{ij}\cdot (g_{\overline{j}\overline{k}}-\frac 1{g(\xi _{\perp },\xi
_{\perp })}\cdot g_{\overline{j}\overline{l}}\cdot \xi _{\perp }^l\cdot g_{%
\overline{k}\overline{r}}\cdot \xi _{\perp }^r)\cdot g^{kl}\cdot d_{%
\overline{l}\overline{m}}\cdot n_{\perp }^m\cdot \partial _i=  \notag \\
&=&(g_k^i-\frac 1{g(\xi _{\perp },\xi _{\perp })}\cdot g_l^i\cdot \xi
_{\perp }^l\cdot g_{\overline{k}\overline{r}}\cdot \xi _{\perp }^r)\cdot
g^{kl}\cdot d_{\overline{l}\overline{m}}\cdot n_{\perp }^m\cdot \partial _i=
\notag \\
&=&(g^{il}-\frac 1{g(\xi _{\perp },\xi _{\perp })}\cdot \xi _{\perp }^i\cdot
g_r^l\cdot \xi _{\perp }^r)\cdot d_{\overline{l}\overline{m}}\cdot n_{\perp
}^m\cdot \partial _i=  \notag \\
&=&(g^{il}-\frac 1{g(\xi _{\perp },\xi _{\perp })}\cdot \xi _{\perp }^i\cdot
\xi _{\perp }^l)\cdot d_{\overline{l}\overline{m}}\cdot n_{\perp }^m\cdot
\partial _i=  \notag \\
&=&h^{\xi _{\perp }}[d(n_{\perp })]=h^{n_{\perp }}[d(n_{\perp })]\,\,\,\,\,%
\text{.}  \label{3.142}
\end{eqnarray}

Therefore, 
\begin{eqnarray}
\overline{g}[d(n_{\perp })] &=&\mp g(\overline{g}[d(n_{\perp })],n_{\perp
})\cdot n_{\perp }+\overline{g}[h_{\xi _{\perp }}(\overline{g}[d(n_{\perp
})])]\,=  \notag \\
&=&\mp d(n_{\perp },n_{\perp })\cdot n_{\perp }\,+h^{n_{\perp }}[d(n_{\perp
})]\,\,\,\,\,\text{.}  \label{3.143}
\end{eqnarray}

On the other side, the term $\overline{g}[d(m_{\perp })]$ could be projected
in an analogous way 
\begin{equation}
\overline{g}[d(m_{\perp })]=\mp g(\overline{g}[d(m_{\perp })],n_{\perp
})\cdot n_{\perp }+\overline{g}[h_{\xi _{\perp }}(\overline{g}[d(m_{\perp
})])]\,\,\,\,\,\text{.}  \label{3.144}
\end{equation}

The terms at the right side of the expression could be found in the
corresponding forms by the use of the relations 
\begin{eqnarray}
g(\overline{g}[d(m_{\perp })],n_{\perp }) &=&g_{\overline{i}\overline{j}%
}\cdot g^{ik}\cdot d_{\overline{k}\overline{l}}\cdot m_{\perp }^l\cdot
n_{\perp }^j=  \label{3.145} \\
&=&d_{\overline{j}\overline{l}}\cdot m_{\perp }^l\cdot n_{\perp
}^j=d(n_{\perp },m_{\perp })\,\,\,\,\text{,}  \notag \\
\mp g(\overline{g}[d(n_{\perp })],n_{\perp })\cdot n_{\perp } &=&\mp
d(n_{\perp },m_{\perp })\cdot n_{\perp }\,\,\,\,\text{,}  \label{3.146}
\end{eqnarray}
\begin{eqnarray}
\overline{g}[h_{\xi _{\perp }}(\overline{g}[d(m_{\perp })])] &=&g^{ij}\cdot
(h_{\xi _{\perp }})_{\overline{j}\overline{k}}\cdot g^{kl}\cdot d_{\overline{%
l}\overline{r}}\cdot m_{\perp }^r\cdot \partial _i=  \label{3.147} \\
&=&\overline{g}(h_{\xi _{\perp }})(\overline{g})[d(m_{\perp })]\,\,\,\,\,\,%
\text{,}  \label{3.148} \\
\overline{g}(h_{\xi _{\perp }})(\overline{g}) &=&h^{\xi _{\perp
}}=h^{n_{\perp }}\,\,\,\,\text{,}  \label{3.149} \\
\overline{g}[h_{\xi _{\perp }}(\overline{g}[d(m_{\perp })])] &=&h^{n_{\perp
}}[d(m_{\perp })]\,\,\,\,\text{.}  \label{3.150}
\end{eqnarray}

Therefore, 
\begin{equation}
\overline{g}[d(m_{\perp })]=\mp d(n_{\perp },m_{\perp })\cdot n_{\perp
}+h^{n_{\perp }}[d(m_{\perp })]\,\,\,\,\text{.}  \label{3.151}
\end{equation}

For $\overline{g}[d(\widetilde{n}_{\perp })]$ the expressions follow 
\begin{eqnarray}
\overline{g}[d(\widetilde{n}_{\perp })] &=&\alpha \cdot \overline{g}%
[d(n_{\perp })]+\beta \cdot \overline{g}[d(m_{\perp })]=  \notag \\
&=&\mp \alpha \cdot d(n_{\perp },n_{\perp })\cdot n_{\perp }\,+\alpha \cdot
h^{n_{\perp }}[d(n_{\perp })]\mp  \notag \\
&&\mp \beta \cdot d(n_{\perp },m_{\perp })\cdot n_{\perp }+\beta \cdot
h^{n_{\perp }}[d(m_{\perp })]  \label{3.152}
\end{eqnarray}
\begin{eqnarray}
\overline{g}[d(\widetilde{n}_{\perp })] &=&\mp [\alpha \cdot d(n_{\perp
},n_{\perp })+\beta \cdot d(n_{\perp },m_{\perp })]\cdot n_{\perp }+  \notag
\\
&&+\alpha \cdot h^{n_{\perp }}[d(n_{\perp })]+\beta \cdot h^{n_{\perp
}}[d(m_{\perp })]\,\,\,\,\,\,\text{.}  \label{3.153}
\end{eqnarray}

On the other side, the structures of $_{rel}v=v_z+v_c$ could be represented
under the condition $\pounds _u\xi _{\perp }=0$ in the forms 
\begin{equation}
_{rel}v=l_{\xi _{\perp }}\cdot \overline{g}[d(n_{\perp })]\,\,\,\,\,\text{,}
\label{3.154}
\end{equation}
\begin{equation}
v_z=\mp l_{\xi _{\perp }}\cdot d(n_{\perp },n_{\perp })\cdot n_{\perp }\,\,\,%
\text{,}  \label{3.155}
\end{equation}
\begin{eqnarray}
v_c &=&\overline{g}[h_{\xi _{\perp }}(_{rel}v)]=l_{\xi _{\perp }}\cdot 
\overline{g}(h_{\xi _{\perp }})(\overline{g})[d(n_{\perp })]=l_{\xi _{\perp
}}\cdot h^{n_{\perp }}[d(n_{\perp })]\,\,\,\,\text{,}  \label{3.156} \\
g(v_c,n_{\perp }) &=&0\text{ \thinspace \thinspace ,}  \notag
\end{eqnarray}
\begin{eqnarray}
_{rel}v &=&v_z+v_c=l_{\xi _{\perp }}\cdot \overline{g}[d(n_{\perp })]\,= 
\notag \\
&=&\mp l_{\xi _{\perp }}\cdot d(n_{\perp },n_{\perp })\cdot n_{\perp
}+l_{\xi _{\perp }}\cdot h^{n_{\perp }}[d(n_{\perp })]\,\,\,\,\,\,\,\text{.}
\label{3.157}
\end{eqnarray}

Let us introduce now a vector field $\eta _{\perp }=l_{\xi _{\perp }}\cdot
m_{\perp }$, orthogonal to the vector field $\xi _{\perp }=l_{\xi _{\perp
}}\cdot n_{\perp }$ and $u$, but with the same length as $\xi _{\perp }$,
i.e. 
\begin{equation}
g(\eta _{\perp },\xi _{\perp })=l_{\xi _{\perp }}^2\cdot g(m_{\perp
},n_{\perp })=0\,\,\,\,\text{,\thinspace \thinspace \thinspace \thinspace
\thinspace \thinspace \thinspace \thinspace }g(\eta _{\perp },\eta _{\perp
})=\mp l_{\xi _{\perp }}^2\,\,\,\,\text{.}  \label{3.158}
\end{equation}

The corresponding to $\eta _{\perp }$ relative velocity $_{rel}v_\eta $ and
relative acceleration $_{rel}a_\eta $ have analogous forms as $_{rel}v$ and $%
_{rel}a$. 
\begin{eqnarray}
_{rel}v_\eta &=&\overline{g}[d(\eta _{\perp })]=l_{\xi _{\perp }}\cdot 
\overline{g}[d(m_{\perp })]\,\,\,\,\,\text{,}  \label{3.159} \\
_{rel}a_\eta &=&\overline{g}[A(\eta _{\perp })]=l_{\xi _{\perp }}\cdot 
\overline{g}[A(m_{\perp })]\,\,\,\,\,\text{.}  \label{3.160}
\end{eqnarray}

The decomposition of $_{rel}v_\eta $ has the form 
\begin{equation}
_{rel}v_\eta =v_{\eta z}+v_{\eta c}\,\,\,\,\text{,}  \label{3.161}
\end{equation}
\begin{eqnarray}
_{rel}v_\eta &=&\frac 1{g(\xi _{\perp },\xi _{\perp })}\cdot g(_{rel}v_\eta
,\xi _{\perp })\cdot \xi _{\perp }+\overline{g}[h_{\xi _{\perp
}}(_{rel}v_\eta )]=v_{\eta z}+v_{\eta _c}=  \notag \\
&=&\mp g(_{rel}v_\eta ,n_{\perp })\cdot n_{\perp }+\overline{g}[h_{\xi
_{\perp }}(_{rel}v_\eta )]  \label{3.162}
\end{eqnarray}
where 
\begin{eqnarray}
v_{\eta z} &=&\mp g(_{rel}v_\eta ,n_{\perp })\cdot n_{\perp }\,\,\,\text{,}
\label{3.163} \\
v_{\eta _c} &=&\overline{g}[h_{\xi _{\perp }}(_{rel}v_\eta )]\,\,\,\,\,\,%
\text{.}  \label{3.164}
\end{eqnarray}

The explicit form of $v_{\eta z}$ and $v_{\eta c}$ could be found by the use
of the relations under the condition $\pounds _u\eta _{\perp }=0$%
\begin{eqnarray}
g(_{rel}v_\eta ,n_{\perp }) &=&g(\overline{g}[d(\eta _{\perp })],n_{\perp
})=g_{\overline{i}\overline{j}}\cdot g^{ik}\cdot d_{\overline{k}\overline{l}%
}\cdot l_{\xi _{\perp }}\cdot m_{\perp }^l\cdot n_{\perp }^j=  \notag \\
&=&l_{\xi _{\perp }}\cdot d_{\overline{j}\overline{l}}\cdot n_{\perp
}^j\cdot m_{\perp }^l=l_{\xi _{\perp }}\cdot d(n_{\perp },m_{\perp })\,\,\,\,%
\text{,}  \label{3.165} \\
\overline{g}[h_{\xi _{\perp }}(_{rel}v_\eta )] &=&l_{\xi _{\perp }}\cdot 
\overline{g}[h_{\xi _{\perp }}(\overline{g}[d(m_{\perp })])]=l_{\xi _{\perp
}}\cdot h^{n_{\perp }}[d(m_{\perp })]\,\,\,\,\,\text{,}  \label{3.166}
\end{eqnarray}
as 
\begin{eqnarray}
v_{\eta z} &=&\mp g(_{rel}v_\eta ,n_{\perp })\cdot n_{\perp }=\mp l_{\xi
_{\perp }}\cdot d(n_{\perp },m_{\perp })\cdot n_{\perp }\,\,\,\,\,\,\,\text{,%
}  \label{3.167} \\
v_{\eta _c} &=&\overline{g}[h_{\xi _{\perp }}(_{rel}v_\eta )]\,=l_{\xi
_{\perp }}\cdot h^{n_{\perp }}[d(m_{\perp })]\,\,\,\,\,\,\text{.}
\label{3.168}
\end{eqnarray}

Now we can find the relations between the relative velocities $_{rel}v$, $%
_{rel}v_\eta $, and the expression for $\overline{g}[d(\widetilde{n}_{\perp
})]$%
\begin{eqnarray}
\overline{g}[d(\widetilde{n}_{\perp })] &=&\alpha \cdot \overline{g}%
[d(n_{\perp })]+\beta \cdot \overline{g}[d(m_{\perp })]=  \notag \\
&=&\mp \alpha \cdot d(n_{\perp },n_{\perp })\cdot n_{\perp }\,+\alpha \cdot
h^{n_{\perp }}[d(n_{\perp })]\mp  \notag \\
&&\mp \beta \cdot d(n_{\perp },m_{\perp })\cdot n_{\perp }+\beta \cdot
h^{n_{\perp }}[d(m_{\perp })]\,\,\,\,\,\text{,}  \label{3.169}
\end{eqnarray}
\begin{eqnarray}
\overline{g}[d(\widetilde{n}_{\perp })] &=&\mp \alpha \cdot (\mp \frac
1{l_{\xi _{\perp }}}\cdot v_z)+\alpha \cdot \frac 1{l_{\xi _{\perp }}}\cdot
v_c\mp  \notag \\
&&\mp \beta \cdot (\mp \frac 1{l_{\xi _{\perp }}}\cdot v_{\eta z})+\beta
\cdot \frac 1{l_{\xi _{\perp }}}\cdot v_{\eta c}\,\,\,\,\,\text{,}
\label{3.170}
\end{eqnarray}
\begin{eqnarray}
\overline{g}[d(\widetilde{n}_{\perp })] &=&\alpha \cdot \frac 1{l_{\xi
_{\perp }}}\cdot v_z+\alpha \cdot \frac 1{l_{\xi _{\perp }}}\cdot v_c+
\label{3.171} \\
&&+\beta \cdot \frac 1{l_{\xi _{\perp }}}\cdot v_{\eta z}+\beta \cdot \frac
1{l_{\xi _{\perp }}}\cdot v_{\eta c}\,\,\,\,\text{,}  \notag
\end{eqnarray}
\begin{eqnarray}
\overline{g}[d(\widetilde{n}_{\perp })] &=&\frac 1{l_{\xi _{\perp }}}\cdot
[\alpha \cdot (v_z+v_c)+\beta \cdot (v_{\eta z}+v_{\eta c})]=  \notag \\
&=&\frac 1{l_{\xi _{\perp }}}\cdot (\alpha \cdot \,_{rel}v+\beta \cdot
\,_{rel}v_\eta )\,\,\,\,\,\text{.}  \label{3.172}
\end{eqnarray}

4. The orthogonal to $u$ deformation acceleration vector $\overline{g}%
[A(k_{\perp })]$ could be found after projection by the use of the
projective metrics $h_{\xi _{\perp }}$ and $h^{\xi _{\perp }}$ in parts
collinear to the vector field $\xi _{\perp }$ and orthogonal to $\xi _{\perp
}$. At the same time both the parts are orthogonal to the vector field $u$.
Since $k_{\perp }=\mp l_{k_{\perp }}\cdot \widetilde{n}_{\perp }$ we have
the relations

\begin{eqnarray}
A(k_{\perp }) &=&A(\mp l_{k_{\perp }}\cdot \widetilde{n}_{\perp })=\mp
l_{k_{\perp }}\cdot A(\widetilde{n}_{\perp })=  \notag \\
&=&\mp l_{k_{\perp }}\cdot A(\alpha \cdot n_{\perp }+\beta \cdot m_{\perp })=
\notag \\
&=&\mp l_{k_{\perp }}\cdot [\alpha \cdot A(n_{\perp })+\beta \cdot
A(m_{\perp })\,]\,\,\,\text{,}  \label{3.173}
\end{eqnarray}

\begin{equation}
\overline{g}[A(n_{\perp })]=\frac{g(\overline{g}[A(n_{\perp })],\xi _{\perp
})}{g(\xi _{\perp },\xi _{\perp })}\cdot \xi _{\perp }+\overline{g}[h_{\xi
_{\perp }}(\overline{g}[A(n_{\perp })])]\,\,\,\,\,\,\text{,}  \label{3.174}
\end{equation}
\begin{eqnarray}
\overline{g}[A(n_{\perp })] &=&\frac{g(\overline{g}[A(n_{\perp })],n_{\perp
})}{\mp l_{\xi _{\perp }}^2}\cdot l_{\xi _{\perp }}^2\cdot n_{\perp }+%
\overline{g}[h_{\xi _{\perp }}(\overline{g}[A(n_{\perp })])]\,\,=  \notag \\
&=&\mp g(\overline{g}[A(n_{\perp })],n_{\perp })\cdot n_{\perp }+\overline{g}%
[h_{\xi _{\perp }}(\overline{g}[A(n_{\perp })])]\,\,\,\,\text{,}
\label{3.175}
\end{eqnarray}
\begin{equation}
g(\overline{g}[A(n_{\perp })],n_{\perp })=g_{\overline{i}\overline{j}}\cdot
g^{ik}\cdot A_{\overline{k}\overline{l}}\cdot n_{\perp }^l\cdot n_{\perp
}^j=A_{\overline{j}\overline{l}}\cdot n_{\perp }^j\cdot n_{\perp
}^l=A(n_{\perp },n_{\perp })\,\,\,\text{,}  \label{3.176}
\end{equation}
\begin{equation}
\mp g(\overline{g}[A(n_{\perp })],n_{\perp })\cdot n_{\perp }=\mp A(n_{\perp
},n_{\perp })\cdot n_{\perp }\,\,\,\,\text{.}  \label{3.177}
\end{equation}

Therefore, 
\begin{eqnarray}
\overline{g}[A(n_{\perp })] &=&\mp g(\overline{g}[A(n_{\perp })],n_{\perp
})\cdot n_{\perp }+\overline{g}[h_{\xi _{\perp }}(\overline{g}[A(n_{\perp
})])]\,=  \notag \\
&=&\mp A(n_{\perp },n_{\perp })\cdot n_{\perp }\,+h^{n_{\perp }}[A(n_{\perp
})]\,\,\,\,\,\text{.}  \label{3.178}
\end{eqnarray}

On the other side, the term $\overline{g}[A(m_{\perp })]$ could be projected
in an analogous way 
\begin{equation}
\overline{g}[A(m_{\perp })]=\mp g(\overline{g}[A(m_{\perp })],n_{\perp
})\cdot n_{\perp }+\overline{g}[h_{\xi _{\perp }}(\overline{g}[A(m_{\perp
})])]\,\,\,\,\,\text{.}  \label{3.179}
\end{equation}

The terms at the left side of the expression could be found in the
corresponding forms by the use of the relations 
\begin{eqnarray}
g(\overline{g}[A(m_{\perp })],n_{\perp }) &=&g_{\overline{i}\overline{j}%
}\cdot g^{ik}\cdot A_{\overline{k}\overline{l}}\cdot m_{\perp }^l\cdot
n_{\perp }^j=  \notag \\
&=&A_{\overline{j}\overline{l}}\cdot m_{\perp }^l\cdot n_{\perp
}^j=A(n_{\perp },m_{\perp })\,\,\,\,\text{,}  \label{3.180} \\
\mp g(\overline{g}[A(n_{\perp })],n_{\perp })\cdot n_{\perp } &=&\mp
A(n_{\perp },m_{\perp })\cdot n_{\perp }\,\,\,\,\text{,}  \label{3.181}
\end{eqnarray}
\begin{eqnarray}
\overline{g}[h_{\xi _{\perp }}(\overline{g}[A(m_{\perp })])] &=&g^{ij}\cdot
(h_{\xi _{\perp }})_{\overline{j}\overline{k}}\cdot g^{kl}\cdot A_{\overline{%
l}\overline{r}}\cdot m_{\perp }^r\cdot \partial _i=  \label{3.182} \\
&=&\overline{g}(h_{\xi _{\perp }})(\overline{g})[A(m_{\perp })]\,\,\,\,\,\,%
\text{,}  \label{3.183} \\
\overline{g}(h_{\xi _{\perp }})(\overline{g}) &=&h^{\xi _{\perp
}}=h^{n_{\perp }}\,\,\,\,\text{,}  \label{3.184} \\
\overline{g}[h_{\xi _{\perp }}(\overline{g}[A(m_{\perp })])] &=&h^{n_{\perp
}}[A(m_{\perp })]\,\,\,\,\text{.}  \label{3.185}
\end{eqnarray}

Therefore, 
\begin{equation}
\overline{g}[A(m_{\perp })]=\mp A(n_{\perp },m_{\perp })\cdot n_{\perp
}+h^{n_{\perp }}[A(m_{\perp })]\,\,\,\,\text{.}  \label{3.186}
\end{equation}

For $\overline{g}[A(\widetilde{n}_{\perp })]$ the expressions follow 
\begin{eqnarray}
\overline{g}[A(\widetilde{n}_{\perp })] &=&\alpha \cdot \overline{g}%
[A(n_{\perp })]+\beta \cdot \overline{g}[A(m_{\perp })]=  \notag \\
&=&\mp \alpha \cdot A(n_{\perp },n_{\perp })\cdot n_{\perp }\,+\alpha \cdot
h^{n_{\perp }}[A(n_{\perp })]\mp  \notag \\
&&\mp \beta \cdot A(n_{\perp },m_{\perp })\cdot n_{\perp }+\beta \cdot
h^{n_{\perp }}[A(m_{\perp })]  \label{3.187}
\end{eqnarray}
\begin{eqnarray}
\overline{g}[A(\widetilde{n}_{\perp })] &=&\mp [\alpha \cdot A(n_{\perp
},n_{\perp })+\beta \cdot A(n_{\perp },m_{\perp })]\cdot n_{\perp }+  \notag
\\
&&+\alpha \cdot h^{n_{\perp }}[A(n_{\perp })]+\beta \cdot h^{n_{\perp
}}[A(m_{\perp })]\,\,\,\,\,\,\text{.}  \label{3.188}
\end{eqnarray}

On the other side, the structures of $_{rel}a=a_z+a_c$ could be represented
under the condition $\pounds _u\xi _{\perp }=0$ in the forms 
\begin{equation}
_{rel}a=l_{\xi _{\perp }}\cdot \overline{g}[A(n_{\perp })]\,\,\,\,\,\text{,}
\label{3.189}
\end{equation}
\begin{equation}
a_z=\mp l_{\xi _{\perp }}\cdot A(n_{\perp },n_{\perp })\cdot n_{\perp }\,\,\,%
\text{,}  \label{3.190}
\end{equation}
\begin{eqnarray}
a_c &=&\overline{g}[h_{\xi _{\perp }}(_{rel}a)]=l_{\xi _{\perp }}\cdot 
\overline{g}(h_{\xi _{\perp }})(\overline{g})[A(n_{\perp })]=l_{\xi _{\perp
}}\cdot h^{n_{\perp }}[A(n_{\perp })]\,\,\,\,\text{,}  \label{3.191} \\
g(a_c,n_{\perp }) &=&0\text{ \thinspace \thinspace ,}  \notag
\end{eqnarray}
\begin{eqnarray}
_{rel}a &=&a_z+a_c=l_{\xi _{\perp }}\cdot \overline{g}[A(n_{\perp })]\,=
\label{4.1} \\
&=&\mp l_{\xi _{\perp }}\cdot A(n_{\perp },n_{\perp })\cdot n_{\perp
}+l_{\xi _{\perp }}\cdot h^{n_{\perp }}[A(n_{\perp })]\,\,\,\,\,\,\,\text{.}
\label{4.2}
\end{eqnarray}

Let us introduce now a vector field $\eta _{\perp }=l_{\xi _{\perp }}\cdot
m_{\perp }$, orthogonal to the vector field $\xi _{\perp }=l_{\xi _{\perp
}}\cdot n_{\perp }$ and $u$, but with the same length as $\xi _{\perp }$,
i.e. 
\begin{equation}
g(\eta _{\perp },\xi _{\perp })=l_{\xi _{\perp }}^2\cdot g(m_{\perp
},n_{\perp })=0\,\,\,\,\text{,\thinspace \thinspace \thinspace \thinspace
\thinspace \thinspace \thinspace \thinspace }g(\eta _{\perp },\eta _{\perp
})=\mp l_{\xi _{\perp }}^2\,\,\,\,\text{.}  \label{4.3}
\end{equation}

The corresponding to $\eta _{\perp }$ relative velocity $_{rel}v_\eta $ and
relative acceleration $_{rel}a_\eta $ have analogous forms as $_{rel}v$ and $%
_{rel}a$. 
\begin{eqnarray}
_{rel}v_\eta &=&\overline{g}[d(\eta _{\perp })]=l_{\xi _{\perp }}\cdot 
\overline{g}[d(m_{\perp })]\,\,\,\,\,\text{,}  \label{4.4} \\
_{rel}a_\eta &=&\overline{g}[A(\eta _{\perp })]=l_{\xi _{\perp }}\cdot 
\overline{g}[A(m_{\perp })]\,\,\,\,\,\text{.}  \label{4.5}
\end{eqnarray}

The decomposition of $_{rel}a_\eta $ has the form 
\begin{equation}
_{rel}a_\eta =a_{\eta z}+a_{\eta c}\,\,\,\,\text{,}  \label{4.6}
\end{equation}
\begin{eqnarray}
_{rel}a_\eta &=&\frac 1{g(\xi _{\perp },\xi _{\perp })}\cdot g(_{rel}a_\eta
,\xi _{\perp })\cdot \xi _{\perp }+\overline{g}[h_{\xi _{\perp
}}(_{rel}a_\eta )]=a_{\eta z}+a_{\eta _c}=  \notag \\
&=&\mp g(_{rel}a_\eta ,n_{\perp })\cdot n_{\perp }+\overline{g}[h_{\xi
_{\perp }}(_{rel}a_\eta )]  \label{4.7}
\end{eqnarray}
where 
\begin{eqnarray}
a_{\eta z} &=&\mp g(_{rel}a_\eta ,n_{\perp })\cdot n_{\perp }\,\,\,\text{,}
\label{4.8} \\
a_{\eta _c} &=&\overline{g}[h_{\xi _{\perp }}(_{rel}a_\eta )]\,\,\,\,\,\,%
\text{.}  \label{4.9}
\end{eqnarray}

The explicit form of $a_{\eta z}$ and $a_{\eta c}$ could be found by the use
of the relations under the condition $\pounds _u\eta _{\perp }=0$%
\begin{eqnarray}
g(_{rel}a_\eta ,n_{\perp }) &=&g(\overline{g}[A(\eta _{\perp })],n_{\perp
})=g_{\overline{i}\overline{j}}\cdot g^{ik}\cdot A_{\overline{k}\overline{l}%
}\cdot l_{\xi _{\perp }}\cdot m_{\perp }^l\cdot n_{\perp }^j=  \notag \\
&=&l_{\xi _{\perp }}\cdot A_{\overline{j}\overline{l}}\cdot n_{\perp
}^j\cdot m_{\perp }^l=l_{\xi _{\perp }}\cdot A(n_{\perp },m_{\perp })\,\,\,\,%
\text{,}  \label{4.10} \\
\overline{g}[h_{\xi _{\perp }}(_{rel}a_\eta )] &=&l_{\xi _{\perp }}\cdot 
\overline{g}[h_{\xi _{\perp }}(\overline{g}[A(m_{\perp })])]=l_{\xi _{\perp
}}\cdot h^{n_{\perp }}[A(m_{\perp })]\,\,\,\,\,\text{,}  \label{4.11}
\end{eqnarray}
as 
\begin{eqnarray}
a_{\eta z} &=&\mp g(_{rel}a_\eta ,n_{\perp })\cdot n_{\perp }=\mp l_{\xi
_{\perp }}\cdot A(n_{\perp },m_{\perp })\cdot n_{\perp }\,\,\,\,\,\,\,\text{,%
}  \label{4.12} \\
a_{\eta _c} &=&\overline{g}[h_{\xi _{\perp }}(_{rel}a_\eta )]\,=l_{\xi
_{\perp }}\cdot h^{n_{\perp }}[A(m_{\perp })]\,\,\,\,\,\,\text{.}
\label{4.13}
\end{eqnarray}

Now we can find the relations between the relative velocities $_{rel}a$, $%
_{rel}a_\eta $, and the expression for $\overline{g}[A(\widetilde{n}_{\perp
})]$%
\begin{eqnarray}
\overline{g}[A(\widetilde{n}_{\perp })] &=&\alpha \cdot \overline{g}%
[A(n_{\perp })]+\beta \cdot \overline{g}[A(m_{\perp })]=  \notag \\
&=&\mp \alpha \cdot A(n_{\perp },n_{\perp })\cdot n_{\perp }\,+\alpha \cdot
h^{n_{\perp }}[A(n_{\perp })]\mp  \notag \\
&&\mp \beta \cdot A(n_{\perp },m_{\perp })\cdot n_{\perp }+\beta \cdot
h^{n_{\perp }}[A(m_{\perp })]\,\,\,\,\,\text{,}  \label{4.14}
\end{eqnarray}
\begin{eqnarray}
\overline{g}[A(\widetilde{n}_{\perp })] &=&\mp \alpha \cdot (\mp \frac
1{l_{\xi _{\perp }}}\cdot a_z)+\alpha \cdot \frac 1{l_{\xi _{\perp }}}\cdot
a_c\mp  \notag \\
&&\mp \beta \cdot (\mp \frac 1{l_{\xi _{\perp }}}\cdot a_{\eta z})+\beta
\cdot \frac 1{l_{\xi _{\perp }}}\cdot a_{\eta c}\,\,\,\,\,\text{,}
\label{4.15}
\end{eqnarray}
\begin{eqnarray}
\overline{g}[A(\widetilde{n}_{\perp })] &=&\alpha \cdot \frac 1{l_{\xi
_{\perp }}}\cdot a_z+\alpha \cdot \frac 1{l_{\xi _{\perp }}}\cdot a_c+ 
\notag \\
&&+\beta \cdot \frac 1{l_{\xi _{\perp }}}\cdot a_{\eta z}+\beta \cdot \frac
1{l_{\xi _{\perp }}}\cdot a_{\eta c}\,\,\,\,\text{,}  \label{4.16}
\end{eqnarray}
\begin{eqnarray}
\overline{g}[A(\widetilde{n}_{\perp })] &=&\frac 1{l_{\xi _{\perp }}}\cdot
[\alpha \cdot (a_z+a_c)+\beta \cdot (a_{\eta z}+a_{\eta c})]=  \notag \\
&=&\frac 1{l_{\xi _{\perp }}}\cdot (\alpha \cdot \,_{rel}a+\beta \cdot
\,_{rel}a_\eta )\,\,\,\,\,\text{.}  \label{4.17}
\end{eqnarray}

5. After the consideration and finding out of the explicit forms of the
terms in $_{rel}k_{\perp }$

\begin{eqnarray}
_{rel}k_{\perp } &=&-d\tau \cdot (\nabla _{u\,}{}\widetilde{k})_{\perp
}+\frac 12\cdot d\tau ^2\cdot (\nabla _u\nabla _u\widetilde{k})_{\perp }= 
\notag \\
&=&-d\tau \cdot \{\pm \frac \omega {l_u^2}\cdot a_{\perp }+\overline{g}%
[d(k_{\perp })]\}+  \notag \\
&&+\frac 12\cdot d\tau ^2\cdot \{\pm \frac \omega {l_u^2}\cdot (\nabla
_ua)_{\perp }+\overline{g}[A(k_{\perp })]\}\,\,\,\text{,}  \label{4.18}
\end{eqnarray}
we can find the explicit forms of the change $_{rel}k_{\perp }$ of the
vector $k_{\perp }$along the world line of the observer 
\begin{eqnarray}
_{rel}k_{\perp } &=&-d\tau \cdot \{\pm \frac \omega {l_u^2}\cdot [(a_{\perp
})_z+(a_{\perp })_c]\mp  \notag \\
&&\mp l_{k_{\perp }}\cdot \overline{g}[d(\widetilde{n}_{\perp })]\}+  \notag
\\
&&+\frac 12\cdot d\tau ^2\cdot \{\pm \frac \omega {l_u^2}\cdot [(\nabla
_ua)_{\perp z}+(\nabla _ua)_{\perp c}]\mp  \notag \\
&&\mp l_{k_{\perp }}\cdot \overline{g}[A(\widetilde{n}_{\perp })]\}\,\,\,%
\text{,}  \label{4.19}
\end{eqnarray}
\begin{eqnarray}
_{rel}k_{\perp } &=&-d\tau \cdot \{\pm \frac \omega {l_u^2}\cdot [(a_{\perp
})_z+(a_{\perp })_c]\mp  \notag \\
&&\mp l_{k_{\perp }}\cdot [\frac 1{l_{\xi _{\perp }}}\cdot (\alpha \cdot
\,_{rel}v+\beta \cdot \,_{rel}v_\eta )]\}+  \notag \\
&&+\frac 12\cdot d\tau ^2\cdot \{\pm \frac \omega {l_u^2}\cdot [(\nabla
_ua)_{\perp z}+(\nabla _ua)_{\perp c}]\mp  \notag \\
&&\mp l_{k_{\perp }}\cdot [\frac 1{l_{\xi _{\perp }}}\cdot (\alpha \cdot
\,_{rel}a+\beta \cdot \,_{rel}a_\eta )\,]\}\,\,\,\,\text{,}  \label{4.20}
\end{eqnarray}
\begin{eqnarray}
_{rel}k_{\perp } &=&-d\tau \cdot \{\pm \frac \omega {l_u^2}\cdot [(a_{\perp
})_z+(a_{\perp })_c]\mp  \notag \\
&&\mp \frac{l_{k_{\perp }}}{l_{\xi _{\perp }}}\cdot (\alpha \cdot
\,_{rel}v+\beta \cdot \,_{rel}v_\eta )\}+  \notag \\
&&+\frac 12\cdot d\tau ^2\cdot \{\pm \frac \omega {l_u^2}\cdot [(\nabla
_ua)_{\perp z}+(\nabla _ua)_{\perp c}]\mp  \notag \\
&&\mp \frac{l_{k_{\perp }}}{l_{\xi _{\perp }}}\cdot (\alpha \cdot
\,_{rel}a+\beta \cdot \,_{rel}a_\eta )\}\,\,\,\,\text{.}  \label{4.21}
\end{eqnarray}

Since 
\begin{equation}
l_{k_{\perp }}=\frac \omega {l_u}=\text{\thinspace }l_{k_{\parallel
}}\,\,\,\,\,\text{,}  \label{4.22}
\end{equation}
we obtain the final form of $_{rel}k_{\perp }$ with respect to the relative
velocity and relative acceleration 
\begin{eqnarray}
_{rel}k_{\perp } &=&\mp d\tau \cdot l_{k_{\perp }}\cdot \{\frac 1{l_u}\cdot
[(a_{\perp })_z+(a_{\perp })_c]-  \notag \\
&&-\frac 1{l_{\xi _{\perp }}}\cdot (\alpha \cdot \,_{rel}v+\beta \cdot
\,_{rel}v_\eta )\}\pm  \notag \\
&&\pm \frac 12\cdot d\tau ^2\cdot l_{k_{\perp }}\cdot \{\frac 1{l_u}\cdot
[(\nabla _ua)_{\perp z}+(\nabla _ua)_{\perp c}]-  \notag \\
&&-\frac 1{l_{\xi _{\perp }}}\cdot (\alpha \cdot \,_{rel}a+\beta \cdot
\,_{rel}a_\eta )\}\,\,\,\,\,\text{.}  \label{4.23}
\end{eqnarray}

If we, further, express the time interval $d\tau $ by its equivalent
relations 
\begin{equation}
d\tau =\mp \frac{l_{\xi _{\perp }}\cdot d\lambda }{l_u}=\frac{dl}{l_u}%
\,\,\,\,  \label{4.24}
\end{equation}
the relation of $_{rel}k_{\perp }$ to the relative velocity and relative
acceleration could also be written in the forms 
\begin{eqnarray}
_{rel}k_{\perp } &=&\frac{l_{k_{\perp }}}{l_u}\cdot \{\frac{l_{\xi _{\perp
}}\cdot d\lambda }{l_u}\cdot [(a_{\perp })_z+(a_{\perp })_c]-d\lambda \cdot
(\alpha \cdot \,_{rel}v+\beta \cdot \,_{rel}v_\eta )\}\pm  \notag \\
&&\pm \frac 12\cdot \frac{l_{\xi _{\perp }}^2\cdot d\lambda ^2}{l_u^2}\cdot
l_{k_{\perp }}\cdot \{\frac 1{l_u}\cdot [(\nabla _ua)_{\perp z}+(\nabla
_ua)_{\perp c}]-  \notag \\
&&-\frac 1{l_{\xi _{\perp }}}\cdot (\alpha \cdot \,_{rel}a+\beta \cdot
\,_{rel}a_\eta )\}\,\,\,\,\,\,\,\,\text{,}\,\text{.}  \label{4.25}
\end{eqnarray}
\begin{eqnarray}
_{rel}k_{\perp } &=&\frac{l_{k_{\perp }}}{l_u}\cdot \{\mp \frac{dl}{l_u}%
\cdot [(a_{\perp })_z+(a_{\perp })_c]\pm \frac{dl}{l_{\xi _{\perp }}}\cdot
(\alpha \cdot \,_{rel}v+\beta \cdot \,_{rel}v_\eta )\}\pm  \notag \\
&&\pm \frac 12\cdot \frac{dl^2}{l_u^2}\cdot l_{k_{\perp }}\cdot \{\frac
1{l_u}\cdot [(\nabla _ua)_{\perp z}+(\nabla _ua)_{\perp c}]-  \notag \\
&&-\frac 1{l_{\xi _{\perp }}}\cdot (\alpha \cdot \,_{rel}a+\beta \cdot
\,_{rel}a_\eta )\}\,\,\,\,\,\,\,\,\text{.}  \label{4.26}
\end{eqnarray}

By the use of the explicit forms of $\overline{k}_{\perp }$ and $k_{\perp }$%
\begin{eqnarray}
\overline{k}_{\perp } &=&\mp l_{\overline{k}_{\perp }}\cdot \widetilde{n}%
_{\perp }^{\prime }=\mp \frac{\overline{\omega }}{l_u}\,\cdot \widetilde{n}%
_{\perp }^{\prime }\,\,\,\,\,\,\text{,}  \label{4.27} \\
k_{\perp } &=&\mp l_{k_{\perp }}\cdot \widetilde{n}_{\perp }=\mp \frac
\omega {l_u}\,\cdot \widetilde{n}_{\perp }\,\,\,\,\text{,}  \label{4.28}
\end{eqnarray}
we can find the explicit forms of the expressions 
\begin{equation}
\overline{S}:=\frac{g(_{rel}k_{\perp },m_{\perp })}{g(\widetilde{n}_{\perp
},k_{\perp })}=\frac 1{l_{k_{\perp }}}\cdot g(_{rel}k_{\perp },m_{\perp
})\,\,\,\,\text{,}  \label{4.29}
\end{equation}
\begin{equation}
\overline{C}:=\frac{g(_{rel}k_{\perp },n_{\perp })}{g(\widetilde{n}_{\perp
},k_{\perp })}=\frac 1{l_{k_{\perp }}}\cdot g(_{rel}k_{\perp },n_{\perp
})\,\,\,\,\text{,}  \label{4.30}
\end{equation}
because of the relation 
\begin{equation}
g(\widetilde{n}_{\perp },k_{\perp })=g(\widetilde{n}_{\perp },\,\mp
l_{k_{\perp }}\cdot \widetilde{n}_{\perp })=\mp l_{k_{\perp }}\cdot g(\,%
\widetilde{n}_{\perp },\,\widetilde{n}_{\perp })=l_{k_{\perp }}\,\,\,\,\,%
\text{.}  \label{4.31}
\end{equation}

6. By the use of the relations 
\begin{equation}
g((a_{\perp })_z+(a_{\perp })_c,n_{\perp })=g((a_{\perp })_z,n_{\perp
})\,\,\,\,\,\text{,\thinspace \thinspace \thinspace \thinspace \thinspace
\thinspace \thinspace \thinspace \thinspace \thinspace \thinspace \thinspace
\thinspace }g((a_{\perp })_c,n_{\perp })=0\,\,\,\,\text{,}  \label{4.32}
\end{equation}
\begin{equation}
g(_{rel}v,n_{\perp })=g(v_z,n_{\perp })\,\,\,\,\,\,\text{,\thinspace
\thinspace \thinspace \thinspace \thinspace \thinspace \thinspace \thinspace
\thinspace \thinspace }g(_{rel}v_\eta ,n_{\perp })=g(v_{\eta z},n_{\perp
})\,\,\,\text{,}  \label{4.33}
\end{equation}
\begin{equation}
g((\nabla _ua)_{\perp z}+(\nabla _ua)_{\perp c},n_{\perp })=g((\nabla
_ua)_{\perp z},n_{\perp })\,\,\,\text{,}  \label{4.34}
\end{equation}
\begin{equation}
g(\,_{rel}a,n_{\perp })=g(a_z,n_{\perp })\,\,\,\,\,\,\,\,\text{,\thinspace
\thinspace \thinspace \thinspace \thinspace \thinspace \thinspace \thinspace
\thinspace \thinspace \thinspace }g(_{rel}a_\eta ,n_{\perp })=g(a_{\eta
z},n_{\perp })\,\,\,\,\,\text{,}  \label{4.35}
\end{equation}
\begin{eqnarray}
g(_{rel}k_{\perp },n_{\perp }) &=&\frac{l_{k_{\perp }}}{l_u}\cdot \{\mp 
\frac{dl}{l_u}\cdot g((a_{\perp })_z,n_{\perp })\pm \frac{dl}{l_{\xi _{\perp
}}}\cdot [\alpha \cdot g(v_z,n_{\perp })+\beta \cdot \,g(v_{\eta z},n_{\perp
})]\}\pm  \notag \\
&&\pm \frac 12\cdot \frac{dl^2}{l_u^2}\cdot l_{k_{\perp }}\cdot \{\frac
1{l_u}\cdot g((\nabla _ua)_{\perp z},n_{\perp })-  \notag \\
&&-\frac 1{l_{\xi _{\perp }}}\cdot [\alpha \cdot \,g(a_z,n_{\perp })+\beta
\cdot \,g(a_{\eta z},n_{\perp })]\}\,\,\,\,\,\,\,\,\text{.}  \label{4.36}
\end{eqnarray}
we can find the explicit form of the quantity $\overline{C}$.

On the other side, we can use the relations 
\begin{equation}
(a_{\perp })_z=\mp l_{(a_{\perp })_z}\cdot n_{\perp }\,\,\,\,\text{,}
\label{4.37}
\end{equation}
\begin{equation}
g((a_{\perp })_z,n_{\perp })=\mp l_{(a_{\perp })_z}\cdot g(n_{\perp
},n_{\perp })=l_{(a_{\perp })_z}\,\,\,\,\text{,}  \label{4.38}
\end{equation}
\begin{eqnarray}
v_z &=&\mp l_{v_z}\cdot n_{\perp }\,\,\,\,\,\text{,}  \label{4.39} \\
g(v_z,n_{\perp }) &=&\mp l_{v_z}\cdot g(n_{\perp },n_{\perp })=l_{v_z}\,\,\,%
\text{,}  \label{4.40}
\end{eqnarray}
\begin{eqnarray}
v_{\eta z} &=&\mp l_{v_{\eta z}}\cdot n_{\perp }\,\,\,\,\,\text{,}
\label{4.41} \\
\,g(v_{\eta z},n_{\perp }) &=&\mp l_{v_{\eta z}}\cdot g(n_{\perp },n_{\perp
})=l_{v_{\eta z}}\,\,\,\text{,}  \label{4.42}
\end{eqnarray}
\begin{eqnarray}
(\nabla _ua)_{\perp z} &=&\mp l_{(\nabla _ua)_{\perp z}}\cdot n_{\perp
}\,\,\,\,\text{,}  \label{4.43} \\
g((\nabla _ua)_{\perp z},n_{\perp }) &=&\mp l_{(\nabla _ua)_{\perp z}}\cdot
g(n_{\perp },n_{\perp })=l_{(\nabla _ua)_{\perp z}}\,\,\,\,\text{,}
\label{4.44} \\
g((\nabla _ua)_{\perp z},m_{\perp }) &=&\mp l_{(\nabla _ua)_{\perp z}}\cdot
g(n_{\perp },m_{\perp })=0\,\,\,\,\,\text{,}  \label{4.45}
\end{eqnarray}
\begin{eqnarray}
a_z &=&\mp l_{a_z}\cdot n_{\perp }\,\,\,\,\,\,\text{,}  \label{4.46} \\
g(a_z,n_{\perp }) &=&\mp l_{a_z}\cdot g(n_{\perp },n_{\perp
})=l_{a_z}\,\,\,\,\,\text{,}  \label{4.47} \\
g(a_z,m_{\perp }) &=&\mp l_{a_z}\cdot g(n_{\perp },m_{\perp
})=0\,\,\,\,\,\,\,\text{,}  \label{4.48}
\end{eqnarray}
\begin{eqnarray}
a_{\eta z} &=&\mp l_{a_{\eta z}}\cdot n_{\perp }\,\,\,\,\,\,\text{,}
\label{4.49} \\
g(a_{\eta z},n_{\perp }) &=&\mp l_{a_{\eta z}}\cdot g(n_{\perp },n_{\perp
})=l_{a_{\eta z}}\,\,\,\,\,\text{,}  \label{4.50} \\
g(a_{\eta z},m_{\perp }) &=&\mp l_{a_{\eta z}}\cdot g(n_{\perp },m_{\perp
})=0\,\,\,\,\,\,\text{,}  \label{4.51}
\end{eqnarray}
for finding out the explicit form of $g(_{rel}k_{\perp },n_{\perp })$ and $%
g(_{rel}k_{\perp },m_{\perp })$. 
\begin{equation}
g(_{rel}k_{\perp },n_{\perp })=\frac{l_{k_{\perp }}}{l_u}\cdot \{\mp \frac{dl%
}{l_u}\cdot l_{(a_{\perp })_z}\pm \frac{dl}{l_{\xi _{\perp }}}\cdot [\alpha
\cdot l_{v_z}+\beta \cdot \,l_{v_{\eta z}}]\}\pm  \label{4.52}
\end{equation}
\begin{equation}
\pm \frac 12\cdot \frac{dl^2}{l_u^2}\cdot l_{k_{\perp }}\cdot \{\frac
1{l_u}\cdot l_{(\nabla _ua)_{\perp z}}-\frac 1{l_{\xi _{\perp }}}\cdot
[\alpha \cdot l_{a_z}+\beta \cdot \,l_{a_{\eta z}}]\}\,\,\,\,\,\,\,\,\text{.}
\label{4.53}
\end{equation}
or the form

\begin{eqnarray}
g(_{rel}k_{\perp },n_{\perp }) &=&\frac{l_{k_{\perp }}}{l_u}\cdot \{\pm 
\frac{dl}{l_{\xi _{\perp }}}\cdot [\alpha \cdot l_{v_z}+\beta \cdot
\,l_{v_{\eta z}}]\mp \frac{dl}{l_u}\cdot l_{(a_{\perp })_z}\}\mp  \notag \\
&&\mp \frac 12\cdot \frac{dl^2}{l_u^2}\cdot l_{k_{\perp }}\cdot \{\frac
1{l_{\xi _{\perp }}}\cdot [\alpha \cdot l_{a_z}+\beta \cdot \,l_{a_{\eta
z}}]-\frac 1{l_u}\cdot l_{(\nabla _ua)_{\perp z}}\}\,\,\,\text{.}
\label{4.54}
\end{eqnarray}

The explicit form of $\overline{C}$ could now be found as 
\begin{eqnarray}
\overline{C} &=&\frac 1{l_{k_{\perp }}}\cdot g(_{rel}k_{\perp },n_{\perp })=
\notag \\
&=&\pm \frac 1{l_u}\cdot \{\frac{dl}{l_{\xi _{\perp }}}\cdot [\alpha \cdot
l_{v_z}+\beta \cdot \,l_{v_{\eta z}}]-\frac{dl}{l_u}\cdot l_{(a_{\perp
})_z}\}\mp  \notag \\
&&\mp \frac 12\cdot \frac{dl^2}{l_u^2}\cdot \{\frac 1{l_{\xi _{\perp
}}}\cdot [\alpha \cdot l_{a_z}+\beta \cdot \,l_{a_{\eta z}}]-\frac
1{l_u}\cdot l_{(\nabla _ua)_{\perp z}}\}\,\,\,\text{.}  \label{4.55}
\end{eqnarray}

If we introduce the abbreviations 
\begin{eqnarray}
\overline{l}_{v_{z}} &=&\frac{dl}{l_{\xi _{\perp }}}\cdot l_{v_{z}}\,\,\,\,%
\text{,\thinspace \thinspace \thinspace \thinspace \thinspace \thinspace
\thinspace \thinspace \thinspace }\overline{l}_{v_{\eta z}}=\frac{dl}{l_{\xi
_{\perp }}}\cdot \,l_{v_{\eta z}}\,\,\,\text{,\thinspace \thinspace
\thinspace \thinspace \thinspace \thinspace \thinspace \thinspace \thinspace 
}\overline{l}_{(a_{\perp })_{z}}=\frac{dl}{l_{\xi _{\perp }}}\cdot
l_{(a_{\perp })_{z}}\,\,\,\text{,}  \label{4.56} \\
\overline{l}_{a_{z}} &=&\frac{dl}{l_{\xi _{\perp }}}\cdot
l_{a_{z}}\,\,\,\,\,\,\text{,\thinspace \thinspace \thinspace \thinspace }\,%
\overline{l}_{a_{\eta z}}=\frac{dl}{l_{\xi _{\perp }}}\cdot \,l_{a_{\eta
z}}\,\,\,\,\text{,\thinspace \thinspace \thinspace \thinspace }\overline{l}%
_{(\nabla _{u}a)_{\perp z}}=\frac{dl}{l_{\xi _{\perp }}}\cdot l_{(\nabla
_{u}a)_{\perp z}}\,\text{,}  \label{4.57}
\end{eqnarray}%
then $\overline{C}$ could be represented in the form 
\begin{eqnarray}
\overline{C} &=&\pm \frac{1}{l_{u}}\cdot \lbrack (\alpha \cdot \overline{l}%
_{v_{z}}+\beta \cdot \,\overline{l}_{v_{\eta z}})-\frac{l_{\xi _{\perp }}}{%
l_{u}}\cdot \overline{l}_{(a_{\perp })_{z}}]\mp   \notag \\
&&\mp \frac{1}{2}\cdot \frac{dl}{l_{u}^{2}}\cdot \lbrack (\alpha \cdot 
\overline{l}_{a_{z}}+\beta \cdot \,\overline{l}_{a_{\eta z}})-\frac{l_{\xi
_{\perp }}}{l_{u}}\cdot \overline{l}_{(\nabla _{u}a)_{\perp z}}]\,\,\,\text{.%
}  \label{4.58}
\end{eqnarray}

In analogous way, we can find the explicit form of $\overline{S\text{.}}$

7. By the use of the relations

\begin{eqnarray}
g(_{rel}k_{\perp },m_{\perp }) &=&\frac{l_{k_{\perp }}}{l_u}\cdot \{\mp 
\frac{dl}{l_u}\cdot [g((a_{\perp })_z,m_{\perp })+g((a_{\perp })_c,m_{\perp
})]\pm  \notag \\
&&\pm \frac{dl}{l_{\xi _{\perp }}}\cdot [\alpha \cdot g(\,_{rel}v,m_{\perp
})+\beta \cdot \,g(_{rel}v_\eta ,m_{\perp })]\}\pm  \notag \\
&&\pm \frac 12\cdot \frac{dl^2}{l_u^2}\cdot l_{k_{\perp }}\cdot \{\frac
1{l_u}\cdot [g((\nabla _ua)_{\perp z},m_{\perp })+g((\nabla _ua)_{\perp
c},m_{\perp })]-  \notag \\
&&-\frac 1{l_{\xi _{\perp }}}\cdot [\alpha \cdot g(\,_{rel}a,m_{\perp
})+\beta \cdot \,g(_{rel}a_\eta ,m_{\perp })]\}\,\,\,\,\,\,\,\,\text{,}
\label{4.59}
\end{eqnarray}
\begin{eqnarray}
g(_{rel}k_{\perp },m_{\perp }) &=&\frac{l_{k_{\perp }}}{l_u}\cdot \{\mp 
\frac{dl}{l_u}\cdot g((a_{\perp })_c,m_{\perp })\pm  \notag \\
&&\pm \frac{dl}{l_{\xi _{\perp }}}\cdot [\alpha \cdot g(\,_{rel}v,m_{\perp
})+\beta \cdot \,g(_{rel}v_\eta ,m_{\perp })]\}\pm  \notag \\
&&\pm \frac 12\cdot \frac{dl^2}{l_u^2}\cdot l_{k_{\perp }}\cdot \{\frac
1{l_u}\cdot g((\nabla _ua)_{\perp c},m_{\perp })-  \notag \\
&&-\frac 1{l_{\xi _{\perp }}}\cdot [\alpha \cdot g(\,_{rel}a,m_{\perp
})+\beta \cdot \,g(_{rel}a_\eta ,m_{\perp })]\}\,\,\,\,\,\,\,\,\text{,}
\label{4.60}
\end{eqnarray}
\begin{eqnarray}
(a_{\perp })_c &=&\mp l_{(a_{\perp })_c}\cdot m_{\perp }\,\,\,\,\text{,}
\label{4.61} \\
g((a_{\perp })_c,m_{\perp }) &=&\mp l_{(a_{\perp })_c}\cdot g(m_{\perp
},m_{\perp })=l_{(a_{\perp })_c}\,\,\,\,\,\text{,}  \label{4.62}
\end{eqnarray}
\begin{equation}
v_c=\mp l_{v_c}\cdot m_{\perp }\,\,\,\,\,\text{,}  \label{4.63}
\end{equation}

\begin{eqnarray}
g(\,_{rel}v,m_{\perp }) &=&g(v_{z}+v_{c},m_{\perp })=  \notag \\
&=&g(v_{c},m_{\perp })=\mp l_{v_{c}}\cdot g(m_{\perp },m_{\perp
})=l_{v_{c}}\,\,\,\text{,}  \label{4.64}
\end{eqnarray}%
\begin{eqnarray}
v_{\eta c} &=&\mp l_{v_{\eta c}}\cdot m_{\perp }\,\,\,\,\,\,\text{,}
\label{4.65} \\
\,g(_{rel}v_{\eta },m_{\perp }) &=&g(v_{\eta z}+v_{\eta c},m_{\perp
})=g(v_{\eta c},m_{\perp })=\mp l_{v_{\eta c}}\cdot g(m_{\perp },m_{\perp })
\label{4.66} \\
&=&l_{v_{\eta c}}\,\,\,\text{,}  \label{4.66a}
\end{eqnarray}%
\begin{eqnarray}
(\nabla _{u}a)_{\perp c} &=&\mp l_{(\nabla _{u}a)_{\perp c}}\cdot m_{\perp
}\,\,\,\,\text{,}  \label{4.67} \\
g((\nabla _{u}a)_{\perp c},m_{\perp }) &=&\mp l_{(\nabla _{u}a)_{\perp
c}}\cdot g(m_{\perp },m_{\perp })=l_{(\nabla _{u}a)_{\perp c}}\,\,\,\,\text{,%
}  \label{4.68}
\end{eqnarray}%
\begin{equation}
_{rel}a=a_{z}+a_{c}\,\,\,\,\text{,\thinspace \thinspace \thinspace
\thinspace \thinspace }a_{z}=\mp l_{\text{\thinspace }a_{z}}\cdot n_{\perp
}\,\,\,\,\,\text{,\thinspace \thinspace \thinspace \thinspace \thinspace
\thinspace \thinspace }a_{c}=\mp l_{\text{\thinspace }a_{c}}\cdot m_{\perp
}\,\,\,\text{,}  \label{4.69}
\end{equation}%
\begin{equation}
g(\,_{rel}a,m_{\perp })=g(a_{z}+a_{c},m_{\perp })=g(a_{c},m_{\perp })=\mp l_{%
\text{\thinspace }a_{c}}\cdot g(m_{\perp },m_{\perp })=l_{\text{\thinspace }%
a_{c}}\,\,\,\text{,}  \label{4.70}
\end{equation}%
\begin{equation}
_{rel}a_{\eta }=a_{\eta z}+a_{\eta c}\,\,\,\,\text{,\thinspace \thinspace
\thinspace \thinspace \thinspace \thinspace \thinspace }a_{\eta z}=\mp
l_{a_{\eta z}}\cdot n_{\perp }\,\,\,\,\,\,\text{,\thinspace \thinspace
\thinspace \thinspace \thinspace \thinspace }a_{\eta c}=\mp l_{a_{\eta
c}}\cdot m_{\perp }\,\,\,\,\text{,}  \label{4.71}
\end{equation}%
\begin{equation}
g(_{rel}a_{\eta },m_{\perp })=g(a_{\eta c},m_{\perp })=\mp l_{a_{\eta
c}}\cdot g(m_{\perp },m_{\perp })=l_{a_{\eta c}}\,\,\,\,\,\text{.}
\label{4.72}
\end{equation}%
we can find the explicit forms of $g(_{rel}v_{\eta },m_{\perp })$ related to
the centrifugal (centripetal) and Coriolis velocities and accelerations:

\begin{equation*}
g(_{rel}k_{\perp },m_{\perp })=\frac{l_{k_{\perp }}}{l_{u}}\cdot \{\mp \frac{%
dl}{l_{u}}\cdot l_{(a_{\perp })_{c}}\pm \frac{dl}{l_{\xi _{\perp }}}\cdot
\lbrack \alpha \cdot l_{v_{c}}+\beta \cdot \,l_{v_{\eta c}}]\}\pm 
\end{equation*}%
\begin{equation}
\pm \frac{1}{2}\cdot \frac{dl^{2}}{l_{u}^{2}}\cdot l_{k_{\perp }}\cdot \{%
\frac{1}{l_{u}}\cdot l_{(\nabla _{u}a)_{\perp c}}-\frac{1}{l_{\xi _{\perp }}}%
\cdot \lbrack \alpha \cdot l_{\text{\thinspace }a_{c}}+\beta \cdot
\,l_{a_{\eta c}}]\}\,\,\,\,\,\,\,\,\text{,}  \label{4.73}
\end{equation}

\begin{equation*}
g(_{rel}k_{\perp },m_{\perp })=\pm \frac{l_{k_{\perp }}}{l_{u}}\cdot \lbrack 
\frac{dl}{l_{\xi _{\perp }}}\cdot (\alpha \cdot l_{v_{c}}+\beta \cdot
\,l_{v_{\eta c}})-\frac{dl}{l_{u}}\cdot l_{(a_{\perp })_{c}}]\mp 
\end{equation*}%
\begin{equation}
\mp \frac{1}{2}\cdot \frac{dl^{2}}{l_{u}^{2}}\cdot l_{k_{\perp }}\cdot
\lbrack \frac{1}{l_{\xi _{\perp }}}\cdot (\alpha \cdot l_{\text{\thinspace }%
a_{c}}+\beta \cdot \,l_{a_{\eta c}})-\frac{1}{l_{u}}\cdot l_{(\nabla
_{u}a)_{\perp c}}]\,\,\,\,\text{.}  \label{4.74}
\end{equation}

The explicit form of $\overline{S}$ could be found as 
\begin{eqnarray}
\overline{S} &:&=\frac 1{l_{k_{\perp }}}\cdot g(_{rel}k_{\perp },m_{\perp
})\,=  \notag \\
&=&\pm \frac 1{l_u}\cdot [\frac{dl}{l_{\xi _{\perp }}}\cdot (\alpha \cdot
l_{v_c}+\beta \cdot \,l_{v_{\eta c}})-\frac{dl}{l_u}\cdot l_{(a_{\perp
})_c}]\mp  \notag \\
&&\mp \frac 12\cdot \frac{dl^2}{l_u^2}\cdot [\frac 1{l_{\xi _{\perp }}}\cdot
(\alpha \cdot l_{\text{\thinspace }a_c}+\beta \cdot \,l_{a_{\eta c}})-\frac
1{l_u}\cdot l_{(\nabla _ua)_{\perp c}}]\,\,\,\,\text{.}  \label{4.75}
\end{eqnarray}

If we introduce the abbreviations 
\begin{eqnarray}
\overline{l}_{v_c} &=&\frac{dl}{l_{\xi _{\perp }}}\cdot l_{v_c}\,\,\,\,\,\,%
\text{,\thinspace \thinspace \thinspace \thinspace \thinspace \thinspace
\thinspace \thinspace \thinspace }\overline{l}_{v_{\eta c}}=\frac{dl}{l_{\xi
_{\perp }}}\cdot \,l_{v_{\eta c}}\,\,\,\,\,\text{,\thinspace \thinspace
\thinspace \thinspace \thinspace \thinspace \thinspace \thinspace \thinspace 
}\overline{l}_{(a_{\perp })\,c}=\frac{dl}{l_{\xi _{\perp }}}\cdot
l_{(a_{\perp })_c}\,\,\,\,\,\,\text{,}  \label{4.76} \\
\overline{l}_{a_c} &=&\frac{dl}{l_{\xi _{\perp }}}\cdot l_{a_c}\,\,\,\,\,\,%
\text{,\thinspace \thinspace \thinspace \thinspace \thinspace \thinspace
\thinspace \thinspace }\,\overline{l}_{a_{\eta c}}=\frac{dl}{l_{\xi _{\perp
}}}\cdot \,l_{a_{\eta c}}\,\,\,\,\text{,\thinspace \thinspace \thinspace
\thinspace \thinspace \thinspace \thinspace }\overline{l}_{(\nabla
_ua)_{\perp c}}=\frac{dl}{l_{\xi _{\perp }}}\cdot l_{(\nabla _ua)_{\perp
c}}\,\,\,\,\text{,}  \label{4.77} \\
\overline{l}_{\diamond } &\lesseqgtr &0\,\,\,\text{.}  \label{4.78}
\end{eqnarray}
then $\overline{S}$ could be represented in the form 
\begin{eqnarray}
\overline{S} &=&\pm \frac 1{l_u}\cdot [(\alpha \cdot \overline{l}%
_{v_c}+\beta \cdot \overline{\,l}_{v_{\eta c}})-\frac{l_{\xi _{\perp }}}{l_u}%
\cdot \overline{l}_{(a_{\perp })_c}]\mp  \notag \\
&&\mp \frac 12\cdot \frac{dl}{l_u^2}\cdot [(\alpha \cdot \overline{l}_{\text{%
\thinspace }a_c}+\beta \cdot \,\overline{l}_{a_{\eta c}})-\frac{l_{\xi
_{\perp }}}{l_u}\cdot \overline{l}_{(\nabla _ua)_{\perp c}}]\,\,\,\,\text{.}
\label{4.79}
\end{eqnarray}

The explicit forms of $\overline{C}$ and $\overline{S}$ determine the
relations describing the aberration, the Doppler effect, and the Hubble
effect in spaces with affine connections and metrics.

\section{Aberration}

\textit{Aberration} is the deviation of the direction of the vector field $%
\overline{k}_{\perp }$ from the direction of the vector field $k_{\perp }$.
If $\overline{k}_{\perp }=\mp l_{\overline{k}_{\perp }}\cdot \widetilde{n}%
\,_{\perp }^{\prime }$ and $k_{\perp }=\mp l_{k_{\perp }}\cdot \widetilde{n}%
_{\perp }$ then the difference between the angles $\theta ^{\prime }$ for $%
\overline{k}_{\perp }$ and $\theta $ for $k_{\perp }$ with respect to the
direction of the vector field $\xi _{\perp }$ is given by the relations

\begin{eqnarray}
\frac{\overline{\omega }}\omega \cdot cos\,\theta \,^{\prime }
&=&cos\,\theta +\frac 1{l_{k_{\perp }}}\cdot g(_{rel}k_{\perp },n_{\perp })=
\notag \\
&=&cos\,\theta +\overline{C}\,\,\,\,\,\,\text{,}  \label{5.1}
\end{eqnarray}
\begin{eqnarray}
\frac{\overline{\omega }}\omega \cdot sin\,\theta \,^{\prime }
&=&sin\,\theta +\frac 1{l_{k_{\perp }}}\cdot g(_{rel}k_{\perp },m_{\perp })=
\notag \\
&=&sin\,\theta +\overline{S}\,\,\,\,\,\,\,\,\,\,\,\,\text{.}  \label{5.2}
\end{eqnarray}

From the last (above) two relations, it follows for $tg\theta ^{\prime }$%
\begin{eqnarray}
tg\theta ^{\prime } &=&\frac{sin\,\theta \,^{\prime }}{cos\,\theta
\,^{\prime }}=\frac{sin\,\theta +\frac 1{l_{k_{\perp }}}\cdot
g(_{rel}k_{\perp },m_{\perp })}{cos\,\theta +\frac 1{l_{k_{\perp }}}\cdot
g(_{rel}k_{\perp },n_{\perp })}\,=  \notag \\
&=&\frac{sin\,\theta +\overline{S}}{cos\,\theta +\overline{C}}\,\,\,\,\,%
\text{.}  \label{5.3}
\end{eqnarray}

If there is no relative velocities and no relative accelerations between the
emitter and the observer then $\overline{C}=0$ and $\overline{S}=0$. Then 
\begin{eqnarray}
\frac{\overline{\omega }}\omega \cdot cos\,\theta \,^{\prime }
&=&cos\,\theta \,\,\,\,\,\text{,}  \label{5.4} \\
\frac{\overline{\omega }}\omega \cdot sin\,\theta \,^{\prime }
&=&sin\,\theta \,\,\,\,\,\,\text{,}  \label{5.5} \\
\frac{\overline{\omega }^2}{\omega ^2} &=&1\,\,\,\,\,\text{,\thinspace
\thinspace \thinspace \thinspace \thinspace \thinspace \thinspace \thinspace 
}  \notag \\
\overline{\omega } &=&\omega \,\,\,\,\text{.}  \label{5.6}
\end{eqnarray}

The frequency of the signal, emitted by the emitter, and the frequency of
the signal, detected by the observer, are the same and, at the same time, no
aberration occurs.

\textit{Special case}: Auto-parallel motion $(\nabla _uu=a=0)$ of an
observer detected a signal with emitted frequency $\overline{\omega }$. Then

\begin{equation}
\overline{C}=\pm \frac 1{l_u}\cdot [(\alpha \cdot \overline{l}_{v_z}+\beta
\cdot \,\overline{l}_{v_{\eta z}})]\mp \frac 12\cdot \frac{dl}{l_u^2}\cdot
[(\alpha \cdot \overline{l}_{a_z}+\beta \cdot \,\overline{l}_{a_{\eta
z}})]\,\,\,\text{.}  \label{5.7}
\end{equation}

\begin{equation}
\overline{S}=\pm \frac 1{l_u}\cdot [(\alpha \cdot \overline{l}_{v_c}+\beta
\cdot \overline{\,l}_{v_{\eta c}})]\mp \frac 12\cdot \frac{dl}{l_u^2}\cdot
[(\alpha \cdot \overline{l}_{\text{\thinspace }a_c}+\beta \cdot \,\overline{l%
}_{a_{\eta c}})]\,\,\,\,\text{.}  \label{5.8}
\end{equation}
\begin{equation}
\overline{\omega }\cdot sin\theta ^{\prime }=\{sin\theta \pm \frac
1{l_u}\cdot [(\alpha \cdot \overline{l}_{v_c}+\beta \cdot \overline{\,l}%
_{v_{\eta c}})]\mp \frac 12\cdot \frac{dl}{l_u^2}\cdot [(\alpha \cdot 
\overline{l}_{\text{\thinspace }a_c}+\beta \cdot \,\overline{l}_{a_{\eta
c}})]\}\cdot \omega \,\,\,\text{,}  \label{5.9}
\end{equation}
\begin{equation}
\overline{\omega }\cdot cos\theta ^{\prime }=\{cos\theta \pm \frac
1{l_u}\cdot [(\alpha \cdot \overline{l}_{v_z}+\beta \cdot \,\overline{l}%
_{v_{\eta z}})]\mp \frac 12\cdot \frac{dl}{l_u^2}\cdot [(\alpha \cdot 
\overline{l}_{a_z}+\beta \cdot \,\overline{l}_{a_{\eta z}})]\}\cdot \omega
\,\,\,\text{,}  \label{5.10}
\end{equation}

\begin{equation}
\alpha =\mp \,cos\,\theta \,\,\,\,\text{,\thinspace \thinspace \thinspace
\thinspace \thinspace \thinspace \thinspace \thinspace \thinspace \thinspace
\thinspace \thinspace \thinspace \thinspace \thinspace \thinspace }\beta
=\mp sin\,\theta \,\,\,\text{\thinspace \thinspace .}  \label{5.11}
\end{equation}

\section{Doppler effect}

The \textit{Doppler effect} (Doppler shift) is the shift of signal's
frequency caused by the relative motion between the emitter and the observer.

1. Usually, in classical mechanics, and especially in acoustics, there is a
difference between the shift of the frequency when the observer is moving to
or out of the emitter and the shift of the frequency when the emitter is
moving to or out of the observer. In the first case, the signal is
propagating in a medium at rest with respect to the emitter, and in the
second case, the signal is propagating in a medium moving with respect to
the emitter. It is assumed that the signal is propagating in a continuous
media used as a carrier of the signal.

In relativistic physics, and especially in electrodynamics, there is no
difference between the shifts of the frequency of a signal when the observer
is moving to or out of the emitter and when the emitter is moving to or out
of the observer. The relative motion is the only reason for the shift
frequency.

2. If the shift of the frequency of a signal is considered from the point of
view of an observer then only the relative motions with respect to the
observer could be taken into account. The observer detects the signals in
his proper frame of reference (laboratory) and could make a comparison
between the signals sent by emitter at rest with respect to his proper frame
of reference and by emitter moving relatively to observer's proper frame of
reference.

3. The observer could move in space-time where the space could be filled
with a continuous media or with classical fields with physical
interpretation. Since every classical field theory could be considered as a
theory of a continuous media \cite{Manoff-1}, \cite{Manoff-1a}, \cite%
{Manoff-2a} both type of theories could be used for dynamical description of
propagation of signals in space-time. An observer is interested in finding
out how signals are propagating, how the emitter are moving with respect to
the observer, and how the signals are generated by an emitter. Only the
first two questions are subjects of consideration by the use of kinematic
characteristics of the relative velocity, the relative acceleration, and the
properties of null (isotropic) vector fields. The last question is a matter
of considerations of the corresponding dynamical theory.

4. The Doppler effect could be described in spaces with affine connections
and metrics as models of space or space-time on the basis of the relations
between the emitted frequency $\overline{\omega }$ and detected frequency $%
\omega $ of signals propagating in space or space-time. The same relations
are used for consideration of the aberration of signals. As corollary of
them a relation between $\overline{\omega }$ and $\omega $ follows in the
form 
\begin{equation}
\overline{\omega }=[(sin\,\theta +\overline{S\,})^2+(cos\,\theta +\overline{C%
})^2]^{1/2}\cdot \omega \,\,\,\,\,\,\,\,  \label{6.1}
\end{equation}
appearing as a general formula for the generalized Doppler effect in spaces
with affine connections and metrics.

\subsection{Standard (longitudinal) Doppler effect (Doppler shift)}

1. The standard (longitudinal) Doppler effect appears when all Coriolis
velocities and Coriolis accelerations are compensating each other or do not
exist in the relative motion between emitter and detector (observer), i.e.
if 
\begin{equation}
\overline{S\,}=0\,\,\,\,\,\text{.}  \label{6.2}
\end{equation}

Then 
\begin{eqnarray}
\overline{\omega } &=&[(sin\,^2\theta +(cos\,\theta +\overline{C}%
)^2]^{1/2}\cdot \omega =  \notag \\
&=&[sin^2\theta +cos^2\theta +2\cdot \overline{C}\cdot cos\theta +\overline{C%
}^2]^{1/2}\cdot \omega  \label{6.3}
\end{eqnarray}
\begin{equation}
\overline{\omega }=[1+2\cdot \overline{C}\cdot cos\theta +\overline{C}%
^2]^{1/2}\cdot \omega \,\,\,\,\text{,}  \label{6.4}
\end{equation}
where

\begin{eqnarray*}
\overline{C} &=&\pm \frac 1{l_u}\cdot [(\mp cos\,\theta \cdot \overline{l}%
_{v_z}\mp sin\,\theta \cdot \,\overline{l}_{v_{\eta z}})-\frac{l_{\xi
_{\perp }}}{l_u}\cdot \overline{l}_{(a_{\perp })_z}]\mp \\
&&\mp \frac 12\cdot \frac{dl}{l_u^2}\cdot [(\mp cos\,\theta \cdot \overline{l%
}_{a_z}\mp sin\,\theta \cdot \,\overline{l}_{a_{\eta z}})-\frac{l_{\xi
_{\perp }}}{l_u}\cdot \overline{l}_{(\nabla _ua)_{\perp z}}]\,\,\,\text{,}
\end{eqnarray*}
\begin{eqnarray}
\overline{C} &=&\frac 1{l_u}\cdot [-(cos\,\theta \cdot \overline{l}%
_{v_z}+sin\,\theta \cdot \,\overline{l}_{v_{\eta z}})\mp \frac{l_{\xi
_{\perp }}}{l_u}\cdot \overline{l}_{(a_{\perp })_z}]+  \notag \\
&&+\frac 12\cdot \frac{dl}{l_u^2}\cdot [(cos\,\theta \cdot \overline{l}%
_{a_z}+sin\,\theta \cdot \,\overline{l}_{a_{\eta z}})\pm \frac{l_{\xi
_{\perp }}}{l_u}\cdot \overline{l}_{(\nabla _ua)_{\perp z}}]\,\,\,\text{.}
\label{6.5}
\end{eqnarray}

2. If the vector field $k_{\perp }$ is collinear to the vector field $\xi
_{\perp }$ determining the proper frame of reference, i.e. if 
\begin{equation}
k_{\perp }=\mp l_{k_{\perp }}\cdot n_{\perp }\,\,\,\,\text{,\thinspace
\thinspace \thinspace \thinspace \thinspace \thinspace \thinspace \thinspace
\thinspace \thinspace \thinspace \thinspace \thinspace }cos\,\theta =\pm
1\,\,\,\,\text{, \thinspace \thinspace \thinspace \thinspace \thinspace
\thinspace \thinspace }sin\,\theta =0\,\,\,\,\,\text{,}  \label{6.6}
\end{equation}
then

\begin{eqnarray}
\overline{C} &=&\frac 1{l_u}\cdot (\mp \overline{l}_{v_z}\mp \frac{l_{\xi
_{\perp }}}{l_u}\cdot \overline{l}_{(a_{\perp })_z})+\frac 12\cdot \frac{dl}{%
l_u^2}\cdot (\pm \,\overline{l}_{a_z}\pm \frac{l_{\xi _{\perp }}}{l_u}\cdot 
\overline{l}_{(\nabla _ua)_{\perp z}})\,\,\text{,}  \notag \\
\overline{C} &=&\mp [\frac 1{l_u}\cdot (\overline{l}_{v_z}+\frac{l_{\xi
_{\perp }}}{l_u}\cdot \overline{l}_{(a_{\perp })_z})-\frac 12\cdot \frac{dl}{%
l_u^2}\cdot (\,\overline{l}_{a_z}+\frac{l_{\xi _{\perp }}}{l_u}\cdot 
\overline{l}_{(\nabla _ua)_{\perp z}})]\,\,\,\,\text{,}  \label{6.7}
\end{eqnarray}
\begin{eqnarray}
\overline{\omega } &=&[1\pm 2\cdot \overline{C}+\overline{C}^2]^{1/2}\cdot
\omega =  \notag \\
&=&[(1\pm \overline{C})^2]^{1/2}\cdot \omega =(1\pm \overline{C})\cdot
\omega \,\,\,\text{,}  \label{6.8}
\end{eqnarray}
\begin{equation}
\frac{\overline{\omega }-\omega }\omega =\pm \overline{C}=z\,\,\,\,\text{,}
\label{6.8a}
\end{equation}

\begin{eqnarray}
\overline{\omega } &=&\omega \cdot [1-\frac 1{l_u}\cdot (\overline{l}_{v_z}+%
\frac{l_{\xi _{\perp }}}{l_u}\cdot \overline{l}_{(a_{\perp })_z})+\frac
12\cdot \frac{dl}{l_u^2}\cdot (\overline{l}_{a_z}+\frac{l_{\xi _{\perp }}}{%
l_u}\cdot \overline{l}_{(\nabla _ua)_{\perp z}})]=  \notag \\
&=&\omega \cdot [1-\frac{\overline{l}_{v_z}}{l_u}-\frac{l_{\xi _{\perp }}}{%
l_u^2}\cdot \overline{l}_{(a_{\perp })_z}+\frac 12\cdot \frac{dl}{l_u^2}%
\cdot (\overline{l}_{a_z}+\frac{l_{\xi _{\perp }}}{l_u}\cdot \overline{l}%
_{(\nabla _ua)_{\perp z}})]\,\,\,\,\,\text{.}  \label{6.9}
\end{eqnarray}

\textit{Special case}: Auto-parallel motion of the observer: $\nabla _uu=a=0$%
: $\overline{l}_{(a_{\perp })_z}=0$,\thinspace \thinspace \thinspace
\thinspace $\overline{l}_{(\nabla _ua)_{\perp z}}=0\,\,$, 
\begin{eqnarray}
\overline{\omega } &=&\omega \cdot [1-\frac{\overline{l}_{v_z}}{l_u}+\frac
12\cdot \frac{dl}{l_u^2}\cdot \overline{l}_{a_z}]\,\,\,\,\,\text{,}
\label{6.10} \\
\overline{l}_{v_z} &\lesseqqgtr &0\,\,\,\,\,\text{,\thinspace \thinspace
\thinspace \thinspace \thinspace \thinspace \thinspace \thinspace }\overline{%
l}_{a_z}\lesseqqgtr 0\,\,\,\,\,\text{.}  \notag
\end{eqnarray}

If the world line of an observer is an auto-parallel trajectory and $%
k_{\perp }$ is collinear to $\xi _{\perp }$ then the change of the frequency 
$\overline{\omega }$ of the emitter depends on the centrifugal (centripetal)
velocity $\overline{l}_{v_z}$ and the centrifugal (centripetal) acceleration 
$\overline{l}_{a_z}$.

\textit{Special case}: $\nabla _uu=a=0$,\thinspace \thinspace \thinspace
\thinspace $k_{\perp }=\mp l_{k_{\perp }}\cdot n_{\perp }$,\thinspace $%
\overline{l}_{a_z}=0$: 
\begin{equation}
\overline{\omega }=\omega \cdot (1-\frac{\overline{l}_{v_z}}{l_u}%
)\,\,\,\,\,\,\text{,\thinspace \thinspace \thinspace \thinspace \thinspace
\thinspace \thinspace \thinspace \thinspace \thinspace \thinspace \thinspace
\thinspace \thinspace }\overline{l}_{v_z}\lesseqqgtr 0\,\,\,\,\,\text{%
.\thinspace \thinspace \thinspace \thinspace }  \label{6.11}
\end{equation}

Therefore, if the world line of an observer is an auto-parallel trajectory, $%
k_{\perp }$ is collinear to $\xi _{\perp }$, and no centrifugal
(centripetal) acceleration $\overline{l}_{a_z}$ exists between emitter and
observer then the above expression has the well known form for description
of the standard Doppler effect in classical mechanics in $3$-dimensional
Euclidean space. Here, the relation is valid in every $(\overline{L}_n,g)$%
-space considered as a model of a space or a space-time under the given
preconditions.

\subsection{Transversal Doppler effect}

1. The transversal Doppler effect appears when all centrifugal (centripetal)
velocities and centrifugal (centripetal) accelerations are compensating each
other or do not exist in the relative motion between emitter and detector
(observer), i.e. if 
\begin{equation}
\overline{C}=0\,\,\,\,\,\,\text{.}  \label{6.12}
\end{equation}

Then 
\begin{eqnarray}
\overline{\omega } &=&[(sin\,\theta +\overline{S\,})^2+cos^2\,\theta
]^{1/2}\cdot \omega =  \notag \\
&=&[sin^2\theta +2\cdot \overline{S\,}\cdot sin\theta +\overline{S\,}%
^2+cos^2\,\theta ]^{1/2}\cdot \omega =  \notag \\
&=&[1+\,\,2\cdot \overline{S\,}\cdot sin\theta +\overline{S\,}^2]^{1/2}\cdot
\omega \,\,\,\,\,\text{,}  \label{6.13}
\end{eqnarray}
where

\begin{eqnarray}
\overline{S} &:&=\pm \frac 1{l_u}\cdot [\frac{dl}{l_{\xi _{\perp }}}\cdot
(\alpha \cdot l_{v_c}+\beta \cdot \,l_{v_{\eta c}})-\frac{dl}{l_u}\cdot
l_{(a_{\perp })_c}]\mp  \notag \\
&&\mp \frac 12\cdot \frac{dl^2}{l_u^2}\cdot [\frac 1{l_{\xi _{\perp }}}\cdot
(\alpha \cdot l_{\text{\thinspace }a_c}+\beta \cdot \,l_{a_{\eta c}})-\frac
1{l_u}\cdot l_{(\nabla _ua)_{\perp c}}]\,\,\,\,\text{,}  \label{6.14}
\end{eqnarray}

\begin{equation}
\alpha =\mp \,cos\,\theta \,\,\,\,\text{,\thinspace \thinspace \thinspace
\thinspace \thinspace \thinspace \thinspace }\beta =\mp sin\,\theta \,\,\,%
\text{\thinspace \thinspace ,}  \label{6.15}
\end{equation}

\begin{eqnarray}
\overline{S} &:&=\pm \frac{1}{l_{u}}\cdot \lbrack \frac{dl}{l_{\xi _{\perp }}%
}\cdot (\mp \,cos\,\theta \,\cdot l_{v_{c}}\mp sin\,\theta \cdot
\,l_{v_{\eta c}})-\frac{dl}{l_{u}}\cdot l_{(a_{\perp })_{c}}]\mp   \notag \\
&&\mp \frac{1}{2}\cdot \frac{dl^{2}}{l_{u}^{2}}\cdot \lbrack \frac{1}{l_{\xi
_{\perp }}}\cdot (\mp \,cos\,\theta \cdot l_{\text{\thinspace }a_{c}}\mp
sin\,\theta \,\,\,\cdot \,l_{a_{\eta c}})-\frac{1}{l_{u}}\cdot l_{(\nabla
_{u}a)_{\perp c}}]\,\,\text{,}  \label{6.16}
\end{eqnarray}

\begin{eqnarray}
\overline{S} &:&=\frac{1}{l_{u}}\cdot \lbrack -\frac{dl}{l_{\xi _{\perp }}}%
\cdot (\,cos\,\theta \,\cdot l_{v_{c}}+sin\,\theta \cdot \,l_{v_{\eta
c}})\mp \frac{dl}{l_{u}}\cdot l_{(a_{\perp })_{c}}]+  \notag \\
&&+\frac{1}{2}\cdot \frac{dl^{2}}{l_{u}^{2}}\cdot \lbrack \frac{1}{l_{\xi
_{\perp }}}\cdot (\,cos\,\theta \cdot l_{\text{\thinspace }%
a_{c}}+sin\,\theta \,\,\,\cdot \,l_{a_{\eta c}})\pm \frac{1}{l_{u}}\cdot
l_{(\nabla _{u}a)_{\perp c}}]\,\,\text{.}  \label{6.17}
\end{eqnarray}

2. If the vector field $k_{\perp }$ is orthogonal to the vector field $\xi
_{\perp }$ determining the proper frame of reference, i.e. if 
\begin{equation}
k_{\perp }=\mp l_{k_{\perp }}\cdot m_{\perp }\,\,\,\,\text{,\thinspace
\thinspace \thinspace \thinspace \thinspace \thinspace \thinspace \thinspace
\thinspace \thinspace \thinspace \thinspace \thinspace }sin\,\theta =\pm
1\,\,\,\,\text{, \thinspace \thinspace \thinspace \thinspace \thinspace
\thinspace \thinspace }cos\,\theta =0\,\,\,\,\,\text{,}  \label{6.18}
\end{equation}
then 
\begin{eqnarray}
\overline{\omega } &=&[1+\,\,2\cdot \overline{S\,}\cdot sin\theta +\overline{%
S\,}^2]^{1/2}\cdot \omega =  \notag \\
&=&[1\,\pm 2\cdot \overline{S\,}+\overline{S\,}^2]^{1/2}\cdot \omega = 
\notag \\
&=&(1\pm \overline{S\,})\cdot \omega \,\,\,\,\,\,\text{,}  \label{6.19}
\end{eqnarray}
\begin{equation}
\frac{\overline{\omega }-\omega }\omega =\pm \overline{S}=z_c\,\,\,\,\,\text{%
,}  \label{6.19a}
\end{equation}

\begin{eqnarray*}
\overline{S} &:&=\frac 1{l_u}\cdot [-\frac{dl}{l_{\xi _{\perp }}}\cdot (\pm
\,l_{v_{\eta c}})\mp \frac{dl}{l_u}\cdot l_{(a_{\perp })_c}]+ \\
&&+\frac 12\cdot \frac{dl^2}{l_u^2}\cdot (\frac 1{l_{\xi _{\perp }}}\cdot
(\pm \,l_{a_{\eta c}})\pm \frac 1{l_u}\cdot l_{(\nabla _ua)_{\perp
c}})\,\,\,\,\text{,}
\end{eqnarray*}
\begin{eqnarray*}
\overline{S} &:&=\frac 1{l_u}\cdot (\mp \frac{dl}{l_{\xi _{\perp }}}\cdot
\,l_{v_{\eta c}}\mp \frac{dl}{l_u}\cdot l_{(a_{\perp })_c})+ \\
&&+\frac 12\cdot \frac{dl^2}{l_u^2}\cdot (\pm \frac 1{l_{\xi _{\perp
}}}\cdot \,l_{a_{\eta c}}\pm \frac 1{l_u}\cdot l_{(\nabla _ua)_{\perp
c}})\,\,\,\,\text{,}
\end{eqnarray*}

\begin{eqnarray}
\overline{S} &:&=\mp \frac 1{l_u}\cdot (\frac{dl}{l_{\xi _{\perp }}}\cdot
\,l_{v_{\eta c}}+\frac{dl}{l_u}\cdot l_{(a_{\perp })_c})\pm  \notag \\
&&\pm \frac 12\cdot \frac{dl^2}{l_u^2}\cdot (\frac 1{l_{\xi _{\perp }}}\cdot
\,l_{a_{\eta c}}+\frac 1{l_u}\cdot l_{(\nabla _ua)_{\perp c}})\,\,\,\,\text{,%
}  \label{6.20}
\end{eqnarray}
\begin{eqnarray}
\overline{\omega } &=&\{1-\frac 1{l_u}\cdot (\frac{dl}{l_{\xi _{\perp }}}%
\cdot \,l_{v_{\eta c}}+\frac{dl}{l_u}\cdot l_{(a_{\perp })_c})+  \notag \\
&&+\frac 12\cdot \frac{dl^2}{l_u^2}\cdot (\frac 1{l_{\xi _{\perp }}}\cdot
\,l_{a_{\eta c}}+\frac 1{l_u}\cdot l_{(\nabla _ua)_{\perp c}})]\}\cdot
\omega \,\,\,\text{.}  \label{6.21}
\end{eqnarray}

If we introduce the abbreviations 
\begin{eqnarray}
\,\overline{l}_{v_{\eta c}} &=&\frac{dl}{l_{\xi _{\perp }}}\cdot
\,l_{v_{\eta c}}\,\,\,\,\,\,\,\text{,\thinspace \thinspace \thinspace
\thinspace \thinspace \thinspace \thinspace \thinspace \thinspace \thinspace
\thinspace }\overline{l}_{(a_{\perp })_{c}}=\frac{dl}{l_{\xi _{\perp }}}%
\cdot l_{(a_{\perp })_{c}}\,\,\,\,\,\,\,\,\,\text{,}  \label{6.22} \\
\,\overline{l}_{a_{\eta c}} &=&\frac{dl}{l_{\xi _{\perp }}}\cdot l_{a_{\eta
c}}\,\,\,\,\,\,\,\,\text{,\thinspace \thinspace \thinspace \thinspace
\thinspace \thinspace \thinspace \thinspace \thinspace \thinspace \thinspace 
}\overline{l}_{(\nabla _{u}a)_{\perp c}})=\frac{dl}{l_{\xi _{\perp }}}\cdot
l_{(\nabla _{u}a)_{\perp c}}\,\,\,\,\,\,\text{,}  \label{6.23} \\
\overline{l}_{\diamond } &\lesseqqgtr &0\,\,\,\,\,\text{,}  \notag
\end{eqnarray}%
then the expressions for $\overline{S}$ and $\overline{\omega }$ will have
the forms 
\begin{eqnarray*}
\overline{S} &=&\mp \frac{1}{l_{u}}\cdot (\,\overline{l}_{v_{\eta c}}+\frac{%
l_{\xi _{\perp }}}{l_{u}}\cdot \overline{l}_{(a_{\perp })_{c}})\pm \frac{1}{2%
}\cdot \frac{dl}{l_{u}^{2}}\cdot (\,\overline{l}_{a_{\eta c}}+\frac{l_{\xi
_{\perp }}}{l_{u}}\cdot l_{(\nabla _{u}a)_{\perp c}})= \\
&=&\mp \lbrack \frac{1}{l_{u}}\cdot (\,\overline{l}_{v_{\eta c}}+\frac{%
l_{\xi _{\perp }}}{l_{u}}\cdot \overline{l}_{(a_{\perp })_{c}})-\frac{1}{2}%
\cdot \frac{dl}{l_{u}^{2}}\cdot (\,\overline{l}_{a_{\eta c}}+\frac{l_{\xi
_{\perp }}}{l_{u}}\cdot l_{(\nabla _{u}a)_{\perp c}})]\,\,\,\,\,\text{,}
\end{eqnarray*}%
\begin{eqnarray}
\pm \overline{S} &=&-[\frac{1}{l_{u}}\cdot (\,\overline{l}_{v_{\eta c}}+%
\frac{l_{\xi _{\perp }}}{l_{u}}\cdot \overline{l}_{(a_{\perp })_{c}})-\frac{1%
}{2}\cdot \frac{dl}{l_{u}^{2}}\cdot (\,\overline{l}_{a_{\eta c}}+\frac{%
l_{\xi _{\perp }}}{l_{u}}\cdot l_{(\nabla _{u}a)_{\perp c}})]=  \notag \\
&=&-\frac{1}{l_{u}}\cdot (\,\overline{l}_{v_{\eta c}}+\frac{l_{\xi _{\perp }}%
}{l_{u}}\cdot \overline{l}_{(a_{\perp })_{c}})+\frac{1}{2}\cdot \frac{dl}{%
l_{u}^{2}}\cdot (\,\overline{l}_{a_{\eta c}}+\frac{l_{\xi _{\perp }}}{l_{u}}%
\cdot l_{(\nabla _{u}a)_{\perp c}})\,\,\text{,}  \label{6.24}
\end{eqnarray}

\begin{eqnarray}
\overline{\omega } &=&\omega \cdot \lbrack 1-\frac{1}{l_{u}}\cdot (\,%
\overline{l}_{v_{\eta c}}+\frac{l_{\xi _{\perp }}}{l_{u}}\cdot \overline{l}%
_{(a_{\perp })_{c}})+\frac{1}{2}\cdot \frac{dl}{l_{u}^{2}}\cdot (\,\overline{%
l}_{a_{\eta c}}+\frac{l_{\xi _{\perp }}}{l_{u}}\cdot l_{(\nabla
_{u}a)_{\perp c}})]=  \notag \\
&=&\omega \cdot \lbrack 1-\frac{1}{l_{u}}\cdot \overline{l}_{v_{\eta c}}-%
\frac{l_{\xi _{\perp }}}{l_{u}^{2}}\cdot \overline{l}_{(a_{\perp })_{c}}+%
\frac{1}{2}\cdot \frac{dl}{l_{u}^{2}}\cdot (\,\overline{l}_{a_{\eta c}}+%
\frac{l_{\xi _{\perp }}}{l_{u}}\cdot l_{(\nabla _{u}a)_{\perp c}})]\,\,\text{%
.}  \label{6.25}
\end{eqnarray}

\textit{Remark}. The expression for $\overline{S}$ has the same form as the
expression for $\overline{C}$ under the change of $\overline{l}_{vz}$ with $%
\overline{l}_{v\eta c}$, $\overline{l}_{(a_{\perp })_z}$ with $\overline{l}%
_{(a_{\perp })_c}$, and $\overline{l}_{(\nabla _ua)_{\perp z}}$ with $%
l_{(\nabla _ua)_{\perp c}}$.

\textit{Special case}: Auto-parallel motion of the observer: $\nabla _uu=a=0$%
: $\overline{l}_{(a_{\perp })_c}=0$,\thinspace \thinspace \thinspace
\thinspace $\overline{l}_{(\nabla _ua)_{\perp c}}=0\,\,$, 
\begin{equation}
\overline{\omega }=\omega \cdot [1-\frac 1{l_u}\cdot \overline{l}_{v_{\eta
c}}+\frac 12\cdot \frac{dl}{l_u^2}\cdot \,\overline{l}_{a_{\eta c}}]\,\,\,\,%
\text{.}  \label{6.27}
\end{equation}

If the world line of an observer is an auto-parallel trajectory and $%
k_{\perp }$ is orthogonal to $\xi _{\perp }$ then the change of the
frequency $\overline{\omega }$ of the emitter depends on the Coriolis
velocity $\overline{l}_{v_{\eta c}}$ and the Coriolis acceleration $%
\overline{l}_{a_{\eta c}}$.

\textit{Special case}: $\nabla _uu=a=0$,\thinspace \thinspace \thinspace
\thinspace $k_{\perp }=\mp l_{k_{\perp }}\cdot m_{\perp }$,\thinspace $%
\overline{l}_{a_{\eta c}}=0$: 
\begin{equation}
\overline{\omega }=\omega \cdot (1-\frac{\overline{l}_{v_{\eta c}}}{l_u}%
)\,\,\,\,\,\,\text{,\thinspace \thinspace \thinspace \thinspace \thinspace
\thinspace \thinspace \thinspace \thinspace \thinspace \thinspace \thinspace
\thinspace \thinspace }\overline{l}_{v_z}\lesseqqgtr 0\,\,\,\,\,\text{%
.\thinspace \thinspace \thinspace \thinspace }  \label{6.28}
\end{equation}

Therefore, if the world line of an observer is an auto-parallel trajectory, $%
k_{\perp }$ is orthogonal to $\xi _{\perp }$, and no Coriolis acceleration $%
\overline{l}_{a_{\eta c}}$ exists between emitter and observer then the
above expression has analogous form for description of the transversal
Doppler effect as the standard (longitudinal) Doppler effect in classical
mechanics in $3$-dimensional Euclidean space. Here, the relation for the
transversal Doppler effect is valid in every $(\overline{L}_n,g)$-space
considered as a model of a space or a space-time under the given
preconditions.

\section{Hubble effect}

The \textit{Hubble effect} (Hubble shift) is the Doppler shift (Doppler
effect) of signal's frequency caused by the relative motion between the
emitter and the observer when the explicit form of the relative velocities
and of the relative accelerations are given. The \textit{Hubble law (law of
redshift)} is defined as the linear dependence of the distances to galaxies
on their red shift. In more general sense, the Hubble law is the statement
that the relative velocity between an observer and a particle (from the
point of view of the proper frame of reference of the observer) is
proportional to the distance between the observer and the particle. Usually,
the Hubble effect is defined as the change of the frequency $\overline{%
\omega }$ under the motion of an emitter with centrifugal (centripetal)
velocity $v_z$ relatively to an observer.

Usually, the Hubble effect is related only to the centrifugal velocity of an
emitter with respect to an observer on the basis of the Hubble
distance-redshift relation \cite{Weinberg}, \cite{Unzicker} discovered in
1929 and interpreted as a result of the expansion of the universe. The
explicit form of the kinematic characteristics of the centrifugal
(centripetal) and Coriolis velocities and accelerations determine uniquely
the Hubble effect \cite{Manoff-6}.

\subsection{Explicit forms of the centrifugal (centripetal) and Coriolis
velocities and accelerations}

Let us now consider the explicit forms of the relative velocities and of the
relative accelerations determining a Doppler shift (Doppler effect).

The vector fields generating a Doppler effect could be represented into two
groups with respect to their lengths $\overline{l}_{\diamond }$:

(a) Vector fields generating a standard (longitudinal) Doppler effect

\begin{itemize}
\item centrifugal (centripetal) part $\overline{l}_{v_z}$ of relative
centrifugal (centripetal) velocity $\overline{l}_v$ ,

\item centrifugal (centripetal) part $\overline{l}_{v_{\eta z}}$ of relative
velocity $\overline{l}_{_{rel}v_\eta }$ ,

\item centrifugal part (centripetal) $\overline{l}_{(a_{\perp })_z}$ of
acceleration $\overline{l}_{a_{\perp }}$,

\item centrifugal (centripetal) part $\overline{l}_{a_z}$ of the relative
acceleration $\overline{l}_{_{rel}a}$,

\item centrifugal (centripetal) part $\overline{l}_{a_{\eta z}}$ of the
relative acceleration $\overline{l}_{_{rel}a_\eta }$,

\item centrifugal (centripetal) part $\overline{l}_{(\nabla _ua)_{\perp z}}$
of the change of the acceleration $\overline{l}_{a_{\perp }}$.
\end{itemize}

(b) Vector fields generating a transversal Doppler effect

\begin{itemize}
\item Coriolis part $\overline{l}_{v_c}$ of relative velocity $\overline{l}%
_v $ ,

\item Coriolis part $\overline{l}_{v_{\eta c}}$ of a relative velocity $%
\overline{l}_{_{rel}v_\eta }$ ,

\item Coriolis part $\overline{l}_{(a_{\perp })_c}$ of acceleration $%
\overline{l}_a$,

\item Coriolis part $\overline{l}_{a_c}$ of the relative acceleration $%
\overline{l}_{_{rel}a}$,

\item Coriolis part $\overline{l}_{a_{\eta c}}$ of the relative acceleration 
$\overline{l}_{_{rel}a_\eta }$,

\item Coriolis part $\overline{l}_{(\nabla _ua)_{\perp c}}$ of the change $%
\overline{l}_{(\nabla _ua)_{\perp }}$ of the acceleration $\overline{l}_a$.
\end{itemize}

\subsubsection{Explicit form of the relative velocities and accelerations
generating a standard (longitudinal) Doppler effect}

The relative velocities have two essential components: $_{rel}v=v_z+v_c$, $%
_{rel}v_\eta =v_{\eta z}+v_{\eta c}$.

(a) Relative centrifugal (centripetal) velocity $v_z$%
\begin{equation}
v_z=\frac{g(_{rel}v,\xi _{\perp })}{g(\xi _{\perp },\xi _{\perp })}\cdot \xi
_{\perp }\,\,\,\text{,}  \label{7.1}
\end{equation}
could be represented in its explicit form as 
\begin{eqnarray}
v_z &=&[\frac 1{n-1}\cdot \theta \mp \sigma (n_{\perp },n_{\perp })]\cdot
\xi _{\perp }\,\,\,\,\,\,\text{,}  \label{7.2} \\
v_z &=&\mp \,l_{v_z}\cdot n_{\perp }=\,H\cdot l_{\xi _{\perp }}\cdot
n_{\perp }=\,H\cdot \xi _{\perp }\,\,\,\,\,\text{,}  \label{7.3}
\end{eqnarray}
where 
\begin{eqnarray}
H &=&\frac 1{n-1}\cdot \theta \mp \sigma (n_{\perp },n_{\perp })\text{%
\thinspace \thinspace \thinspace \thinspace \thinspace \thinspace \thinspace
\thinspace \thinspace ,}  \label{7.4} \\
l_{v_z} &=&\mp \,H\cdot l_{\xi _{\perp }}\,\,\,\,\,\,\text{,\thinspace
\thinspace \thinspace \thinspace \thinspace \thinspace \thinspace \thinspace
\thinspace \thinspace \thinspace \thinspace \thinspace }\overline{l}_{v_z}=%
\frac{dl}{l_{\xi _{\perp }}}\cdot l_{v_z}=\mp \,H\cdot dl\text{ \thinspace
\thinspace \thinspace \thinspace \thinspace \thinspace .}  \label{7.5}
\end{eqnarray}

The last (above) expressions for $\overline{l}_{v_z}$ and for $\overline{v}%
_z $%
\begin{equation}
\overline{l}_{v_z}=\mp \,H\cdot dl\text{ \thinspace \thinspace \thinspace
\thinspace \thinspace \thinspace \thinspace \thinspace \thinspace \thinspace
\thinspace \thinspace ,\thinspace \thinspace \thinspace \thinspace
\thinspace \thinspace \thinspace \thinspace \thinspace \thinspace \thinspace
\thinspace }\overline{v}_z=\frac{dl}{l_{\xi _{\perp }}}\cdot v_z=H\cdot
dl\cdot n_{\perp }\text{\thinspace \thinspace \thinspace \thinspace }
\label{7.6}
\end{equation}
are the well known relations called standard (longitudinal) Hubble law: the
centrifugal (centripetal) relative velocity $\overline{v}_z$ is proportional
to the distance $dl$ between an emitter and a detector (observer). This form
of the Hubble law is a very special form of the low for the case when only
the centrifugal (centripetal) relative velocity is taken into account.

(b) Centrifugal (centripetal) part $\overline{l}_{v_{\eta z}}$ of a relative
velocity $\overline{l}_{_{rel}v_{\eta }}$%
\begin{eqnarray}
v_{\eta z} &=&\mp g(_{rel}v_{\eta },n_{\perp })\cdot n_{\perp }=\mp l_{\xi
_{\perp }}\cdot d(n_{\perp },m_{\perp })\cdot n_{\perp }=  \label{7.7} \\
&=&\mp l_{v_{\eta z}}\cdot n_{\perp }\,\,\,\,\text{,}  \label{7.7a} \\
d(n_{\perp },m_{\perp }) &=&\sigma (n_{\perp },m_{\perp })+\omega (n_{\perp
},m_{\perp })+\frac{1}{n-1}\cdot \theta \cdot h_{u}(n_{\perp },m_{\perp })= 
\notag \\
&=&\sigma (n_{\perp },m_{\perp })+\omega (n_{\perp },m_{\perp
})\,\,\,\,\,\,\,\text{,}  \label{7.8} \\
h_{u}(n_{\perp },m_{\perp }) &=&g(n_{\perp },m_{\perp })-\frac{1}{\pm
l_{u}^{2}}\cdot g(u,n_{\perp })\cdot g(u,m_{\perp })=  \label{7.9} \\
&=&g(n_{\perp },m_{\perp })=0\,\,\,\,\text{,}  \label{7.9a} \\
v_{\eta z} &=&\mp l_{v_{\eta z}}\cdot n_{\perp }=H_{\eta z}\cdot l_{\xi
_{\perp }}\cdot n_{\perp }=\mp l_{\xi _{\perp }}\cdot d(n_{\perp },m_{\perp
})\cdot n_{\perp }\,\,\,\text{,}  \label{7.10}
\end{eqnarray}%
\begin{eqnarray}
H_{\eta z} &=&\mp d(n_{\perp },m_{\perp })=  \notag \\
&=&\mp \lbrack \sigma (n_{\perp },m_{\perp })+\omega (n_{\perp },m_{\perp
})]=H_{c}\,\,\,\text{,}  \label{7.11} \\
H_{\eta z} &=&H_{c}\,\,\,\,\text{,}  \label{7.12}
\end{eqnarray}%
could be represented in its explicit form as 
\begin{equation}
l_{v_{\eta z}}=\mp H_{c}\cdot l_{\xi _{\perp }}\,\,\,\,\,\,\,\text{%
,\thinspace \thinspace \thinspace \thinspace \thinspace \thinspace
\thinspace \thinspace \thinspace \thinspace \thinspace \thinspace \thinspace
\thinspace }\overline{l}_{v_{\eta z}}=\frac{dl}{l_{\xi _{\perp }}}\cdot
l_{v_{\eta z}}=\mp H_{c}\cdot dl\,\,\,\,\text{,}  \label{7.13}
\end{equation}%
\begin{equation}
\text{\thinspace }\overline{l}_{v_{\eta z}}=\mp H_{c}\cdot dl\,\,\,\,\,\,%
\text{\thinspace \thinspace \thinspace \thinspace \thinspace ,\thinspace
\thinspace \thinspace \thinspace \thinspace \thinspace \thinspace \thinspace
\thinspace \thinspace \thinspace \thinspace \thinspace }\overline{v}_{\eta
z}=\frac{dl}{l_{\xi _{\perp }}}\cdot v_{\eta z}=H_{c}\cdot dl\cdot n_{\perp
}\,\,\,\text{.}  \label{7.14}
\end{equation}

(c) Centrifugal part (centripetal) $\overline{l}_{(a_{\perp })_z}$ of
acceleration $\overline{l}_{a_{\perp }}$%
\begin{equation}
a_{\perp }=\mp g(a_{\perp },n_{\perp })\cdot n_{\perp }+\overline{g}[h_{\xi
_{\perp }}(a_{\perp })]\,\,\,\,\,\text{,}  \label{7.15}
\end{equation}
\begin{eqnarray}
(a_{\perp })_z &=&\mp \text{\thinspace }l_{(a_{\perp })_z}\cdot n_{\perp }%
\text{ \thinspace \thinspace \thinspace ,}  \label{7.16} \\
g(a_{\perp },n_{\perp }) &=&l_{(a_{\perp })_z}=g(\overline{g}%
[h_u(a)],n_{\perp })=  \notag \\
&=&g_{\overline{i}\overline{j}}\cdot g^{ik}\cdot h_{\overline{k}\overline{l}%
}\cdot a^l\cdot n_{\perp }^j=g_j^k\cdot h_{\overline{k}\overline{l}}\cdot
a^l\cdot n_{\perp }^j=  \notag \\
&=&h_{\overline{j}\overline{l}}\cdot a^l\cdot n_{\perp }^j=h_u(n_{\perp
},a)=g(n_{\perp },a)\,\,\,\,\text{,}  \label{7.17}
\end{eqnarray}
\begin{equation}
l_{(a_{\perp })_z}=g(a_{\perp },n_{\perp })\,\,\,\,\,\,\text{,\thinspace
\thinspace \thinspace \thinspace \thinspace \thinspace \thinspace \thinspace
\thinspace \thinspace }\overline{l}_{(a_{\perp })_z}=\frac{dl}{l_{\xi
_{\perp }}}\cdot l_{(a_{\perp })_z}=\frac{dl}{l_{\xi _{\perp }}}\cdot
g(a_{\perp },n_{\perp })\,\,\,\,\text{.}  \label{7.18}
\end{equation}

(d) Centrifugal (centripetal) part $\overline{l}_{a_{z}}$ of the relative
acceleration $\overline{l}_{_{rel}a}$%
\begin{eqnarray}
a_{z} &=&\mp l_{\xi _{\perp }}\cdot A(n_{\perp },n_{\perp })\cdot n_{\perp
}=\mp A(n_{\perp },n_{\perp })\cdot \xi _{\perp }=\mp l_{a_{z}}\cdot
n_{\perp }=  \notag \\
&=&\overline{q}\cdot \xi _{\perp }=\overline{q}\cdot l_{\xi _{\perp }}\cdot
n_{\perp }\,\,\,\,\,\,\text{,}  \label{7.19} \\
g(a_{z},n_{\perp }) &=&l_{a_{z}}=l_{\xi _{\perp }}\cdot A(n_{\perp
},n_{\perp })=\mp \overline{q}\cdot l_{\xi _{\perp }}\,\,\,\,\text{%
,\thinspace \thinspace \thinspace }  \label{7.19a}
\end{eqnarray}%
\begin{equation}
\text{\thinspace }A(n_{\perp },n_{\perp })=\mp \,\overline{q}\,\,\,\,\,\text{%
,}  \label{7.19b}
\end{equation}

\begin{eqnarray}
\overline{q} &=&\frac{1}{n-1}\cdot U\mp \,_{s}D(n_{\perp },n_{\perp })\,\,\,%
\text{,}  \label{7.21} \\
a_{z} &=&\overline{q}\cdot l_{\xi _{\perp }}\cdot n_{\perp }=\mp
l_{a_{z}}\cdot n_{\perp }\,\,\,\,\,\text{,\thinspace \thinspace }
\label{7.22} \\
l_{a_{z}} &=&\mp \overline{q}\cdot l_{\xi _{\perp }}\,\,\,\text{,\thinspace
\thinspace \thinspace \thinspace \thinspace \thinspace \thinspace \thinspace 
}  \label{7.23}
\end{eqnarray}%
\begin{eqnarray}
\overline{l}_{a_{z}} &=&\mp \overline{q}\cdot \frac{dl}{l_{\xi _{\perp }}}%
\cdot l_{\xi _{\perp }}=\mp \,\overline{q}\cdot dl\,\,\,\,\,\,\text{,}
\label{7.24} \\
\overline{a}_{z} &=&\frac{dl}{l_{\xi _{\perp }}}\cdot a_{z}=\mp \overline{l}%
_{a_{z}}\cdot n_{\perp }=\overline{q}\cdot dl\cdot n_{\perp }\,\,\,\,\text{.}
\label{7.25}
\end{eqnarray}

The last (above) two relations for $\overline{a}_z$ and $\overline{l}_{a_z}$
represent the part of the Hubble effect generated by the centrifugal
(centripetal) part \thinspace \thinspace $\overline{l}_{a_z}$ of the
acceleration $_{rel}a$. The centrifugal (centripetal) acceleration is
proportional to the distance $dl$ between an emitter and a detector
(observer).

(e) Centrifugal (centripetal) part $\overline{l}_{a_{\eta z}}$ of the
relative acceleration $\overline{l}_{_{rel}a_{\eta }}$%
\begin{eqnarray}
a_{\eta z} &=&\mp g(_{rel}a_{\eta },n_{\perp })\cdot n_{\perp }=\mp l_{\xi
_{\perp }}\cdot A(m_{\perp },n_{\perp })\cdot n_{\perp }=\mp l_{a_{\eta
z}}\cdot n_{\perp }=  \notag \\
&=&\overline{q}_{\eta }\cdot l_{\xi _{\perp }}\cdot n_{\perp }=\overline{q}%
_{\eta }\cdot \xi _{\perp }\,\,\,\,\,\text{,}  \label{7.26} \\
g(a_{\eta z},n_{\perp }) &=&l_{a_{\eta z}}=l_{\xi _{\perp }}\cdot A(m_{\perp
},n_{\perp })=\mp \overline{q}_{\eta }\cdot l_{\xi _{\perp }}\,\,\,\,\text{%
,\thinspace \thinspace }  \label{7.26a}
\end{eqnarray}%
\begin{equation}
A(n_{\perp },m_{\perp })=\mp \overline{q}_{\eta }\,\,\,\,\text{,}
\label{7.26b}
\end{equation}

\begin{eqnarray}
\overline{q}_{\eta } &=&\mp A(m_{\perp },n_{\perp })=\mp \lbrack
_{s}D(m_{\perp },n_{\perp })+W(m_{\perp },n_{\perp })]\,=\overline{q}%
_{c}\,\,\,\,\,\,\,\text{,}  \label{7.28} \\
\overline{l}_{a_{\eta z}} &=&\mp \overline{q}_{c}\cdot \frac{dl}{l_{\xi
_{\perp }}}\cdot l_{\xi _{\perp }}=\mp \overline{q}_{c}\cdot dl\,\,\,\,\,\,%
\text{,}  \label{7.29} \\
\overline{a}_{\eta z} &=&\frac{dl}{l_{\xi _{\perp }}}\cdot a_{\eta z}=\mp 
\overline{l}_{a_{\eta z}}\cdot n_{\perp }=\overline{q}_{c}\cdot dl\cdot
n_{\perp }\,\,\,\,\text{.}  \label{7.30}
\end{eqnarray}

(f) Centrifugal (centripetal) part $\overline{l}_{(\nabla _ua)_{\perp z}}$
of the change of the acceleration $\overline{l}_{a_{\perp }}$

\begin{eqnarray}
(\nabla _ua)_{\perp z} &=&\mp g((\nabla _ua)_{\perp },n_{\perp })\cdot
n_{\perp }=\mp \text{\thinspace }l_{(\nabla _ua)_{\perp z}}\cdot n_{\perp
}\,\,\,\,\text{,}  \label{7.31} \\
\text{\thinspace }l_{(\nabla _ua)_{\perp z}} &=&g((\nabla _ua)_{\perp
},n_{\perp })=g(\overline{g}[h_u(\nabla _ua)],n_{\perp })=  \notag \\
&=&h_u(n_{\perp },\nabla _ua)\,\,\,\,\text{,}  \label{7.32} \\
\text{\thinspace \thinspace }\overline{l}_{(\nabla _ua)_{\perp }{}_z} &=&%
\frac{dl}{l_{\xi _{\perp }}}\cdot l_{(\nabla _ua)_{\perp z}}=\frac{dl}{%
l_{\xi _{\perp }}}\cdot g(\nabla _ua_{\perp },n_{\perp })\,\,\,\,\,\text{.}
\label{7.34}
\end{eqnarray}

\subsubsection{Explicit form of the relative velocities and accelerations
generating a transversal Doppler effect}

(a) Coriolis part $\overline{l}_{v_c}$ of relative Coriolis velocity $%
\overline{l}_v$

\begin{equation}
v_c=\overline{g}[h_{\xi _{\perp }}(_{rel}v)]  \label{7.35}
\end{equation}
could be represented in its explicit form as 
\begin{eqnarray}
v_c &=&\overline{g}[\sigma (\xi _{\perp })]-\frac{\sigma (\xi _{\perp },\xi
_{\perp })}{g(\xi _{\perp },\xi _{\perp })}\cdot \xi _{\perp }=  \notag \\
&=&\mp l_{v_c}\cdot m_{\perp }=  \notag \\
&=&l_{\xi _{\perp }}\cdot \overline{g}[\sigma (n_{\perp })]\pm \sigma
(n_{\perp },n_{\perp })\cdot l_{\xi _{\perp }}\cdot n_{\perp }+l_{\xi
_{\perp }}\cdot \overline{g}[\omega (n_{\perp })]=  \notag \\
&=&H_c\cdot l_{\xi _{\perp }}\cdot m_{\perp }\,\,\,\,\,\text{.}  \label{7.36}
\end{eqnarray}

By the use of the relations 
\begin{eqnarray}
g(v_{c},m_{\perp }) &=&\mp l_{v_{c}}\cdot g(m_{\perp },m_{\perp
})=l_{v_{c}}\,\,\,\text{,}  \label{7.37} \\
g(v_{c},m_{\perp }) &=&l_{\xi _{\perp }}\cdot g(\overline{g}[\sigma
(n_{\perp })],m_{\perp })+l_{\xi _{\perp }}\cdot g(\overline{g}[\omega
(n_{\perp })],m_{\perp })=l_{v_{c}}\,\text{,}  \label{7.38}
\end{eqnarray}%
\begin{eqnarray}
g(\overline{g}[\sigma (n_{\perp })],m_{\perp }) &=&g_{\overline{i}\overline{j%
}}\cdot g^{ik}\cdot \sigma _{kl}\cdot n_{\perp }^{\overline{l}}\cdot
m_{\perp }^{\overline{j}}=  \notag \\
&=&g_{j}^{k}\cdot \sigma _{kl}\cdot n_{\perp }^{\overline{l}}\cdot m_{\perp
}^{\overline{j}}=\sigma _{jl}\cdot m_{\perp }^{\overline{j}}\cdot n_{\perp
}^{\overline{l}}=  \notag \\
&=&\sigma (m_{\perp },n_{\perp })\,\,\,\text{,}  \label{7.39}
\end{eqnarray}%
\begin{equation}
g(\overline{g}[\omega (n_{\perp })],m_{\perp })=\omega (m_{\perp },n_{\perp
})\,\,\,\,\,\text{,}  \label{7.40}
\end{equation}%
we obtain the expressions for $l_{v_{c}}$ and $v_{c}$ respectively 
\begin{equation}
l_{v_{c}}=[\sigma (m_{\perp },n_{\perp })+\omega (m_{\perp },n_{\perp
})]\cdot l_{\xi _{\perp }}\,\,\,\,\text{,}  \label{7.41}
\end{equation}%
\begin{eqnarray}
v_{c} &=&\mp l_{v_{c}}\cdot m_{\perp }=H_{c}\cdot l_{\xi _{\perp }}\cdot
m_{\perp }=  \notag \\
&=&\mp \lbrack \sigma (m_{\perp },n_{\perp })+\omega (m_{\perp },n_{\perp
})]\cdot l_{\xi _{\perp }}\cdot m_{\perp }\,\,\,\,\text{,}  \label{7.42}
\end{eqnarray}%
where 
\begin{equation}
H_{c}=\mp \lbrack \sigma (m_{\perp },n_{\perp })+\omega (m_{\perp },n_{\perp
})]\,\,\,\,\text{.}  \label{7.43}
\end{equation}

Then 
\begin{eqnarray}
\overline{l}_{v_c} &=&\frac{dl}{l_{\xi _{\perp }}}\cdot l_{v_c}=\mp
\,H_c\cdot dl\,\,\,\text{,}  \label{7.44} \\
\overline{l}_{v_c} &=&\mp \,H_c\cdot dl\,\,\,\,\text{.}  \label{7.45}
\end{eqnarray}

The last (above) expressions for $\overline{l}_{v_c}$ and for $\overline{v}%
_c $%
\begin{equation}
\overline{l}_{v_c}=\mp \,H_c\cdot dl\text{ \thinspace \thinspace \thinspace
\thinspace \thinspace \thinspace \thinspace \thinspace \thinspace \thinspace
\thinspace \thinspace ,\thinspace \thinspace \thinspace \thinspace
\thinspace \thinspace \thinspace \thinspace \thinspace \thinspace \thinspace
\thinspace }\overline{v}_c=\frac{dl}{l_{\xi _{\perp }}}\cdot v_c=H_c\cdot
dl\cdot m_{\perp }\text{\thinspace \thinspace \thinspace \thinspace ,}
\label{7.46}
\end{equation}
are the relations describing the transversal Hubble law: the Coriolis
relative velocity $\overline{v}_c$ is proportional to the distance $dl$
between an emitter and a detector (observer). This form of the Hubble law is
a very special form of the low for the case when only the Coriolis relative
velocity is taken into account.

(b) Coriolis part $\overline{l}_{v_{\eta c}}$ of a relative velocity $%
\overline{l}_{_{rel}v_\eta }$%
\begin{equation}
v_{\eta c}=l_{\xi _{\perp }}\cdot h^{n_{\perp }}[d(m_{\perp })]=\mp
l_{v_{\eta c}}\cdot m_{\perp }=\overline{H}_c\cdot l_{\xi _{\perp }}\cdot
m_{\perp }\text{ \thinspace ,}  \label{7.47}
\end{equation}
could be represented in its explicit form by the use of the relations

\begin{eqnarray}
h^{n_{\perp }}[d(m_{\perp })] &=&(\overline{g}-\frac{1}{g(n_{\perp
},n_{\perp })}\cdot n_{\perp }\otimes n_{\perp })[d(m_{\perp })]=  \notag \\
&=&\overline{g}[d(m_{\perp })]-\frac{1}{\mp 1}\cdot (n_{\perp })[d(m_{\perp
})]\cdot n_{\perp }=  \notag \\
&=&\overline{g}[d(m_{\perp })]\pm (n_{\perp })[d(m_{\perp })]\cdot n_{\perp
}=  \label{7.48} \\
&=&\overline{g}[d(m_{\perp })]\pm d(n_{\perp },m_{\perp })\,\cdot n_{\perp
}\,\,\,\,\,\,\,\text{,}  \label{7.48a}
\end{eqnarray}%
\begin{eqnarray}
g(v_{\eta c},m_{\perp }) &=&l_{v_{\eta c}}=l_{\xi _{\perp }}\cdot g(%
\overline{g}[d(m_{\perp })],m_{\perp })=l_{\xi _{\perp }}\cdot g_{\overline{i%
}\overline{j}}\cdot g^{ik}\cdot d_{\overline{k}\overline{l}}\cdot m_{\perp
}^{l}\cdot m_{\perp }^{j}=  \notag \\
&=&l_{\xi _{\perp }}\cdot g_{j}^{k}\cdot d_{\overline{k}\overline{l}}\cdot
m_{\perp }^{l}\cdot m_{\perp }^{j}=l_{\xi _{\perp }}\cdot d_{\overline{j}%
\overline{l}}\cdot m_{\perp }^{l}\cdot m_{\perp }^{j}=l_{\xi _{\perp }}\cdot
d(m_{\perp },m_{\perp })=  \notag \\
&=&l_{\xi _{\perp }}\cdot \{[\sigma +\omega +\frac{1}{n-1}\cdot \theta \cdot
h_{u})](m_{\perp })\}(m_{\perp })=  \notag \\
&=&l_{\xi _{\perp }}\cdot \lbrack \sigma (m_{\perp },m_{\perp })\mp \frac{1}{%
n-1}\cdot \theta ]=  \label{7.49} \\
&=&\mp l_{\xi _{\perp }}\cdot \lbrack \frac{1}{n-1}\cdot \theta \mp \sigma
(m_{\perp },m_{\perp })]\,\,\,\,\text{,}  \label{7.49a}
\end{eqnarray}%
\begin{equation}
l_{v_{\eta c}}=\mp \lbrack \frac{1}{n-1}\cdot \theta \mp \sigma (m_{\perp
},m_{\perp })]\,\cdot l_{\xi _{\perp }}=\mp \overline{H}_{c}\cdot l_{\xi
_{\perp }}\,\,\,\,\text{,}  \label{7.50}
\end{equation}%
\begin{equation}
\overline{H}_{c}=\frac{1}{n-1}\cdot \theta \mp \sigma (m_{\perp },m_{\perp
})\,\,\,\,\,\,\,\text{,}  \label{7.51}
\end{equation}%
\begin{eqnarray}
v_{\eta c} &=&\mp l_{v_{\eta c}}\cdot m_{\perp }=\overline{H}_{c}\cdot
l_{\xi _{\perp }}\cdot m_{\perp }\,\,\text{,}  \label{7.52} \\
\overline{l}_{v_{\eta c}} &=&\frac{dl}{l_{\xi _{\perp }}}\cdot l_{v_{\eta
c}}=\mp \overline{H}_{c}\cdot dl\,\,\,\,\,\text{,\thinspace \thinspace
\thinspace \thinspace \thinspace \thinspace \thinspace \thinspace \thinspace
\thinspace \thinspace \thinspace \thinspace \thinspace }\overline{v}_{\eta
c}=\frac{dl}{l_{\xi _{\perp }}}\cdot v_{\eta c}=\overline{H}_{c}\cdot
dl\cdot m_{\perp }\,\,\,\,\text{.}  \label{7.53}
\end{eqnarray}

(c) Coriolis part $\overline{l}_{(a_{\perp })_c}$ of acceleration $\overline{%
l}_a$. From the relations

\begin{eqnarray}
(a_{\perp })_{c}\, &=&\overline{g}[h_{\xi _{\perp }}(a_{\perp })]\,\,\,\,\,%
\text{,}  \label{7.55} \\
g(\xi _{\perp },(a_{\perp })_{c}) &=&0\text{ \thinspace \thinspace
,\thinspace \thinspace \thinspace \thinspace \thinspace \thinspace
\thinspace \thinspace }(a_{\perp })_{c}=\mp \text{\thinspace }l_{(a_{\perp
})_{c}}\cdot m_{\perp }\,\,\,\,\,\,\,\,\text{,}  \label{7.56}
\end{eqnarray}%
\begin{eqnarray}
\text{\thinspace }l_{(a_{\perp })_{c}} &=&g(\text{\thinspace }(a_{\perp
})_{c},m_{\perp })=  \notag \\
&=&g(\overline{g}[h_{\xi _{\perp }}(a_{\perp })],m_{\perp })=g_{\overline{i}%
\overline{j}}\cdot g^{ik}\cdot (h_{\xi _{\perp }})_{\overline{k}\overline{l}%
}\cdot a_{\perp }^{l}\cdot m_{\perp }^{j}=  \notag \\
&=&h_{\xi _{\perp }}(m_{\perp },a_{\perp })=h_{n_{\perp }}(m_{\perp
},a_{\perp })=g(m_{\perp },a_{\perp })\,\,\,\,\,\text{,}  \label{7.57} \\
h_{n_{\perp }}(m_{\perp },a_{\perp }) &=&[g\pm g(n_{\perp })\otimes
g(n_{\perp })](m_{\perp },a_{\perp })=  \notag \\
&=&g(m_{\perp },a_{\perp })\pm g(n_{\perp },m_{\perp })\cdot g(n_{\perp
},a_{\perp })=  \label{7.58} \\
&=&g(m_{\perp },a_{\perp })\,\,\text{,}  \label{7.58a}
\end{eqnarray}%
it follows the form of $\overline{l}_{(a_{\perp })_{c}}$%
\begin{eqnarray}
\text{\thinspace }l_{(a_{\perp })_{c}} &=&g(m_{\perp },a_{\perp })\,\,\text{%
,\thinspace \thinspace \thinspace }(a_{\perp })_{c}=\mp \text{\thinspace }%
l_{(a_{\perp })_{c}}\cdot m_{\perp }=\mp \text{\thinspace }g(m_{\perp
},a_{\perp })\text{\thinspace }\cdot m_{\perp }\,\text{,}  \label{7.59} \\
\text{\thinspace }\overline{l}_{(a_{\perp })_{c}} &=&\frac{dl}{l_{\xi
_{\perp }}}\cdot \text{\thinspace }l_{(a_{\perp })_{c}}\,\,\text{,\thinspace
\thinspace \thinspace \thinspace }(\overline{a}_{\perp })_{c}=\frac{dl}{%
l_{\xi _{\perp }}}\cdot \text{\thinspace }(a_{\perp })_{c}=\mp \text{%
\thinspace }\overline{l}_{(a_{\perp })_{c}}\cdot m_{\perp }\,\,\text{.}
\label{7.60}
\end{eqnarray}

(d) Coriolis part $\overline{l}_{a_{c}}$ of the relative acceleration $%
\overline{l}_{_{rel}a}$. By means of the expressions 
\begin{eqnarray}
a_{c} &=&\overline{g}[h_{\xi _{\perp }}(_{rel}a)]=\overline{g}[h_{n_{\perp
}}(_{rel}a)]=l_{\xi _{\perp }}\cdot \overline{g}(h_{\xi _{\perp }})(%
\overline{g})[A(n_{\perp })]=  \notag \\
&=&l_{\xi _{\perp }}\cdot h^{n_{\perp }}[A(n_{\perp })]=\mp l_{a_{c}}\cdot
m_{\perp }=\overline{q}\,_{c}\cdot l_{\xi _{\perp }}\cdot m_{\perp }\text{ ,}
\notag \\
a_{c} &=&l_{\xi _{\perp }}\cdot \{\overline{g}[_{s}D(n_{\perp })]\mp
\,_{s}D(n_{\perp },n_{\perp })\cdot n_{\perp }+\overline{g}[W(n_{\perp })]\}=
\label{7.61} \\
&=&\mp l_{a_{c}}\cdot m_{\perp }\,\,\,\text{,}  \label{7.61a}
\end{eqnarray}%
\begin{eqnarray}
g(a_{c},m_{\perp }) &=&l_{a_{c}}=l_{\xi _{\perp }}\cdot \{g(\overline{g}%
[_{s}D(n_{\perp })],m_{\perp })+g(\overline{g}[W(n_{\perp })],m_{\perp
})\}\,\,\,\text{,}  \notag \\
g(\overline{g}[_{s}D(n_{\perp })],m_{\perp }) &=&\,g_{\overline{i}\overline{j%
}}\cdot g^{ik}\cdot \,_{s}D_{\overline{k}\overline{l}}\cdot n_{\perp
}^{l}\cdot m_{\perp }^{j}=  \notag \\
&=&\,_{s}D_{\overline{j}\overline{l}}\cdot m_{\perp }^{j}\cdot n_{\perp
}^{l}=\,_{s}D(m_{\perp },n_{\perp })\,\,\,\,\text{,}  \label{7.62}
\end{eqnarray}%
\begin{equation}
g(\overline{g}[W(n_{\perp })],m_{\perp })=W(m_{\perp },n_{\perp })\,\,\,\,%
\text{,}  \label{7.63}
\end{equation}%
the explicit form of $\overline{l}_{a_{c}}$ follows 
\begin{equation}
l_{a_{c}}=l_{\xi _{\perp }}\cdot \lbrack _{s}D(m_{\perp },n_{\perp
})+W(m_{\perp },n_{\perp })\,]\,=\mp l_{\xi _{\perp }}\cdot \overline{q}%
_{c}\,\,\text{,}  \label{7.64}
\end{equation}%
\begin{eqnarray}
a_{c} &=&\mp l_{a_{c}}\cdot m_{\perp }=\overline{q}_{c}\cdot l_{\xi _{\perp
}}\cdot m_{\perp }\,\,\,\,\text{,}  \label{7.65} \\
\overline{q}_{c} &=&\mp \lbrack _{s}D(m_{\perp },n_{\perp })+W(m_{\perp
},n_{\perp })\,]\,\,\,\,\,\text{,}  \label{7.66}
\end{eqnarray}%
\begin{equation}
\overline{l}_{a_{c}}=\frac{dl}{l_{\xi _{\perp }}}\cdot l_{a_{c}}=\mp 
\overline{q}_{c}\cdot dl\,\,\,\,\,\,\,\,\text{,\thinspace \thinspace
\thinspace \thinspace \thinspace \thinspace \thinspace \thinspace \thinspace
\thinspace \thinspace \thinspace \thinspace \thinspace }\overline{a}_{c}=%
\frac{dl}{l_{\xi _{\perp }}}\cdot a_{c}=\overline{q}_{c}\cdot dl\cdot
m_{\perp }\,\,\,\,\text{.}  \label{7.67}
\end{equation}

(e) Coriolis part $\overline{l}_{a_{\eta c}}$ of the relative acceleration $%
\overline{l}_{_{rel}a_{\eta }}$. By the use of the relations 
\begin{eqnarray}
a_{\eta _{c}} &=&\overline{g}[h_{\xi _{\perp }}(_{rel}a_{\eta })]\,=%
\overline{g}[h_{n_{\perp }}(_{rel}a_{\eta })]=l_{\xi _{\perp }}\cdot
h^{n_{\perp }}[A(m_{\perp })]=  \notag \\
&=&\mp l_{a_{\eta c}}\cdot m_{\perp }\,\,\,\,\text{,}  \label{7.68} \\
a_{\eta _{c}} &=&l_{\xi _{\perp }}\cdot h^{n_{\perp }}[A(m_{\perp })]\,= 
\notag \\
&=&\mp l_{a_{\eta c}}\cdot m_{\perp }\text{ ,}  \label{7.69} \\
h^{n_{\perp }}[A(m_{\perp })]\, &=&(\overline{g}\pm n_{\perp }\otimes
n_{\perp })[_{s}D(m_{\perp })+W(m_{\perp })+\frac{1}{n-1}\cdot U\cdot
h_{u}(m_{\perp })]=  \notag \\
&=&\overline{g}[_{s}D(m_{\perp })]+\overline{g}[W(m_{\perp })]\pm (n_{\perp
})_{s}D(m_{\perp })\cdot n_{\perp }\pm   \notag \\
&&\pm (n_{\perp })[W(m_{\perp })]\cdot n_{\perp }\,\,\text{,}  \label{7.70}
\end{eqnarray}%
\begin{eqnarray}
g(a_{\eta c},m_{\perp }) &=&l_{a_{\eta c}}=  \notag \\
&=&l_{\xi _{\perp }}\cdot g(\overline{g}[_{s}D(m_{\perp })],m_{\perp })+g(%
\overline{g}[W(m_{\perp })],m_{\perp })=  \notag \\
&=&\,l_{\xi _{\perp }}\cdot \lbrack _{s}D(m_{\perp },m_{\perp })+W(m_{\perp
},m_{\perp })]=  \notag \\
&=&l_{\xi _{\perp }}\cdot \,_{s}D(m_{\perp },m_{\perp })\,\,\,\,\text{,}\,
\label{7.71}
\end{eqnarray}%
\begin{eqnarray}
a_{\eta _{c}} &=&\mp l_{a_{\eta c}}\cdot m_{\perp }=\overline{q}_{\eta
c}\cdot l_{\xi _{\perp }}\cdot m_{\perp }\,\,\,\,\text{,}  \label{7.72} \\
\overline{q}_{\eta c} &=&\,\mp \,_{s}D(m_{\perp },m_{\perp })\,\,\,\,\text{%
,\thinspace \thinspace \thinspace \thinspace \thinspace \thinspace }%
l_{a_{\eta c}}=\mp \,\overline{q}_{\eta c}\cdot l_{\xi _{\perp }}\,\,\,\text{%
,}  \label{7.73} \\
\overline{l}_{a_{\eta c}} &=&\frac{dl}{l_{\xi _{\perp }}}\cdot l_{a_{\eta
c}}=\mp \,\overline{q}_{\eta c}\cdot dl\,\,\,\,\,\text{,\thinspace
\thinspace \thinspace \thinspace \thinspace }\overline{a}_{\eta _{c}}=\text{%
\thinspace }\frac{dl}{l_{\xi _{\perp }}}\cdot a_{\eta _{c}}=\overline{q}%
_{\eta c}\cdot dl\cdot m_{\perp }\,\,\text{.}  \label{7.74}
\end{eqnarray}

(f) Coriolis part $\overline{l}_{(\nabla _ua)_{\perp c}}$ of the change $%
\overline{l}_{(\nabla _ua)_{\perp }}$ of the acceleration $\overline{l}_a$.
By means of the expressions

\begin{eqnarray}
(\nabla _ua)_{\perp c} &=&\overline{g}[h_{\xi _{\perp }}(\nabla _ua)_{\perp
}]=\mp \text{\thinspace }l_{(\nabla _ua)_{\perp c}}\cdot m_{\perp
}\,\,\,\,\,\,\,\,\text{,}  \label{7.75} \\
g(\xi _{\perp },(\nabla _ua)_{\perp c}) &=&0\,\,\,\,\,\text{,\thinspace
\thinspace \thinspace \thinspace \thinspace \thinspace \thinspace \thinspace
\thinspace \thinspace }  \label{7.76} \\
\overline{g}[h_{\xi _{\perp }}(\nabla _ua)_{\perp }] &=&\overline{g}%
[h_{n_{\perp }}(\nabla _ua)_{\perp }]=\mp \text{\thinspace }l_{(\nabla
_ua)_{\perp c}}\cdot m_{\perp }\,\,\,\,\text{,}  \label{7.77}
\end{eqnarray}
\begin{eqnarray}
g((\nabla _ua)_{\perp c},m_{\perp }) &=&\text{\thinspace }l_{(\nabla
_ua)_{\perp c}}=g(\overline{g}[h_{n_{\perp }}(\nabla _ua)_{\perp }],m_{\perp
})=  \notag \\
&=&g(\overline{g}[h_{n_{\perp }}\overline{g}[h_u(\nabla _ua)]],m_{\perp })= 
\notag \\
&=&g(\overline{g}[h_{n_{\perp }}\overline{g}[h_{n_{\parallel }}(\nabla
_ua)]],m_{\perp })=  \notag \\
&=&g_{\overline{i}\overline{j}}\cdot g^{ik}\cdot (h_{n_{\perp }})_{\overline{%
k}\overline{l}}\cdot g^{lm}\cdot (h_{n_{\parallel }})_{\overline{m}\overline{%
n}}\cdot a^n\,_{;r}\cdot u^r\cdot m_{\perp }^j=  \notag \\
&=&(h_{n_{\perp }})_{\overline{j}\overline{l}}\cdot m_{\perp }^j\cdot
g^{lm}\cdot (h_{n_{\parallel }})_{\overline{m}\overline{n}}\cdot
a^n\,_{;r}\cdot u^r=  \notag \\
&=&g_{\overline{j}\overline{l}}\cdot g^{lm}\cdot m_{\perp }^j\cdot
(h_{n_{\parallel }})_{\overline{m}\overline{n}}\cdot a^n\,_{;r}\cdot u^r= 
\notag \\
&=&(h_{n_{\parallel }})_{\overline{j}\overline{n}}\cdot m_{\perp }^j\cdot
a^n\,_{;r}\cdot u^r=  \notag \\
&=&h_{n_{\parallel }}(m_{\perp },\nabla _ua)=h_u\,(m_{\perp },\nabla
_ua)=g(m_{\perp },\nabla _ua)\,\,\,\text{,}  \label{7.78}
\end{eqnarray}
\begin{eqnarray}
\text{\thinspace }l_{(\nabla _ua)_{\perp c}} &=&h_{n_{\perp }}(m_{\perp
},\nabla _ua)=g(m_{\perp },\nabla _ua)\,\,\,\,\,\,\,\text{,\thinspace
\thinspace \thinspace \thinspace \thinspace \thinspace \thinspace \thinspace 
}  \label{7.79} \\
\text{\thinspace }(\nabla _ua)_{\perp c} &=&\text{\thinspace }\mp \text{%
\thinspace }l_{(\nabla _ua)_{\perp c}}\cdot m_{\perp }=\mp h_{n_{\perp
}}(m_{\perp },\nabla _ua)\cdot m_{\perp }=  \notag \\
&=&\mp g(m_{\perp },\nabla _ua)\,\,\,\,\text{,}\,  \label{7.80}
\end{eqnarray}
\begin{equation}
\overline{\text{\thinspace }l}_{(\nabla _ua)_{\perp c}}=\frac{dl}{l_{\xi
_{\perp }}}\cdot l\text{\thinspace }_{(\nabla _ua)_{\perp c}}\,\,\,\,\,\text{%
,\thinspace \thinspace \thinspace \thinspace }(\nabla _u\overline{a})_{\perp
}=\frac{dl}{l_{\xi _{\perp }}}\cdot (\nabla _ua)_{\perp c}\,\,\,\text{%
.\thinspace \thinspace }  \label{7.81}
\end{equation}

\subsection{Standard (longitudinal) Hubble effect (Hubble shift)}

The standard (longitudinal) Hubble effect (Hubble shift) corresponds to the
standard (longitudinal) Doppler effect (Doppler shift). Only the different
types of velocities and accelerations generating the standard Doppler effect
are given in their explicit form by means of the corresponding Hubble
functions and acceleration parameters.

If the vector field $k_{\perp }$ is collinear to the vector field $\xi
_{\perp }$ determining the proper frame of reference, i.e. if 
\begin{equation}
k_{\perp }=\mp l_{k_{\perp }}\cdot n_{\perp }\,\,\,\,\text{,\thinspace
\thinspace \thinspace \thinspace \thinspace \thinspace \thinspace \thinspace
\thinspace \thinspace \thinspace \thinspace \thinspace }cos\,\theta =\pm
1\,\,\,\,\text{, \thinspace \thinspace \thinspace \thinspace \thinspace
\thinspace \thinspace }sin\,\theta =0\,\,\,\,\,\text{,}  \label{7.82}
\end{equation}
the frequency of the emitter $\overline{\omega }$ and the frequency $\omega $
detected by the observer are related to each other by the expression 
\begin{equation}
\overline{\omega }=\omega \cdot [1-\frac{\overline{l}_{v_z}}{l_u}-\frac{%
l_{\xi _{\perp }}}{l_u^2}\cdot \overline{l}_{(a_{\perp })_z}+\frac 12\cdot 
\frac{dl}{l_u^2}\cdot (\overline{l}_{a_z}+\frac{l_{\xi _{\perp }}}{l_u}\cdot 
\overline{l}_{(\nabla _ua)_{\perp z}})]\,\,\,\,\,\text{.}  \label{7.83}
\end{equation}

If we now replace the velocity $\overline{l}_{v_z}$ and the accelerations $%
\overline{l}_{(a_{\perp })_z}$, $\overline{l}_{a_z}$, and $\overline{l}%
_{(\nabla _ua)_{\perp z}}$ with their corresponding explicit forms 
\begin{eqnarray}
\overline{l}_{v_z} &=&\mp \,H\cdot dl\text{ \thinspace \thinspace \thinspace
\thinspace \thinspace \thinspace \thinspace \thinspace ,\thinspace
\thinspace \thinspace \thinspace \thinspace }  \label{7.84} \\
\text{\thinspace \thinspace \thinspace \thinspace \thinspace }\overline{l}%
_{a_z} &=&\mp \,\overline{q}\cdot dl\,\,\,\,\,\,\,\,\,\,\,\text{,\thinspace
\thinspace \thinspace \thinspace \thinspace \thinspace \thinspace \thinspace
\thinspace }  \label{7.85} \\
\text{\thinspace }\overline{l}_{(a_{\perp })_z} &=&\frac{dl}{l_{\xi _{\perp
}}}\cdot g(a_{\perp },n_{\perp })\text{\thinspace \thinspace \thinspace
\thinspace \thinspace \thinspace \thinspace \thinspace \thinspace \thinspace
\thinspace \thinspace ,\thinspace }  \label{7.86} \\
\text{\thinspace \thinspace \thinspace \thinspace \thinspace \thinspace }%
\overline{l}_{(\nabla _ua)_{\perp }{}_z} &=&\frac{dl}{l_{\xi _{\perp }}}%
\cdot g(\nabla _ua_{\perp },n_{\perp })\text{\thinspace \thinspace
\thinspace \thinspace \thinspace \thinspace \thinspace \thinspace ,}
\label{7.87}
\end{eqnarray}
then we obtain the relation between $\overline{\omega }$ and $\omega $ in
the form 
\begin{eqnarray}
\overline{\omega } &=&\omega \cdot [1-\frac{\overline{l}_{v_z}}{l_u}-\frac{%
l_{\xi _{\perp }}}{l_u^2}\cdot \overline{l}_{(a_{\perp })_z}+\frac 12\cdot 
\frac{dl}{l_u^2}\cdot (\overline{l}_{a_z}+\frac{l_{\xi _{\perp }}}{l_u}\cdot 
\overline{l}_{(\nabla _ua)_{\perp z}})]=  \notag \\
&=&\omega \cdot \{1-\frac 1{l_u}\cdot [\overline{l}_{v_z}+\frac{l_{\xi
_{\perp }}}{l_u}\cdot \overline{l}_{(a_{\perp })_z}-\frac 12\cdot \frac{dl}{%
l_u}\cdot (\overline{l}_{a_z}+\frac{l_{\xi _{\perp }}}{l_u}\cdot \overline{l}%
_{(\nabla _ua)_{\perp z}})]\}=  \notag \\
&=&\omega \cdot \{1-\frac 1{l_u}\cdot [\mp \,H\cdot dl+\frac{l_{\xi _{\perp
}}}{l_u}\cdot \frac{dl}{l_{\xi _{\perp }}}\cdot g(a_{\perp },n_{\perp })- 
\notag \\
&&-\frac 12\cdot \frac{dl}{l_u}\cdot (\text{\thinspace }\mp \overline{q}%
\cdot dl\,+\frac{l_{\xi _{\perp }}}{l_u}\cdot \frac{dl}{l_{\xi _{\perp }}}%
\cdot g(\nabla _ua_{\perp },n_{\perp }))]\}\,\,\,\,\text{,}  \label{7.88}
\end{eqnarray}
\begin{eqnarray}
\overline{\omega } &=&\omega \cdot \{1-\frac 1{l_u}\cdot [\mp \,H\cdot dl+%
\frac{dl}{l_u}\cdot g(a_{\perp },n_{\perp })-  \notag \\
&&-\frac 12\cdot \frac{dl}{l_u}\cdot (\text{\thinspace }\mp \overline{q}%
\cdot dl\,+\frac{dl}{l_u}\cdot g(\nabla _ua_{\perp },n_{\perp }))]\}\,\,\,\,%
\text{,}\,  \label{7.89}
\end{eqnarray}
\begin{eqnarray}
\overline{\omega } &=&\omega \cdot \{1+\frac 1{l_u}\cdot [\pm \,H\cdot dl-%
\frac{dl}{l_u}\cdot g(a_{\perp },n_{\perp })-  \notag \\
&&-\frac 12\cdot \frac{dl}{l_u}\cdot (\text{\thinspace }\pm \overline{q}%
\cdot dl\,-\frac{dl}{l_u}\cdot g(\nabla _ua_{\perp },n_{\perp }))]\}\,
\label{7.90}
\end{eqnarray}

\begin{equation}
\overline{\omega }=\omega \cdot \{1\pm \frac 1{l_u}\cdot [H\cdot dl\mp \frac{%
dl}{l_u}\cdot g(a_{\perp },n_{\perp })-\frac 12\cdot \frac{dl}{l_u}\cdot (%
\overline{q}\cdot dl\mp \frac{dl}{l_u}\cdot g(\nabla _ua_{\perp },n_{\perp
}))]\}\,\,\,\text{.}  \label{7.91}
\end{equation}

\textit{Special case}: Auto-parallel motion of the observer: $\nabla _uu=a=0$%
: $\overline{l}_{(a_{\perp })_z}=0$,\thinspace \thinspace \thinspace
\thinspace $\overline{l}_{(\nabla _ua)_{\perp z}}=0\,\,$, 
\begin{equation*}
\overline{\omega }=\omega \cdot [1-\frac{\overline{l}_{v_z}}{l_u}+\frac
12\cdot \frac{dl}{l_u^2}\cdot \overline{l}_{a_z}]\,\,\,\,\,\text{,}
\end{equation*}

\begin{eqnarray}
\overline{\omega } &=&\omega \cdot [1\pm \frac 1{l_u}\cdot (H\cdot dl-\frac
12\cdot \frac{dl}{l_u}\cdot \overline{q}\cdot dl)]=  \notag \\
&=&\omega \cdot [1\pm \frac 1{l_u}\cdot (H-\frac 12\cdot \frac{dl}{l_u}\cdot 
\overline{q})\cdot dl]\,\,\,\,\,\text{.}  \label{7.92}
\end{eqnarray}

If the world line of an observer is an auto-parallel trajectory and $%
k_{\perp }$ is collinear to $\xi _{\perp }$ then the change of the frequency
of the emitter $\overline{\omega }$ depends on the Hubble function $H$ and
the acceleration parameter $\overline{q}$.

\textit{Special case}: $\nabla _uu=a=0$,\thinspace \thinspace \thinspace
\thinspace $k_{\perp }=\mp l_{k_{\perp }}\cdot n_{\perp }$,\thinspace $%
\overline{l}_{a_z}=0$: 
\begin{equation*}
\overline{\omega }=\omega \cdot (1-\frac{\overline{l}_{v_z}}{l_u}%
)\,\,\,\,\,\,\text{,\thinspace \thinspace \thinspace \thinspace \thinspace
\thinspace \thinspace \thinspace \thinspace \thinspace \thinspace \thinspace
\thinspace \thinspace }\overline{l}_{v_z}\lesseqqgtr 0\,\,\,\,\,\text{%
.\thinspace \thinspace \thinspace \thinspace }
\end{equation*}
\begin{equation}
\overline{\omega }=\omega \cdot (1\pm \frac 1{l_u}\cdot H\cdot dl)\,\,\text{.%
}  \label{7.93}
\end{equation}

Therefore, if the world line of an observer is an auto-parallel trajectory, $%
k_{\perp }$ is collinear to $\xi _{\perp }$, and no centrifugal
(centripetal) acceleration $\overline{l}_{a_z}$ exists between emitter and
observer then the above expression has the well known form for description
of the standard Hubble effect in relativistic astrophysics . Here, this
relation is valid in every $(\overline{L}_n,g)$-space considered as a model
of a space or a space-time under the given preconditions.

\subsubsection{Standard (longitudinal) Hubble shift frequency parameter $z$}

The relative difference between both the frequencies (emitted $\overline{%
\omega }\,$\thinspace and detected $\omega $) 
\begin{equation}
\frac{\overline{\omega }-\omega }\omega :=z=\pm \overline{C}\,\,\,\,\text{,}
\label{7.94}
\end{equation}
under the condition $k_{\perp }=\mp l_{k_{\perp }}\cdot n_{\perp }$ appears
in the form

\begin{eqnarray}
\frac{\overline{\omega }-\omega }\omega &:&=z=\pm \overline{C}=\pm \frac
1{l_u}\cdot [H\cdot dl\mp \frac{dl}{l_u}\cdot g(a_{\perp },n_{\perp })- 
\notag \\
&&-\frac 12\cdot \frac{dl}{l_u}\cdot (\overline{q}\cdot dl\mp \frac{dl}{l_u}%
\cdot g(\nabla _ua_{\perp },n_{\perp }))]\,\,\,\text{,}  \label{7.95}
\end{eqnarray}
where 
\begin{equation}
\overline{\omega }=(1+z)\cdot \omega \,\,\,\,\,\text{,}  \label{7.96}
\end{equation}
\begin{equation}
z=\pm \frac 1{l_u}\cdot [H\cdot dl\mp \frac{dl}{l_u}\cdot g(a_{\perp
},n_{\perp })-\frac 12\cdot \frac{dl}{l_u}\cdot (\overline{q}\cdot dl\mp 
\frac{dl}{l_u}\cdot g(\nabla _ua_{\perp },n_{\perp }))]\,\,\,\text{.}
\label{7.97}
\end{equation}

The quantity $z$ could be denoted as \textit{observed standard
(longitudinal) Hubble shift frequency parameter}. If $z=0$ then there will
be no difference between the emitted and the detected frequencies, i. e. $%
\overline{\omega }=\omega $. This will be the case when the emitter and
observer (detector) are at rest to each other, i.e. when no relative
velocities and relative accelerations occur, when centrifugal (centripetal)
relative velocities and accelerations do not exist, or when the centripetal
(centrifugal) velocities and accelerations compensate each other under the
condition 
\begin{equation}
H\mp \frac 1{l_u}\cdot g(a_{\perp },n_{\perp })-\frac 12\cdot \frac
1{l_u}\cdot (\overline{q}\mp \frac 1{l_u}\cdot g(\nabla _ua_{\perp
},n_{\perp }))\cdot dl=0\,\,\,\,\text{.}  \label{7.98}
\end{equation}

If $z>0$ the observed Hubble shift frequency parameter is called \textit{%
longitudinal} \textit{Hubble red shift}. If $z<0$ the observed Hubble shift
frequency parameter is called \textit{longitudinal} \textit{Hubble blue shift%
}. If $\overline{\omega }$ and $\omega $ are known the observed Hubble shift
frequency parameter $z$ could be found. If $\omega $ and $z$ are given then
the corresponding $\overline{\omega }$ could be estimated.

On the other side, from the explicit form of $z$%
\begin{equation}
z=\pm \frac 1{l_u}\cdot [H\cdot dl\mp \frac{dl}{l_u}\cdot g(a_{\perp
},n_{\perp })-\frac 12\cdot \frac{dl}{l_u}\cdot (\overline{q}\cdot dl\mp 
\frac{dl}{l_u}\cdot g(\nabla _ua_{\perp },n_{\perp }))]  \label{7.99}
\end{equation}
if we consider the explicit form of the Hubble function $H$ and of the
acceleration parameter $\overline{q}$ we could find the relation between the
observed shift frequency parameter $z$ and the kinematic characteristics of
the relative velocity such as expansion and shear velocities and
accelerations.

\textit{Special case}: Auto-parallel motion of the observer: $\nabla _uu=a=0$%
: $\overline{l}_{(a_{\perp })_z}=0$,\thinspace \thinspace \thinspace
\thinspace $\overline{l}_{(\nabla _ua)_{\perp z}}=0\,\,$, 
\begin{equation*}
\overline{\omega }=\omega \cdot [1-\frac{\overline{l}_{v_z}}{l_u}+\frac
12\cdot \frac{dl}{l_u^2}\cdot \overline{l}_{a_z}]\,\,\,\,\,\text{,}
\end{equation*}

\begin{equation}
z=\pm \frac 1{l_u}\cdot (H\cdot dl-\frac 12\cdot \frac{dl}{l_u}\cdot 
\overline{q}\cdot dl)=\pm \frac 1{l_u}\,\cdot (H-\frac 12\cdot \frac{dl}{l_u}%
\cdot \overline{q})\cdot dl\,\,\text{.}  \label{7.100}
\end{equation}

If the world line of an observer is an auto-parallel trajectory and $%
k_{\perp }$ is collinear to $\xi _{\perp }$ then the observed Hubble shift
frequency parameter $z$ depends on the Hubble function $H$ and the
acceleration parameter $\overline{q}$ as well as on the absolute value $l_u$
of the velocity of the signal and on the space distance $dl$ propagated by
the signal.

\textit{Special case}: $\nabla _uu=a=0$,\thinspace \thinspace \thinspace
\thinspace $k_{\perp }=\mp l_{k_{\perp }}\cdot n_{\perp }$,\thinspace $%
\overline{l}_{a_z}=0$: 
\begin{equation*}
\overline{\omega }=\omega \cdot (1-\frac{\overline{l}_{v_z}}{l_u}%
)\,\,\,\,\,\,\text{,\thinspace \thinspace \thinspace \thinspace \thinspace
\thinspace \thinspace \thinspace \thinspace \thinspace \thinspace \thinspace
\thinspace \thinspace }\overline{l}_{v_z}\lesseqqgtr 0\,\,\,\,\,\text{%
.\thinspace \thinspace \thinspace \thinspace }
\end{equation*}
\begin{equation}
z=\pm \frac 1{l_u}\cdot H\cdot dl\,\,\text{.}  \label{7.101}
\end{equation}

\textit{Remark}. In relativistic physics ($l_u=c$, $1$) the last (above)
relation is also called Hubble law.

If we express in this special case the observed longitudinal Hubble shift
frequency parameter $z$ in its infinitesimal form 
\begin{equation}
z=\frac{\overline{\omega }-\omega }\omega =\frac{d\omega }\omega =\pm \frac
1{l_u}\cdot H\cdot dl\,\,\,\,\,\,  \label{7.102}
\end{equation}
then we can find the change of the frequency $\omega $ for a global distance 
$l$ propagated by a signal for finite proper time interval of an observer 
\begin{equation*}
\int \frac{d\omega }\omega =\pm \int \frac 1{l_u}\cdot H\cdot dl\,\,\,\,\,%
\text{, }
\end{equation*}
\begin{eqnarray}
log\omega &=&\pm \int \frac 1{l_u}\cdot H\cdot dl\,\,+const.\,\,\text{,} 
\notag \\
\omega &=&\omega _0\cdot exp(\pm \int \frac 1{l_u}\cdot H\cdot
dl\,)\,\,\,\,\,\text{,\thinspace \thinspace \thinspace \thinspace \thinspace
\thinspace \thinspace \thinspace \thinspace }\omega _0=\,\text{const.}
\label{7.103}
\end{eqnarray}

In the relativistic astrophysics, it is assumed that $l_u=\,$const. $=c$ or $%
1$. The Hubble function is also assumed to be a constant function $H=H_0=$
const.

Then 
\begin{equation}
\overline{\omega }=\omega _0\cdot exp(\pm \frac 1{l_u}\cdot H_0\cdot
l\,)=\omega _0\cdot exp(\pm H_0\cdot \frac l{l_u}\,)=\omega _0\cdot exp(\pm
H_0\cdot \tau \,)\,\,\text{,}  \label{7.104}
\end{equation}
where $l_u$ is the absolute value of the velocity of a signal, $\tau $ is
the proper time interval of the observer for which a signal propagates from
the emitter to the observer, and $l$ is the space distance covered by the
signal from the emitter to the observer.

Therefore, if the world line of an observer is an auto-parallel trajectory, $%
k_{\perp }$ is collinear to $\xi _{\perp }$, and no centrifugal
(centripetal) acceleration $\overline{l}_{a_z}$ exists between an emitter
and an observer then the above expression has the well known form for
description of the standard Hubble effect in relativistic astrophysics .
Here, the relation is valid in every $(\overline{L}_n,g)$-space considered
as a model of a space or a space-time under the given preconditions.

\subsection{Transversal Hubble effect (Hubble shift)}

The transversal Hubble effect (Hubble shift) corresponds to the transversal
Doppler effect (Doppler shift). Only the different types of velocities and
accelerations generating the transversal Doppler effect are given in their
explicit form by means of the corresponding Hubble functions and
acceleration parameters.

If the vector field $k_{\perp }$ is collinear to the vector field $\xi
_{\perp }$ determining the proper frame of reference, i.e. if 
\begin{equation}
k_{\perp }=\mp l_{k_{\perp }}\cdot n_{\perp }\,\,\,\,\text{,\thinspace
\thinspace \thinspace \thinspace \thinspace \thinspace \thinspace \thinspace
\thinspace \thinspace \thinspace \thinspace \thinspace }cos\,\theta =\pm
1\,\,\,\,\text{, \thinspace \thinspace \thinspace \thinspace \thinspace
\thinspace \thinspace }sin\,\theta =0\,\,\,\,\,\text{,}  \label{8.1}
\end{equation}%
the frequency of the emitter $\overline{\omega }$ and the frequency $\omega $
detected by the observer are related to each other by the expression 
\begin{eqnarray}
\overline{\omega } &=&\omega \cdot \lbrack 1-\frac{1}{l_{u}}\cdot \overline{l%
}_{v_{\eta c}}-\frac{l_{\xi _{\perp }}}{l_{u}^{2}}\cdot \overline{l}%
_{(a_{\perp })_{c}}+\frac{1}{2}\cdot \frac{dl}{l_{u}^{2}}\cdot (\,\overline{l%
}_{a_{\eta c}}+\frac{l_{\xi _{\perp }}}{l_{u}}\cdot l_{(\nabla _{u}a)_{\perp
c}})]=  \notag \\
&=&\omega \cdot \{1-\frac{1}{l_{u}}\cdot \lbrack \overline{l}_{v_{\eta c}}+%
\frac{l_{\xi _{\perp }}}{l_{u}}\cdot \overline{l}_{(a_{\perp })_{c}}-  \notag
\\
&&-\frac{1}{2}\cdot \frac{dl}{l_{u}}\cdot (\,\overline{l}_{a_{\eta c}}+\frac{%
l_{\xi _{\perp }}}{l_{u}}\cdot l_{(\nabla _{u}a)_{\perp c}})]\}\,\,\,\text{.}
\label{8.2}
\end{eqnarray}

If we now replace the velocity $\overline{l}_{v_{\eta c}}$ and the
accelerations $\overline{l}_{(a_{\perp })_c}$, $\overline{l}_{a_{\eta c}}$,
and $\overline{l}_{(\nabla _ua)_{\perp c}}$ with their corresponding
explicit forms 
\begin{eqnarray}
\overline{l}_{v_{\eta c}} &=&\mp \overline{H}_c\cdot dl\,\,\,\,\,\,\,\text{,}
\label{8.3} \\
\text{\thinspace }\overline{l}_{(a_{\perp })_c} &=&\frac{dl}{l_{\xi _{\perp
}}}\cdot \text{\thinspace }l_{(a_{\perp })_c}=\frac{dl}{l_{\xi _{\perp }}}%
\cdot g(m_{\perp },a_{\perp })\,\,\,\,\,\,\text{,}  \label{8.4} \\
\overline{l}_{a_{\eta c}} &=&\mp \overline{q}_{\eta c}\cdot
dl\,\,\,\,\,\,\,\,\,\,\text{,}  \label{8.5} \\
\overline{\text{\thinspace }l}_{(\nabla _ua)_{\perp c}} &=&\frac{dl}{l_{\xi
_{\perp }}}\cdot \text{\thinspace }l_{(\nabla _ua)_{\perp c}}=\frac{dl}{%
l_{\xi _{\perp }}}\cdot g(m_{\perp },\nabla _ua)\,\,\,\,\,\text{,}
\label{8.6}
\end{eqnarray}
then we obtain the relation between $\overline{\omega }$ and $\omega $ in
the form 
\begin{eqnarray}
\overline{\omega } &=&\omega \cdot \{1-\frac 1{l_u}\cdot [\overline{l}%
_{v_{\eta c}}+\frac{l_{\xi _{\perp }}}{l_u}\cdot \overline{l}_{(a_{\perp
})_c}-\frac 12\cdot \frac{dl}{l_u}\cdot (\,\overline{l}_{a_{\eta c}}+\frac{%
l_{\xi _{\perp }}}{l_u}\cdot l_{(\nabla _ua)_{\perp c}})]\}=  \notag \\
&=&\omega \cdot \{1-\frac 1{l_u}\cdot [\mp \overline{H}_c\cdot dl+\frac{%
l_{\xi _{\perp }}}{l_u}\cdot \frac{dl}{l_{\xi _{\perp }}}\cdot g(m_{\perp
},a_{\perp })]\,-  \notag \\
&&-\frac 12\cdot \frac{dl}{l_u}\cdot [\mp \overline{q}_{\eta c}\cdot dl+%
\frac{l_{\xi _{\perp }}}{l_u}\cdot \frac{dl}{l_{\xi _{\perp }}}\cdot
g(m_{\perp },\nabla _ua)]\}\,\,\,\,\text{,}  \label{8.7}
\end{eqnarray}
\begin{equation}
\overline{\omega }=\omega \cdot \{1\pm \frac 1{l_u}\cdot [\overline{H}%
_c\cdot dl\mp \frac{dl}{l_u}\cdot g(m_{\perp },a_{\perp })-\frac 12\cdot 
\frac{dl}{l_u}\cdot (\overline{q}_{\eta c}\cdot dl\mp \frac{dl}{l_u}\cdot
g(m_{\perp },\nabla _ua))]\}\,\,\,\,\,\text{,}  \label{8.8}
\end{equation}
compared with the case of the standard (longitudinal) Hubble effect

\begin{equation*}
\overline{\omega }=\omega \cdot \{1\pm \frac 1{l_u}\cdot [H\cdot dl\mp \frac{%
dl}{l_u}\cdot g(a_{\perp },n_{\perp })-\frac 12\cdot \frac{dl}{l_u}\cdot (%
\overline{q}\cdot dl\mp \frac{dl}{l_u}\cdot g(\nabla _ua_{\perp },n_{\perp
}))]\}\,\,\,\text{.}
\end{equation*}

\textit{Special case}: Auto-parallel motion of the observer: $\nabla _uu=a=0$%
: $\overline{l}_{(a_{\perp })_c}=0$,\thinspace \thinspace \thinspace
\thinspace $\overline{l}_{(\nabla _ua)_{\perp c}}=0\,\,$, 
\begin{equation*}
\overline{\omega }=\omega \cdot [1-\frac 1{l_u}\cdot \overline{l}_{v_{\eta
c}}+\frac 12\cdot \frac{dl}{l_u^2}\cdot \,\overline{l}_{a_{\eta c}}]\,\,\,\,%
\text{,}
\end{equation*}
\begin{eqnarray}
\overline{\omega } &=&\omega \cdot [1\pm \frac 1{l_u}\cdot (\overline{H}%
_c\cdot dl-\frac 12\cdot \frac{dl}{l_u}\cdot \overline{q}_{\eta c}\cdot dl)]=
\notag \\
&=&\omega \cdot [1\pm \frac 1{l_u}\cdot (\overline{H}_c-\frac 12\cdot \frac{%
dl}{l_u}\cdot \overline{q}_{\eta c})\cdot dl]\,\,\,\,\text{.}  \label{8.9}
\end{eqnarray}

If the world line of an observer is an auto-parallel trajectory and $%
k_{\perp }$ is orthogonal to $\xi _{\perp }$ then the change of the
frequency of the emitter $\overline{\omega }$ depends on the Coriolis
velocity $\overline{l}_{v_{\eta c}}$ and the Coriolis acceleration $%
\overline{l}_{a_{\eta c}}$.

\textit{Special case}: $\nabla _uu=a=0$,\thinspace \thinspace \thinspace
\thinspace $k_{\perp }=\mp l_{k_{\perp }}\cdot m_{\perp }$,\thinspace $%
\overline{l}_{a_{\eta c}}=0$: 
\begin{eqnarray}
\overline{\omega } &=&\omega \cdot (1-\frac{\overline{l}_{v_{\eta c}}}{l_u}%
)\,\,\,\,\,\,\text{,\thinspace \thinspace \thinspace \thinspace \thinspace
\thinspace \thinspace \thinspace \thinspace \thinspace \thinspace \thinspace
\thinspace \thinspace }\overline{l}_{v_{\eta z}}\lesseqqgtr 0\,\,\,\,\,\text{%
,}  \notag \\
\overline{\omega } &=&\omega \cdot (1\pm \frac 1{l_u}\cdot \overline{H}%
_c\cdot dl)\,\,\,\,\,\,\,\,\text{.}\,\text{\thinspace \thinspace }
\label{8.10}
\end{eqnarray}

Therefore, if the world line of an observer is an auto-parallel trajectory, $%
k_{\perp }$ is orthogonal to $\xi _{\perp }$, and no Coriolis acceleration $%
\overline{l}_{a_{\eta c}}$ exists between emitter and observer then the
above expression has analogous form for description of the transversal
Hubble effect as the standard (longitudinal) Hubble effect in relativistic
astrophysics. Here, this relation for the transversal Hubble effect is valid
in every $(\overline{L}_n,g)$-space considered as a model of a space or a
space-time under the given preconditions.

\subsubsection{Transversal Hubble shift frequency parameter $z_c$}

The relative difference between both the frequencies (emitted $\overline{%
\omega }\,$\thinspace and detected $\omega $) when a transversal Doppler
effect and a transversal Hubble effect correspondingly occur could be
written as 
\begin{equation}
\frac{\overline{\omega }-\omega }\omega :=z_c=\pm \overline{S}  \label{8.11}
\end{equation}
and under the condition $k_{\perp }=\mp l_{k_{\perp }}\cdot n_{\perp }$
appears in the form

\begin{eqnarray}
\frac{\overline{\omega }-\omega }\omega &:&=z_c=\pm \frac 1{l_u}\cdot [%
\overline{H}_c\cdot dl\mp \frac{dl}{l_u}\cdot g(m_{\perp },a_{\perp })- 
\notag \\
&&-\frac 12\cdot \frac{dl}{l_u}\cdot \overline{q}_{\eta c}\cdot dl\mp \frac{%
dl}{l_u}\cdot g(m_{\perp },\nabla _ua)]  \label{8.12}
\end{eqnarray}
where 
\begin{equation}
\overline{\omega }=(1+z_c)\cdot \omega \,\,\,\,\,\text{,}  \label{8.13}
\end{equation}
\begin{eqnarray}
z_c &=&\pm \frac 1{l_u}\cdot [\overline{H}_c\cdot dl\mp \frac{dl}{l_u}\cdot
g(m_{\perp },a_{\perp })-  \notag \\
&&-\frac 12\cdot \frac{dl}{l_u}\cdot (\overline{q}_{\eta c}\cdot dl\mp \frac{%
dl}{l_u}\cdot g(m_{\perp },\nabla _ua))]\,\,\,\text{.}  \label{8.14}
\end{eqnarray}

The quantity $z_c$ could be denoted as \textit{observed transversal Hubble
shift frequency parameter}. If $z_c=0$ then there will be no difference
between the emitted and the detected frequencies, i. e. $\overline{\omega }%
=\omega $. This will be the case when the emitter and observer (detector)
are at rest to each other, i.e. when no relative velocities and relative
accelerations occur, when Coriolis relative velocities and Coriolis relative
accelerations do not exist, or when the centripetal (centrifugal) velocities
and accelerations compensate each other under the condition 
\begin{equation}
\overline{H}_c\mp \frac 1{l_u}\cdot g(m_{\perp },a_{\perp })-\frac 12\cdot
\frac 1{l_u}\cdot (\overline{q}_{\eta c}\mp \frac 1{l_u}\cdot g(m_{\perp
},\nabla _ua))\cdot dl=0\,\,\,\text{.}  \label{8.15}
\end{equation}

If $z_c>0$ the observed transversal Hubble shift frequency parameter is
called \textit{transversal} \textit{Hubble red shift}. If $z_c<0$ the
observed transversal Hubble shift frequency parameter is called \textit{%
transversal} \textit{Hubble's blue shift}. If $\overline{\omega }$ and $%
\omega $ are known the observed transversal Hubble shift frequency parameter 
$z_c$ could be found. If $\omega $ and $z$ are given then the corresponding $%
\overline{\omega }$ could be estimated.

On the other side, from the explicit form of $z_c$%
\begin{eqnarray}
z_c &=&\pm \frac 1{l_u}\cdot [\overline{H}_c\cdot dl\mp \frac{dl}{l_u}\cdot
g(m_{\perp },a_{\perp })-  \notag \\
&&-\frac 12\cdot \frac{dl}{l_u}\cdot (\overline{q}_{\eta c}\cdot dl\mp \frac{%
dl}{l_u}\cdot g(m_{\perp },\nabla _ua))]\,  \label{8.16}
\end{eqnarray}
if we consider the explicit form of the Hubble function $\overline{H}_c$ and
of the acceleration parameter $\overline{q}_{\eta c}$ we could find the
relation between the observed shift frequency parameter $z_c$ and the
kinematic characteristics of the relative velocity such as expansion and
shear velocities and accelerations.

\textit{Special case}: Auto-parallel motion of the observer: $\nabla _uu=a=0$%
: $\overline{l}_{(a_{\perp })_z}=0$,\thinspace \thinspace \thinspace
\thinspace $\overline{l}_{(\nabla _ua)_{\perp z}}=0\,\,$, 
\begin{equation*}
\overline{\omega }=\omega \cdot [1-\frac 1{l_u}\cdot \overline{l}_{v_{\eta
c}}+\frac 12\cdot \frac{dl}{l_u^2}\cdot \,\overline{l}_{a_{\eta c}}]\,\,\,%
\text{,}
\end{equation*}

\begin{equation}
z_c=\pm \frac 1{l_u}\cdot (\overline{H}_c-\frac 12\cdot \frac{dl}{l_u}\cdot 
\overline{q}_{\eta c})\cdot dl\,\,\,\,\,\text{.}  \label{8.17}
\end{equation}

If the world line of an observer is an auto-parallel trajectory and $%
k_{\perp }$ is collinear to $\xi _{\perp }$ then the observed transversal
Hubble shift frequency parameter $z_c$ depends on the Hubble function $%
\overline{H}_c$ and the acceleration parameter $\overline{q}_{\eta c}$ as
well as on the absolute value $l_u$ of the velocity of the signal and on the
space distance $dl$ propagated by the signal.

\textit{Special case}: $\nabla _uu=a=0$,\thinspace \thinspace \thinspace
\thinspace $k_{\perp }=\mp l_{k_{\perp }}\cdot n_{\perp }$,\thinspace $%
\overline{l}_{a_z}=0$: 
\begin{equation*}
\overline{\omega }=\omega \cdot (1-\frac{\overline{l}_{v_{\eta c}}}{l_u}%
)\,\,\,\,\,\,\text{,\thinspace \thinspace \thinspace \thinspace \thinspace
\thinspace \thinspace \thinspace \thinspace \thinspace \thinspace \thinspace
\thinspace \thinspace }\overline{l}_{v_{\eta z}}\lesseqqgtr 0\,\,\,\,\,\text{%
,\thinspace \thinspace \thinspace }
\end{equation*}
\begin{equation}
z_c=\pm \frac 1{l_u}\cdot \overline{H}_c\cdot dl\,\,\,\,\,\text{.}
\label{8.19}
\end{equation}

If we express in this special case the observed transversal Hubble shift
frequency parameter $z_c$ in its infinitesimal form 
\begin{equation}
z_c=\frac{\overline{\omega }-\omega }\omega =\frac{d\omega }\omega =\pm
\frac 1{l_u}\cdot \overline{H}_c\cdot dl\,\,\,\,\,\,  \label{8.20}
\end{equation}
then we can find the change of the frequency $\omega $ for a global distance 
$l$ propagated by a signal for finite proper time interval of an observer 
\begin{equation*}
\int \frac{d\omega }\omega =\pm \int \frac 1{l_u}\cdot \overline{H}_c\cdot
dl\,\,\,\,\,\text{, }
\end{equation*}
\begin{eqnarray}
log\,\omega &=&\pm \int \frac 1{l_u}\cdot \overline{H}_c\cdot
dl\,\,+const.\,\,\text{,}  \notag \\
\omega &=&\omega _0\cdot exp(\pm \int \frac 1{l_u}\cdot \overline{H}_c\cdot
dl\,)\,\,\,\,\,\text{,\thinspace \thinspace \thinspace \thinspace \thinspace
\thinspace \thinspace \thinspace \thinspace }\omega _0=\,\text{const.}
\label{8.21}
\end{eqnarray}

In the relativistic astrophysics, it is assumed that $l_u=\,$const. $=c$ or $%
1$. If the Hubble function is also assumed to be a constant function $%
\overline{H}_c=\overline{H}_{c0}=$ const.

Then 
\begin{equation}
\overline{\omega }=\omega _0\cdot exp(\pm \frac 1{l_u}\cdot \overline{H}%
_{c0}\cdot l\,)=\omega _0\cdot exp(\pm \overline{H}_{c0}\cdot \frac
l{l_u}\,)=\omega _0\cdot exp(\pm \overline{H}_{c0}\cdot \tau \,)\,\,\text{,}
\label{8.23}
\end{equation}
where $l_u$ is the absolute value of the velocity of a signal, $\tau $ is
the proper time interval of the observer for which a signal propagates from
the emitter to the observer, and $l$ is the space distance covered by the
signal from the emitter to the observer. If $l=\,$const. then the $\overline{%
\omega }$ will change by a constant quantity 
\begin{equation}
\overline{\omega }=K_0\cdot \omega \,\,\,\,\,\text{,\thinspace \thinspace
\thinspace \thinspace \thinspace \thinspace \thinspace \thinspace \thinspace
\thinspace \thinspace \thinspace }K_0=exp(\pm \overline{H}_{c0}\cdot \frac
l{l_u}\,)=exp(\pm \overline{H}_{c0}\cdot \tau \,)=\,\text{const.}
\label{8.24}
\end{equation}
and we will observe a constant shift of the emitted frequency with respect
to the observer (detector) during a time interval. If we further write $K_0$
in the form 
\begin{equation*}
K_0=1\pm \overline{K}_0\,\,\,\,\,\,\,\,\,\,\,\,\,\text{,\thinspace
\thinspace \thinspace \thinspace \thinspace \thinspace \thinspace }\overline{%
K}_0=\text{ const.,\thinspace \thinspace }
\end{equation*}
the change of the frequency $\overline{\omega }$ could be represented in the
forms

\begin{eqnarray}
\overline{\omega } &=&(1\pm \overline{K}_0)\cdot \omega \,\,\,\,\text{,}
\label{8.25} \\
z_c &=&\pm \overline{K}_0=\,\text{const.}  \label{8.26}
\end{eqnarray}

If the space distance between an emitter and an observer does not change but
there is a Coriolis velocity $\overline{l}_{v_{\eta c}}$ between them then
there will be a constant difference $z_c=\pm \overline{K}_0=\,$const.
between the emitted frequency $\overline{\omega }$ and the detected
frequency $\omega $. This could be the case when an emitter sends signals to
a detector and rotates along the detector at a constant space distance from
it. At the same time the detector moves at an auto-parallel world line.

Therefore, if the world line of an observer is an auto-parallel trajectory, $%
k_{\perp }$ is collinear to $\xi _{\perp }$, and no Coriolis acceleration $%
\overline{l}_{a_{\eta c}}$ exists between emitter and observer then the
above expression could be used for description of the transversal Hubble
effect in relativistic astrophysics . Here, the relation is valid in every $(%
\overline{L}_n,g)$-space considered as a model of a space or of a space-time
under the given preconditions.

\section{Conclusion}

In the present paper we have considered the notion of null (isotropic)
vector field in spaces with affine connections and metrics for describing
effects caused by the relative motion between emitters and detectors in
spaces with affine connections and metrics used as models of space or of
space-time. On the basis of the notions of centrifugal (centripetal) and
Coriolis velocities and accelerations the notions of aberration, standard
(longitudinal) and transversal Doppler effects, and standard and transversal
Hubble effect are introduced and considered. It is shown that the reasons
for aberration, Doppler effect, and Hubble effect could be not only relative
velocities between an emitter and a detector but also relative accelerations
between them. It is shown that the Hubble effect is nothing more than the
Doppler effect with explicitly given structures of the relative velocities
and relative accelerations. By the use of the Hubble law, leading to the
introduction of the Hubble effect, some connections between the kinematic
characteristics of the relative velocity and the relative acceleration, on
the one side, and the Doppler effects, the Hubble effect, and the
aberration, on the other side, are investigated.

The aberration, the Doppler effects, and the Hubble effects are considered
on the grounds of purely kinematic considerations. It should be stressed
that the Hubble functions $H$ and $\overline{H}_c$ are introduced on a
purely kinematic basis related to the notions of relative centrifugal
(centripetal) velocity and to the notions of Coriolis velocities
respectively. Their dynamic interpretations in a theory of gravitation
depend on the structures of the theory and the relations between the field
equations and on both the functions. In this paper it is shown that notions
the specialists use to apply in theories of gravitation and cosmological
models could have a good kinematic grounds independent of any concrete
classical field theory. Aberration, Doppler effects, and Hubble effects
could be used in mechanics of continuous media and in other classical field
theories in the same way as the standard Doppler effect is used in classical
and relativistic mechanics.


\begin{thebibliography}{99}
\bibitem{Manoff-1} Manoff S., \textit{Invariant projections of
energy-momentum tensors for field theories in spaces with affine connection
and metric}.\textbf{\ }J. Math. Phys. \textbf{32} (1991) 3, 728-734

\bibitem{Manoff-1a} Manoff S., Lazov R.,\textit{\ Invariant projections and
covariant divergency of the energy-momentum tensors.} In \textit{Aspects of
Complex Analysis, Differential Geometry and Mathematical Physics}. Eds. S.
Dimiev, K. Sekigava, (World Scientific, Singapore, 1999), pp. 289-314
[Extended version: E-print (1999) arXiv: gr-qc/99 07 085]

\bibitem{Manoff-2} Manoff S., \textit{Geometry and Mechanics in Different
Models of Space-Time}: \textit{Geometry and Kinematics}. (Nova Science
Publishers, New York, 2002)

\bibitem{Manoff-2a} Manoff S., \textit{Geometry and Mechanics in Different
Models of Space-Time}: \textit{Dynamics and Applications}. (Nova Science
Publishers, New York, 2002)

\bibitem{Manoff-7} Manoff S., \textit{Null vector fields in spaces with
affine connections and metrics. Doppler effect, Hubble effect, and
aberration effect}. E-print gr-qc/03 05 028

\bibitem{Stephani} Stephani H., \textit{Allgemeine Relativitaetstheorie}
(VEB Deutscher Verlag d. Wissenschaften, Berlin, 1977), pp. 75-76

\bibitem{Ehlers} Ehlers J., \textit{Beitraege zur relativistischen Mechanik
kontinuierlicher Medien}. Abhandlungen d. Mainzer Akademie d.
Wissenschaften, Math.-Naturwiss. Kl. Nr. \textbf{11 }(1961)

\bibitem{Misner} Misner Ch. W., Thorne K. S., Wheeler J. A., \textit{%
Gravitation. }(W. H. Freeman and Company, San Francisco, 1973). Russian
translation: Vol 1., Vol. 2., Vol. 3. (Mir, Moscow, 1977)

\bibitem{Manoff-3} Manoff S., \textit{Kinematics of vector fields}. In 
\textit{Complex Structures and Vector Fields}. eds. Dimiev St., Sekigawa K.
(World Sci. Publ., Singapore, 1995), pp. 61-113

\bibitem{Manoff-3a} Manoff S., \textit{Spaces with contravariant and
covariant affine connections and metrics.} Physics of elementary particles
and atomic nucleus (Physics of Particles and Nuclei) [Russian Edition: 
\textbf{30} (1999) 5, 1211-1269], [English Edition: \textbf{30} (1999) 5,
527-549]

\bibitem{Manoff-4} Manoff S., \textit{Frames of reference in spaces with
affine connections and metrics}. Class. Quantum Grav. \textbf{18} (2001) 6,
1111-1125. E-print (1999) gr-qc/99 08 061

\bibitem{Manoff-6} Manoff S., \textit{Centrifugal (centripetal), Coriolis
velocities, accelerations, and Hubble law in spaces with affine connections
and metrics}. E-print (2002) gr-qc/0212038; Central European J. of Physics
(to appear)

\bibitem{Weinberg} Weinberg S., \textit{Gravitation and cosmology:
Principles and applications of the general theory of relativity}. (John
Wiley and Sons, New York, 1972)

\bibitem{Unzicker} Unzicker A., Galaxies as Rotating Buckets - a Hypothesis
on the Gravitational Constant Based on Mach's Principle. E-print gr-qc/03 08
087
\end{thebibliography}
\end{document}